\documentclass[aps, prb, reprint]{revtex4-2} 
\usepackage{amsmath}
\usepackage{amssymb}

\usepackage{graphicx} 
\usepackage{bm}
\usepackage{hyperref}

\usepackage{multirow}

\usepackage{txfonts} 

\usepackage{color}

\begin{document}

\title{
Spin-Point-Group Classification of Multipoles for Nonrelativistic Collinear Magnets
}

\date{\today}

\author{Yuuki Ogawa and Satoru Hayami}
\affiliation{Graduate School of Science, Hokkaido University, Sapporo 060-0810, Japan}

\begin{abstract}
Multipole moments provide a unified symmetry language for describing electronic degrees of freedom and their associated physical responses.
Although active multipoles have been systematically classified for crystallographic and magnetic point groups, their classification in spin point groups, which naturally describe magnetic systems in the absence of spin--orbit coupling, has remained unexplored.
In this work, we present a complete classification of active orbital and spin multipoles for all 32 nonmagnetic and 122 collinear spin point groups.
Treating orbital and spin degrees of freedom independently, we identify the symmetry-allowed multipoles associated with nonmagnetic and collinear magnetic orderings and clarify their hierarchy through comparisons among the corresponding spin point groups.
The resulting classification provides a symmetry-based database that distinguishes structural and magnetic contributions to active multipoles and directly identifies microscopic order parameters, including those responsible for $d$-, $g$-, and $i$-wave altermagnetism.
Furthermore, by classifying response tensors within the same framework, we establish a correspondence between active multipoles and symmetry-allowed physical responses.
This correspondence systematically distinguishes nonrelativistic responses that survive without spin--orbit coupling from those requiring relativistic effects, providing a unified framework for understanding and predicting spin-dependent electromagnetic and transport phenomena in magnetic materials.
\end{abstract}

\maketitle

\section{Introduction  \label{sec:intro}}

Multipoles provide a unified framework for describing symmetry breaking in condensed matter systems and have become an indispensable tool for understanding electronic and magnetic orders and their associated physical responses~\cite{kuramoto2009multipole, Santini_RevModPhys.81.807, suzuki2018first, kusunose2022generalization, hayami2024unified}.
Originally introduced to characterize electronic and magnetic degrees of freedom, the multipole concept has subsequently been generalized to a wide variety of symmetry-breaking phenomena, including spin-orbital ordering~\cite{Hayami_PhysRevB.98.165110, Watanabe_PhysRevB.98.245129, kuniyoshi2026sixteenfold} and unconventional superconductivity~\cite{Nomoto_PhysRevB.94.174513, Sumita_PhysRevResearch.2.033225, kirikoshi2024classification}.
The symmetry properties of multipoles determine the allowed electric, magnetic, transport, optical, and mechanical responses, establishing a direct connection between microscopic electronic degrees of freedom and macroscopic observables. 
Based on this concept, systematic classifications of multipoles have been developed for crystallographic and magnetic point groups, providing a powerful framework for identifying symmetry-allowed physical responses in condensed matter systems~\cite{Hayami_PhysRevB.98.165110, Watanabe_PhysRevB.98.245129, Yatsushiro_PhysRevB.104.054412}. 
Such classifications have successfully explained a broad range of cross-correlated phenomena, such as the anomalous Hall effect~\cite{Suzuki_PhysRevB.95.094406, yamasaki2020augmented, hayami2021essential, kimata2021x}, magnetoelectric effect~\cite{Spaldin_0953-8984-20-43-434203, Hayami_PhysRevB.90.081115, thole2018magnetoelectric, Shitade_PhysRevB.98.020407, Gao_PhysRevB.97.134423}, piezomagnetic effect~\cite{Watanabe_PhysRevB.96.064432, Ogawa2025}, and nonlinear transport responses~\cite{Watanabe_PhysRevResearch.2.043081, Yatsushiro_PhysRevB.105.155157, Kirikoshi_PhysRevB.107.155109, miyamoto2025nonlinear, sudo2026large}, as well as in guiding the search for functional materials.

The conventional multipole classification is formulated within the framework of magnetic point groups~\cite{Yatsushiro_PhysRevB.104.054412}, where spin and spatial degrees of freedom are coupled and assumed to rotate simultaneously. 
This description is appropriate when spin--orbit coupling (SOC) is taken into account. 
In the absence of SOC, however, spin rotations become independent of spatial symmetry operations, magnetic point groups no longer provide a complete symmetry description. 
Consequently, although magnetic-point-group symmetry determines whether a physical response is symmetry allowed in systems with SOC, it cannot determine whether the response survives in the nonrelativistic limit. 
More fundamentally, the existing multipole classification itself does not distinguish whether an active multipole represents a relativistic response requiring SOC or a genuine nonrelativistic response.
A symmetry framework capable of treating spin and spatial degrees of freedom independently is therefore required for systematically understanding physical responses without SOC.

The significance of this issue has become particularly evident with the recent emergence of altermagnetism and related nonrelativistic spin-dependent phenomena. 
Altermagnets exhibit momentum-dependent spin-split electronic band structures despite vanishing net magnetization, and many unconventional physical responses have been shown to persist even without SOC~\cite{Noda2016, Okugawa2018, Ahn2019, Naka2019, Hayami_YK2019, Hayami_PhysRevB.102.144441, Yuan2020, Naka_PhysRevB.103.125114, Yuan2021PRMat, Yuan2021PRB, Gonzalez-Hernandez2021, Smejkal2022PRX1, Smejkal2022PRX3, Smejkal2022PRX4}. 
Within the multipole framework, the characteristic $d$-wave altermagnetic state has been identified as a magnetic octupole order, demonstrating that multipoles provide a natural descriptor of nonrelativistic magnetic order~\cite{Bhowal_PhysRevX.14.011019, McClarty_PhysRevLett.132.176702, schiff2025collinear, oike2025thermodynamic,sato2026quantum, sato2025orbital}.
More recently, the scope of nonrelativistic spin-dependent phenomena has expanded beyond the original concept of altermagnetism to a broader class of collinear magnets preserving symmetries such as $\mathcal{T}$~\cite{Matsuda2025} or combined $\mathcal{PT}$ symmetry~\cite{Hayami_PhysRevB.106.024405, Kondo_PhysRevResearch.4.013186, ogawa2026nonrelativistic}, 
where a variety of unconventional transport and spin responses have been predicted.
These developments have demonstrated that nonrelativistic spin-dependent functionalities are much broader than originally anticipated and have highlighted the necessity of identifying the symmetry principles governing such phenomena.

In the absence of SOC, the appropriate symmetry description is provided by spin point groups (SPGs), which treat spin and spatial operations independently. 
Although spin point groups have long been established in magnetic group theory, they were initially developed primarily for the classification and characterization of magnetic structures, spin waves, and neutron-scattering properties, rather than for electronic band structures or transport phenomena~\cite{Naish1962,Kitz1965,Brinkman1966_spinwave,Brinkman1966,Litvin1973,Litvin_Opechowski1974,Litvin1977}. 
Motivated by the recent discovery of spin-dependent phenomena without SOC, SPGs have only been applied to the symmetry analysis of physical responses~\cite{Liu2022, Watanabe2024, Xiao2024, Chen2024,Jiang2024, Chen2025, SciPostPhys.18.3.109,Etxebarria2025, Elcoro2026Automatic,Elcoro2026_arXiv:2606.19254}.
However, a comprehensive multipole classification based on SPGs, together with its connection to physical responses, has not yet been established. 
Such a classification would provide a direct symmetry criterion for identifying responses originating from magnetic order that remain active even in the nonrelativistic limit.

In this work, we complete the multipole classification for all 122 collinear SPGs and all 32 nonmagnetic (paramagnetic) SPGs. 
By comparing the active multipoles of a collinear SPG with those of the corresponding nonmagnetic SPG obtained by removing the magnetic order, we identify the multipoles induced solely by magnetic ordering. 
This comparison establishes a direct symmetry criterion for determining which physical responses survive in the absence of SOC.
Our framework unifies the symmetry analysis of collinear magnetic structures in the nonrelativistic limit and provides a systematic basis for exploring nonrelativistic functionalities in magnetic materials.

The remainder of this paper is organized as follows.
In Sec.~\ref{sec:M_MT_representation}, we introduce nonmagnetic and collinear SPGs and summarize their symmetry classification.
Section~\ref{sec:multipole} presents the classification of four types of orbital multipoles together with spin multipoles within the framework of SPGs.
Based on this formulation, Sec.~\ref{sec:active_MP} identifies the active multipoles for all nonmagnetic and collinear SPGs.
Section~\ref{sec:res} establishes the correspondence between active multipoles and symmetry-allowed physical responses, thereby providing a symmetry criterion for identifying responses that remain active in the absence of SOC.
Finally, Sec.~\ref{sec:summary} summarizes the main results and discusses their implications.
In Appendix~\ref{sec:ap_real}, we show the multipole notation under cubic and hexagonal systems.
Appendix~\ref{sec:ap_hexagonal} presents tensor expression in hexagonal/trigonal systems.
Collinear SPGs with non-unitary nontrivial groups and their halving unitary subgroups are listed in Appendix~\ref{sec:ap_SPG_subgroup}.
Appendix~\ref{sec:ap_classification} provides multipole classification based on 122 collinear SPGs and 32 nonmagnetic SPGs.
At the end, Appendix~\ref{sec:ap_Laue} shows the correspondence between collinear SPGs and collinear spin Laue groups.

\section{Nonmagnetic and collinear spin point groups \label{sec:M_MT_representation}} 
In this section, we overview SPGs, which provide the appropriate symmetry framework for magnetic systems in the absence of SOC.
Unlike conventional magnetic point groups, 
SPGs treat spin and orbital (real-space) degrees of freedom independently, reflecting the fact that spin rotations remain continuous symmetries when SOC is absent.
This independent treatment enables one to distinguish physical responses originating solely from magnetic ordering from those that rely on SOC.

An element of a SPG is represented by $[S  \,\|\, R]$, where $S$ acts on spin space and $R$ acts on orbital space. 
For example, the spatial inversion $\mathcal{P}$ operation is represented as $[E \,\|\, I = \bar{1} = -\bm{1}]$ ($E$ denotes the identity operation) and the time reversal $\mathcal{T}$ operation as an ``inversion" in spin space $[\theta = 1' =-\bm{1} \,\|\, E]$. 
Although both operations reverse crystal momentum, they act on different degrees of freedom: $[E \,\|\, I]$ changes only the orbital coordinates, while $[\theta \,\|\,  E]$ changes only the spin direction.
This distinction is absent in conventional magnetic point groups but becomes essential in the nonrelativistic limit.

In general, the overall SPG $\mathcal{G}$ is expressed as a direct product of the spin-only group $\mathcal{G}_{\rm SO}$ and the nontrivial spin point group $\mathcal{G}_{\rm NT}$~\cite{Litvin_Opechowski1974};
\begin{align}
\mathcal{G} = \mathcal{G}_{\rm SO} \times \mathcal{G}_{\rm NT},
\end{align}
where $\mathcal{G}_{\rm SO}$ contains only spin-space operations, and has four types corresponding to nonmagnetic (paramagnetic), collinear, coplanar, and noncoplanar magnetic orders~\cite{Kitz1965, Litvin_Opechowski1974}.
There are 598 nontrivial SPGs $\mathcal{G}_{\rm NT}$ enumerated by Litvin~\cite{Litvin1977}.
However, subsequent work has shown that the Litvin--Opechowski classification of spin-only groups can be extended to include additional cases relevant to collinear magnetic structures with the time-reversal symmetry~\cite{Elcoro2026_arXiv:2606.19254}.
In place of modifying spin-only groups, in this work, we add $\mathcal{T}$-symmetric nontrivial SPGs for such cases.
Restricting ourselves to nonmagnetic and collinear magnetic orders, there are 32 unitary nontrivial SPGs that are common to both types of order, whereas additional 58 non-unitary nontrivial SPG and 32 $\mathcal{T}$-symmetric nontrivial SPG are unique to collinear magnetic order~\cite{Litvin1977, SciPostPhys.18.3.109, Elcoro2026_arXiv:2606.19254}:
\begin{itemize}
\item[(I)] nonmagnetic SPGs (32),
\item[(II)] collinear SPGs with $\mathcal{T}$-symmetric nontrivial groups (32),
\item[(III)] collinear SPGs with non-unitary nontrivial groups (58),
\item[(IV)] collinear SPGs with unitary nontrivial groups (32),
\end{itemize}
where the numbers in parentheses represent the number of SPGs.
Nonmagnetic SPGs refer to genuinely nonmagnetic (paramagnetic) systems and collinear SPGs with $\mathcal{T}$-symmetric nontrivial groups correspond to collinear magnetic systems preserving an effective time-reversal symmetry, such as the combined operation of time reversal and a lattice translation ($\mathcal{T}\tau$).
The key difference between the two SPGs lies in their rotational symmetry in spin space (i.e., spin-only group symmetry): while nonmagnetic systems exhibit arbitrary spin rotations about any axis, corresponding to SO(3), the spin-rotation symmetry is reduced to SO(2) in collinear systems because the collinear magnetic order selects a specific spin-polarization axis.
In the following, we use the Litvin's notation for unitay and non-unitary nontrivial groups such as 
${}^1 2/{}^1  m {}^1 \bar{3}$ and ${}^1 4/{}^1 m {}^1 \bar{3} {}^1 2/{}^1 m$, 
although they are also often denoted more compactly as ${}^1  m {}^1 \bar{3}$ and ${}^1 m {}^1 \bar{3} {}^1 m$, respectively.
For $\mathcal{T}$-symmetric nontrivial groups, we add ${}^{\bar{1}} 1$ at the right of the corresponding unitary nontrivial groups. 
For example, we denote them as
${}^1 2/{}^1  m {}^1 \bar{3} {}^{\bar{1}} 1$ and ${}^1 4/{}^1 m {}^1 \bar{3} {}^1 2/{}^1 m {}^{\bar{1}} 1$.

Supposing ${\bm G}$ is the underlying crystallographic point group. 
For a nonmagnetic system, spin rotations remain fully independent of spatial symmetry operations, and time-reversal symmetry is preserved.
Accordingly, the nonmagnetic SPG is written as 
\begin{align}
\label{eq:nonmagSPG}
\mathcal{G}
&= 
\mathcal{G}_{\mathrm{SO}}^{\rm NM} \times \mathcal{G}_{\rm NT} 
\nonumber\\ 
&= 
\mathrm{SO(3)} \times
( [E \,\|\, \bm{G}]  + [\theta \,\|\, \bm{G}] ),
\end{align}
where the spin-only group and the nontrivial SPG are 
$\mathcal{G}_{\mathrm{SO}}^{\rm NM} = \mathrm{SO(3)} \times \{ [E \,\|\, E], [\theta \,\|\, E] \}$ 
and 
$\mathcal{G}_{\rm NT} = [E \,\|\, \bm{G}]$, respectively.

Meanwhile, collinear SPGs consist of 32 $\mathcal{T}$-symmetric nontrivial SPGs, 58 non-unitary nontrivial SPGs, and 32 unitary nontrivial SPGs.
The nontrivial SPG $\mathcal{G}_{\rm NT} $ is expressed as
\begin{align}
\mathcal{G}_{\rm NT}
=
\begin{cases}
[E \,\|\, \bm{G}]
+
[\theta \,\|\, \bm{G}],
& \text{($\mathcal{T}$-symmetric)},\\[2mm]
[E \,\|\, \bm{H}]
+
[\theta \,\|\, \bm{G}-\bm{H}],
& \text{(non-unitary)},\\[2mm]
[E \,\|\, \bm{G}],
& \text{(unitary)},
\end{cases}
\label{eq:nontrivial}
\end{align}
where $\bm{H}$ is a halving unitary subgroup of $\bm{G}$.
The correspondence between the 58 non-unitary nontrivial SPGs and ${\bm H}$ is summarized in Table~\ref{table:subgroup} in Appendix~\ref{sec:ap_SPG_subgroup}.
The spin-only group for collinear magnetic order is
\begin{align}
\mathcal{G}_{\rm SO}^{\infty}
=
{\rm SO}(2)
\rtimes
\{ [E \,\|\, E], [\theta  2_{\perp\bm{n}} \,\|\, E] \},
\end{align}
where $2_{\perp\bm n}$ denotes a $\pi$ rotation about an axis perpendicular to the spin-polarization direction $\bm n$.

Combining $\mathcal{G}_{\rm SO}^{\infty}$ with $\mathcal{G}_{\rm NT}$, the overall collinear SPG is written as
\begin{align}
\mathcal{G}
&=
\mathcal{G}_{\mathrm{SO}}^\infty \times \mathcal{G}_{\rm NT} 
\nonumber\\
&=
\mathrm{SO(2)} \times
\begin{cases}
[E \,\|\, \bm{G}] + [\theta \,\|\, \bm{G} ]  \\
\, + [\theta 2_{\perp\bm{n}} \,\|\, E]([E \,\|\, \bm{G}] + [\theta \,\|\, \bm{G} ]), \\
[E \,\|\, \bm{H}] + [\theta  \,\|\, R_0] [E \,\|\, \bm{H}]  \\
\, + [\theta 2_{\perp\bm{n}} \,\|\, E]([E \,\|\, \bm{H}] + [\theta  \,\|\, R_0] [E \,\|\, \bm{H}])  ,\\
[E \,\|\, \bm{G}] + [\theta 2_{\perp\bm{n}} \,\|\, \bm{G} ], 
\end{cases}
\end{align}
where $R_0$ is an arbitrary element belonging to $\bm G-\bm H$.
Since $[\theta \,\|\, \bm{G}-\bm{H}] = [\theta \,\|\, R_0][E \,\|\, \bm H]$, the antiunitary part of the nontrivial SPG is generated by combining a single antiunitary generator $[\theta \,\|\, R_0]$ with all elements of the halving subgroup $\bm H$.

These three classes have a clear physical interpretation.
Unitary SPGs allow a finite uniform spin polarization and thus describe collinear ferromagnets, whereas non-unitary SPGs and $\mathcal{T}$-symmetric SPGs forbid a net spin polarization while allowing spin order, thereby describing collinear antiferromagnets.
Hence, the distinction between ferromagnetic and antiferromagnetic symmetries is naturally captured by the structure of the nontrivial SPG.

Altogether, the present work covers 154 SPGs, consisting of 32 nonmagnetic SPGs, 32 collinear SPGs with $\mathcal{T}$-symmetric nontrivial groups, 58 collinear SPGs with non-unitary nontrivial groups, and 32 collinear SPGs with unitary nontrivial groups. 
In the following sections, we systematically classify orbital and spin multipoles under these SPGs and establish their correspondence with symmetry-allowed physical responses, providing a unified symmetry framework for identifying nonrelativistic responses induced by collinear magnetic ordering.

\section{Four types of orbital multipoles and spin multipoles \label{sec:multipole}}

Before classifying multipoles under SPGs, we introduce orbital and spin multipoles separately.
Unlike magnetic point groups, SPGs treat orbital and spin degrees of freedom as independent symmetry spaces.
This distinction naturally leads to two complementary sets of multipoles, namely orbital and spin multipoles~\cite{ogawa2026nonrelativistic}, which constitute the basis for the symmetry classification presented in the following sections.

\begin{table}[h] 
\begin{center}
\caption{
\label{Parities}
Classification of the four types of multipoles (Electric (E): $Q_{lm}$, Magnetic (M): $M_{lm}$, Magnetic-toroidal (MT): $T_{lm}$, and Electric-toroidal (ET): $G_{lm}$) according to their spatial inversion $(\mathcal{P})$, time-reversal $(\mathcal{T})$, and combined ($\mathcal{PT}$) parities, with those activated in collinear spin point groups (Collinear SPG) indicated by checkmarks $(\checkmark)$~\cite{ogawa2026nonrelativistic}. 
Panels (a) and (b) list the orbital and spin multipoles, respectively. 
}
\begingroup
\renewcommand{\arraystretch}{1.5} 
(a) orbital multipoles $X^{\rm (o)}$
\begin{tabular}{cccccc} \hline\hline
Type
& Notation
& $\mathcal{P}$ & \, \, $\mathcal{T}$ \, \, & $\mathcal{PT}$ & Collinear SPG 
\\ \hline   
E
& $Q_{lm}^{\rm (o)}$
& $(-1)^l$ & $+1$ & $(-1)^l$ & $\checkmark$  
\\ 
M
& $M_{lm}^{\rm (o)}$
& $(-1)^{l+1}$ & $-1$ & $(-1)^l$ &    
\\ 
MT
& $T_{lm}^{\rm (o)}$
& $(-1)^l$ & $-1$ & $(-1)^{l+1}$ &  
\\ 
ET
& $G_{lm}^{\rm (o)}$
& $(-1)^{l+1}$ & $+1$ & $(-1)^{l+1}$ & $\checkmark$ 
\\ \hline\hline
\end{tabular}
\\ \vspace{12pt}
(b) spin multipoles $X^{\rm (s)} \equiv \sigma X^{\rm (o)}$ 
\begin{tabular}{cccccc} \hline\hline
Type  
& Notation
& $\mathcal{P}$ & \, \, $\mathcal{T}$ \, \, & $\mathcal{PT}$ & Collinear SPG
\\ \hline
E
& $Q_{lm}^{\rm (s)}$
& $(-1)^l$ & $-1$ & $(-1)^{l+1}$ & $\checkmark$ 
\\  
M 
& $M_{lm}^{\rm (s)}$
& $(-1)^{l+1}$ & $+1$ & $(-1)^{l+1}$ &   
\\  
MT 
& $T_{lm}^{\rm (s)}$
& $(-1)^l$ & $+1$ & $(-1)^l$ & 
\\ 
ET
& $G_{lm}^{\rm (s)}$
& $(-1)^{l+1}$ & $-1$ & $(-1)^l$ & $\checkmark$ 
\\ \hline\hline
\end{tabular}
\endgroup
\end{center}
\end{table}

The four types of orbital multipoles are defined by the quantum-mechanical operator expressions as~\cite{kusunose2008description,hayami2018microscopic,kusunose2020complete}
\begin{align}
\label{eq:E_def}
\hat{Q}_{lm}^{\mathrm{(o)}} &= -e\sum_jO_{lm}({\bm r}_j), \\
\label{eq:M_def}
\hat{M}_{lm}^{\mathrm{(o)}} &= -\mu_{\rm B} \sum_j {\bm m}_l^{\mathrm{(o)}}({\bm r}_j) \cdot {\bm \nabla} O_{lm}({\bm r}_j), \\ 
\label{eq:MT_def}
\hat{T}_{lm}^{\mathrm{(o)}} &= -\mu_{\rm B} \sum_j {\bm t}_l^{\mathrm{(o)}}({\bm r}_j) \cdot {\bm \nabla} O_{lm}({\bm r}_j), \\ 
\label{eq:ET_def}
\hat{G}_{lm}^{\mathrm{(o)}} &= -e\sum_j \sum_{\alpha\beta}^{x,y,z} g_l^{\alpha\beta \mathrm{(o)}}({\bm r}_j) \nabla_\alpha \nabla_\beta O_{lm}({\bm r}_j),
\end{align}
where $\hat{Q}_{lm}^{\mathrm{(o)}}$, $\hat{M}_{lm}^{\mathrm{(o)}}$, $\hat{T}_{lm}^{\mathrm{(o)}}$, and $\hat{G}_{lm}^{\mathrm{(o)}}$ denote the orbital 
electric (E), magnetic (M), magnetic-toroidal (MT), and electric-toroidal (ET) multipoles with the azimuthal quantum number $l$ (rank of multipoles) and magnetic quantum number $m$, respectively.
In Eqs.~\eqref{eq:E_def}--\eqref{eq:ET_def}, $-e$ and $-\mu_{\rm B}$ are the electron charge and Bohr magneton, respectively, which are taken to be unity hereafter, i.e., $-e, -\mu_{\rm B} \to 1$. 
The function $O_{lm}({\bm r})$ is defined from the spherical harmonics $Y_{lm}(\hat{\bm r})$ with $\hat{\bm r}={\bm r}/|{\bm r}|$ as 
\begin{align}
O_{lm}({\bm r}) = \sqrt{\frac{4\pi}{2l+1}}r^l Y_{lm}^* (\hat{\bm r}). 
\end{align}
The multipoles are denoted as $X_0$ for monopole ($l=0$), $(X_{x},X_{y}, X_{z})$ for dipole ($l=1$), $(X_{u},X_{v}, X_{yz}, X_{zx}, X_{xy})$ for quadrupole ($l=2$), $(X_{xyz}, X^{\alpha}_{x}, X^{\alpha}_{y}, X^{\alpha}_z, X^{\beta}_{x}, X^{\beta}_{y}, X^{\beta}_z)$ for octupole ($l=3$), and $(X_{4}, X_{4u}, X_{4v}, X_{4x}^\alpha. X_{4y}^\alpha, X_{4z}^\alpha, X_{4x}^\beta, X_{4y}^\beta, X_{4z}^\beta)$ for hexadecapole ($l=4$) following Ref.~\cite{Hayami_PhysRevB.98.165110}.
${\bm m}_l^{\mathrm{(o)}}({\bm r}_j)$, ${\bm t}^{\mathrm{(o)}}_l({\bm r}_j)$, and $g_l^{\alpha\beta \mathrm{(o)}} ({\bm r}_j)$ represent the magnetic moment, magnetic-toroidal moment, and electric-toroidal tensor, respectively, which are expressed as
\begin{align}
{\bm m}_l^{\mathrm{(o)}}({\bm r}_j) &= \frac{2{\bm l}_j}{l+1}, \\
{\bm t}_l^{\mathrm{(o)}}({\bm r}_j) &= \frac{2 {\bm r}_j \times {\bm l}_j }{(l+1)(l+2)},  \\
g_l^{\alpha\beta \mathrm{(o)}} ({\bm r}_j) &= m_l^{\alpha \mathrm{(o)}} ({\bm r}_j) \, t_l^{\beta \mathrm{(o)}}({\bm r}_j), 
\end{align}
where ${\bm l}_j$ is the dimensionless orbital angular-momentum operators of an electron at ${\bm r}_j$.
These operators describe purely orbital degrees of freedom and transform only under the orbital part of a SPG. 
The parities under the spatial inversion ($\mathcal P$) and time-reversal ($\mathcal T$) operations are
$(\mathcal P,\mathcal T)
=
[(-1)^l,+1]$
for
$Q^{\rm (o)}_{lm}$,
$[(-1)^{l+1},-1]$
for
$M^{\rm (o)}_{lm}$,
$[(-1)^l,-1]$
for
$T^{\rm (o)}_{lm}$,
and
$[(-1)^{l+1},+1]$
for
$G^{\rm (o)}_{lm}$.

In SPGs, orbital and spin operations belong to independent symmetry spaces.
Consequently, an orbital multipole and its spin counterpart generally transform according to different irreducible representations.
To describe the latter systematically, we introduce the spin multipoles by attaching the spin degree of freedom to each orbital multipole as
\begin{align}
X^{\rm (s)}  \equiv \sigma X^{\mathrm{(o)}},
\end{align}
where
\begin{align}
\sigma \equiv \bm{\sigma}\cdot\bm{n} 
\end{align}
with $\bm{n}$ being the spin-quantization axis in a collinear magnetic state.
Since $\sigma$ is even under spatial inversion and odd under time reversal, the spin multipoles inherit the same spatial-inversion parity as the corresponding orbital multipoles, whereas their time-reversal parity is reversed: 
$(\mathcal P,\mathcal T)
=
[(-1)^l,-1]$
for
$Q^{\rm (s)}_{lm}$,
$[(-1)^{l+1},+1]$
for
$M^{\rm (s)}_{lm}$,
$[(-1)^l,+1]$
for
$T^{\rm (s)}_{lm}$,
and
$[(-1)^{l+1},-1]$
for
$G^{\rm (s)}_{lm}$.
Table~\ref{Parities} summarizes the spatial-inversion, time-reversal, and combined $\mathcal{PT}$ parities of both orbital and spin multipoles.

Unlike the conventional multipole classification based on magnetic point groups, the present formulation distinguishes orbital and spin multipoles explicitly and classifies them independently under SPGs.
As summarized in Table~\ref{Parities}, only electric and electric-toroidal multipoles can become active as symmetry-breaking order parameters in collinear SPGs, whereas magnetic and magnetic-toroidal multipoles are forbidden.
This observation considerably simplifies the subsequent classification of active multipoles.
In the following sections, we classify both orbital and spin multipoles for 154 SPGs and establish their correspondence with symmetry-allowed physical responses.

\section{Active multipoles under collinear and nonmagnetic spin point groups \label{sec:active_MP}}

Having established the symmetry framework of SPGs and introduced orbital and spin multipoles, we now identify the active multipoles under all nonmagnetic and collinear SPGs.
This classification constitutes the central result of the present work.
Unlike the conventional multipole classification based on magnetic point groups~\cite{Yatsushiro_PhysRevB.104.054412}, the present classification distinguishes orbital and spin multipoles and therefore enables us to identify which multipoles emerge solely from magnetic ordering in the absence of SOC.

The key idea is to compare the active multipoles in a collinear SPG with those in the corresponding nonmagnetic SPG obtained by restoring the spin rotational symmetry.
Multipoles that newly become active through this comparison are induced exclusively by magnetic ordering and therefore characterize nonrelativistic symmetry breaking.
Since every linear-response tensor can be decomposed into multipoles with definite symmetry~\cite{Hayami_PhysRevB.98.165110,Yatsushiro_PhysRevB.104.054412}, the present classification immediately provides a symmetry criterion for determining whether a given physical response survives in the absence of SOC.

\begin{table*}[htb!]
\centering
\caption{
Active orbital and spin multipoles for nonmagnetic SPGs ($\mathcal{G}_{\mathrm{SO}}^{\mathrm{NM}} \times \mathcal{G}_{\mathrm{NT}}$) and collinear SPGs 
($\mathcal{G}_{\mathrm{SO}}^{\infty} \times \mathcal{G}_{\mathrm{NT}}$) according to $\mathcal{P}$, $\mathcal{T}$, and $\mathcal{PT}$ symmetries.
``Even'' and ``odd'' parity refer to the spatial-inversion parity of the multipoles.
E and ET stand for electric and electric-toroidal, respectively.
\label{table:type_of_active_MP}
}
\begin{tabular}{lcccccccc}
\hline \hline
SPG type & Nontrivial SPG $\mathcal{G}_{\mathrm{NT}}$ & $\mathcal{P}$ & $\mathcal{T}$ & $\mathcal{PT}$ & Even-parity &  Odd-parity  & Even-parity & Odd-parity \\ \cline{6-9}
& & & & & Orbital E, ET & Orbital E, ET & Spin E, ET & Spin E, ET \\ 
\hline 
Nonmagnetic SPG
& ${}^1 4/{}^1 m {}^1 \bar{3} {}^1 2/{}^1 m$, ${}^1 2/{}^1 m {}^1 \bar{3}$, 
& $\bigcirc$& $\bigcirc$& $\bigcirc$ & $\checkmark$ \rule[0pt]{0pt}{8pt} 
\\
& ${}^1 4/{}^1 m {}^1 m {}^1 m$,  ${}^1 4/{}^1 m$, 
\\
& ${}^1 m {}^1 m {}^1 m$, ${}^1 2/{}^1 m$, ${}^1 \bar{1}$,
\\
& ${}^1 6/{}^1 m {}^1 m {}^1 m$, ${}^1 6/{}^1 m$, ${}^1 \bar{3} {}^1 m$, ${}^1 \bar{3}$
\\
\cline{2-9}
& ${}^1 4 {}^1 3 {}^1 2$, ${}^1 \bar{4} {}^1 3 {}^1 m$, ${}^1 2 {}^1 3$, 
& $\times$& $\bigcirc$& $\times$ & $\checkmark$ & $\checkmark$ \rule[0pt]{0pt}{8pt} 
\\
& ${}^1 4 {}^1 2 {}^1 2$, ${}^1 \bar{4} {}^1 2 {}^1 m$, ${}^1 4 {}^1 m {}^1 m$, 
\\
\hline\hline
\end{tabular}
\end{table*}

Before presenting the detailed classification for 154 SPGs, it is useful to summarize the general relationship between symmetry and active multipoles.
Table~\ref{table:type_of_active_MP} classifies the types of active orbital and spin multipoles according to the presence or absence of spatial inversion ($\mathcal{P}$), time-reversal ($\mathcal{T}$), and combined $\mathcal{PT}$ symmetries.
Rather than listing individual multipoles, the table identifies which classes of multipoles can become active in each symmetry class and therefore provides a global overview of the complete classification presented below.
Since only electric (E) and electric-toroidal (ET) multipoles are active in collinear SPGs (Table~\ref{Parities}), the following discussion focuses exclusively on their orbital and spin counterparts.

For nonmagnetic SPGs, time-reversal symmetry is preserved, and only orbital electric and electric-toroidal multipoles are allowed.
When collinear magnetic ordering develops, time-reversal symmetry is often broken while the orbital symmetry remains unchanged, leading to the activation of spin electric and spin electric-toroidal multipoles.
These spin multipoles therefore represent the symmetry breaking induced solely by magnetic ordering. 
This activation does not occur in collinear SPGs with $\mathcal{T}$-symmetric nontrivial groups, where spin multipoles remain forbidden by time-reversal symmetry.

While time-reversal symmetry determines whether orbital or spin multipoles can become active, spatial inversion determines their parity.
Accordingly, centrosymmetric SPGs allow only even-parity multipoles, whereas noncentrosymmetric SPGs permit both even- and odd-parity multipoles.
For non-unitary SPGs, the combined $\mathcal{PT}$ symmetry provides an additional symmetry constraint in addition to $\mathcal P$ and $\mathcal T$, leading to different combinations of active orbital and spin multipoles.

Table~\ref{table:type_of_active_MP} serves as a roadmap for the detailed classifications presented in the following subsections.
Once the symmetry of a target material is identified, the table immediately indicates the classes of orbital and spin multipoles that can be activated, even before consulting the complete classification tables.
It therefore provides a convenient starting point for identifying possible order parameters, constructing symmetry-allowed Landau free-energy expansions, and predicting nonrelativistic electromagnetic and transport responses.
Most importantly, the comparison between the nonmagnetic and collinear SPGs immediately reveals which spin and orbital multipoles are induced exclusively by magnetic ordering, providing the basis for identifying physical responses that survive in the absence of SOC.

\renewcommand{\arraystretch}{1.4}
\begin{table*}[htb!]
\centering
\caption{
Active even-parity orbital electric (E) and electric-toroidal (ET) multipoles for centrosymmetric nonmagnetic SPGs and $\mathcal{T}$-symmetric collinear SPGs. 
The triclinic point groups $\mathcal{G}_{\mathrm{SO}}^{\mathrm{NM}} \times {}^1 \bar{1}$ and $\mathcal{G}_{\mathrm{SO}}^{\infty} \times {}^1 \bar{1}  {}^{\bar{1}} 1$, in which all the even-parity orbital E and ET multipoles are active, are omitted.
 \label{table:active_MP_nonmagSPG_1}}
\vspace{2mm}
\begin{tabular}{ll cc@{\hspace{3mm}}cccc}
\hline\hline
&&  \multicolumn{6}{c}{Even-parity} 
\\ 
\cline{3-8} 

&& \multicolumn{3}{c}{Orbital E} &&  \multicolumn{2}{c}{Orbital ET}  
\\ 
\cline{3-5} \cline{7-8} 

Nonmagnetic SPG
&
Collinear SPG
& $l=0$ & $2$ & $4$ &&
$1$ & $3$ 
\\ 
\hline

$\mathcal{G}_{\mathrm{SO}}^{\mathrm{NM}} \times {}^1 4/{}^1 m {}^1 \bar{3} {}^1 2/{}^1 m$
&
$\mathcal{G}_{\mathrm{SO}}^{\infty} \times  {}^1 4/{}^1 m {}^1 \bar{3} {}^1 2/{}^1 m {}^{\bar{1}} 1$ 
& $Q_0^{\mathrm{(o)}}$ &  & $Q_4^{\mathrm{(o)}}$ &&
& 
\\

$\mathcal{G}_{\mathrm{SO}}^{\mathrm{NM}} \times {}^1 2/{}^1 m {}^1 \bar{3}$
&
$\mathcal{G}_{\mathrm{SO}}^{\infty} \times {}^1 2/{}^1 m {}^1 \bar{3} {}^{\bar{1}} 1$
& $Q_0^{\mathrm{(o)}}$ & & $Q_4^{\mathrm{(o)}}$ &&
& $G_{xyz}^{\mathrm{(o)}}$
\\

$\mathcal{G}_{\mathrm{SO}}^{\mathrm{NM}} \times {}^1 4/{}^1 m {}^1 m {}^1 m$
&
$\mathcal{G}_{\mathrm{SO}}^{\infty} \times {}^1 4/{}^1 m {}^1 m {}^1 m {}^{\bar{1}} 1$
& $Q_0^{\mathrm{(o)}}$ & $Q_u^{\mathrm{(o)}}$ & $Q_4^{\mathrm{(o)}}, Q_{4u}^{\mathrm{(o)}}$ &&
& 
\\

$\mathcal{G}_{\mathrm{SO}}^{\mathrm{NM}} \times {}^1 4/{}^1 m$
&
$\mathcal{G}_{\mathrm{SO}}^{\infty} \times {}^1 4/{}^1 m {}^{\bar{1}} 1$
& $Q_0^{\mathrm{(o)}}$ & $Q_u^{\mathrm{(o)}}$ & $Q_4^{\mathrm{(o)}}, Q_{4u}^{\mathrm{(o)}}, Q_{4z}^{\alpha \mathrm{(o)}}$ &&
$G_z^{\mathrm{(o)}}$ & $G_z^{\alpha \mathrm{(o)}}$ 
\\

$\mathcal{G}_{\mathrm{SO}}^{\mathrm{NM}} \times {}^1 m {}^1 m {}^1 m$
&
$\mathcal{G}_{\mathrm{SO}}^{\infty} \times {}^1 m {}^1 m {}^1 m {}^{\bar{1}} 1$
& $Q_0^{\mathrm{(o)}}$ & $Q_u^{\mathrm{(o)}}$, $Q_{v}^{\mathrm{(o)}}$ & $Q_4^{\mathrm{(o)}}$, $Q_{4u}^{\mathrm{(o)}}$, $Q_{4v}^{\mathrm{(o)}}$ &&
& $G_{xyz}^{\mathrm{(o)}}$ 
\\
\hline\hline
\end{tabular}
\end{table*}

\renewcommand{\arraystretch}{1.4}
\begin{table*}[htb!]
\centering
\caption{
Active even-parity orbital and spin electric (E) and electric-toroidal (ET) multipoles for centrosymmetric collinear SPGs with unitary nontrivial group.
The triclinic point group $\mathcal{G}_{\mathrm{SO}}^{\infty} \times {}^1 \bar{1}$, in which all the even-parity E and ET multipoles are active, is omitted.
 \label{table:active_MP_collinearSPG_unitary_1}}
\vspace{2mm}
\begin{tabular}{lcc@{\hspace{3mm}}cccc}
\hline\hline
&  \multicolumn{6}{c}{Even-parity} 
\\ 
\cline{2-7} 

& \multicolumn{3}{c}{E} &&  \multicolumn{2}{c}{ET}  
\\ 
\cline{2-4} \cline{6-7} 

Collinear SPG
& $l=0$ & $2$ & $4$ &&
$1$ & $3$ 
\\ 
\hline

$\mathcal{G}_{\mathrm{SO}}^{\infty} \times {}^1 4/{}^1 m {}^1 \bar{3} {}^1 2/{}^1 m$  
& $Q_0^{\mathrm{(o)}}$ &  & $Q_4^{\mathrm{(o)}}$ &&
& 
\\
& $Q_0^{\mathrm{(s)}}$ &  & $Q_4^{\mathrm{(s)}}$ &&
& 
\\

$\mathcal{G}_{\mathrm{SO}}^{\infty} \times {}^1 2/{}^1 m {}^1 \bar{3}$  
& $Q_0^{\mathrm{(o)}}$ & & $Q_4^{\mathrm{(o)}}$ &&
& $G_{xyz}^{\mathrm{(o)}}$
\\
& $Q_0^{\mathrm{(s)}}$ & & $Q_4^{\mathrm{(s)}}$ &&
& $G_{xyz}^{\mathrm{(s)}}$
\\

$\mathcal{G}_{\mathrm{SO}}^{\infty} \times {}^1 4/{}^1 m {}^1 m {}^1 m$  
& $Q_0^{\mathrm{(o)}}$ & $Q_u^{\mathrm{(o)}}$ & $Q_4^{\mathrm{(o)}}, Q_{4u}^{\mathrm{(o)}}$ &&
& 
\\
\hline\hline
\end{tabular}
\end{table*}

\subsection{Centrosymmetric spin point groups}

We first discuss centrosymmetric SPGs, in which spatial inversion symmetry is preserved.
The active multipoles for the corresponding nonmagnetic SPGs and collinear SPGs with $\mathcal{T}$-symmetric nontrivial groups are summarized in Table~\ref{table:active_MP_nonmagSPG_1}, and those for collinear SPGs with unitary nontrivial groups are summarized in Table~\ref{table:active_MP_collinearSPG_unitary_1}.
Since inversion symmetry is retained, only even-parity multipoles are symmetry allowed, making the role of magnetic ordering particularly transparent.

Table~\ref{table:active_MP_nonmagSPG_1} lists the active orbital electric and electric-toroidal multipoles for centrosymmetric nonmagnetic SPGs and collinear SPGs with $\mathcal{T}$-symmetric nontrivial groups. 
Because time-reversal symmetry is preserved, no spin multipoles appear. 
The active orbital multipoles are determined solely by the crystallographic symmetry and therefore describe structural order parameters independent of magnetic ordering.

Table~\ref{table:active_MP_collinearSPG_unitary_1} presents the corresponding classification for centrosymmetric collinear SPGs with unitary nontrivial group.
The orbital multipoles are identical to those in the corresponding nonmagnetic SPGs, while additional spin electric and electric-toroidal multipoles with the same spatial symmetry become active.
These spin multipoles originate exclusively from spontaneous magnetic ordering and thus provide the symmetry-breaking order parameters characteristic of nonrelativistic collinear magnets.

The one-to-one correspondence between the orbital and spin multipoles greatly simplifies the identification of magnetic order parameters.
For a given crystallographic point group, the active orbital multipoles are first determined from the corresponding nonmagnetic SPG.
The additional spin multipoles appearing in the collinear SPG then directly identify the symmetry breaking induced by magnetic ordering.

As a representative example, let us consider the cubic point group $4/m\bar{3}2/m$.
In the nonmagnetic SPG, the active multipoles consist of the orbital electric monopole $Q_0^{\mathrm{(o)}}$ and the hexadecapole $Q_4^{\mathrm{(o)}}$, as shown in Table~\ref{table:active_MP_nonmagSPG_1}.
Upon the onset of collinear ferromagnetic ordering, the corresponding spin multipoles, $Q_0^{\mathrm{(s)}}$ and $Q_4^{\mathrm{(s)}}$, become active, while the orbital multipoles remain unchanged, as shown in Table~\ref{table:active_MP_collinearSPG_unitary_1}.
The same procedure applies to all centrosymmetric SPGs.
Such one-to-one correspondence clearly demonstrates that the spin multipoles provide a direct fingerprint of magnetic symmetry breaking without modifying the underlying crystallographic symmetry.

\renewcommand{\arraystretch}{1.4}
\begin{table*}[htb!]
\centering
\caption{
Active orbital electric (E) and electric-toroidal (ET) multipoles for noncentrosymmetric nonmagnetic SPGs and collinear SPGs with $\mathcal{T}$-symmetric nontrivial groups.
The triclinic point groups $\mathcal{G}_{\mathrm{SO}}^{\mathrm{NM}} \times {}^1 1$ and $\mathcal{G}_{\mathrm{SO}}^{\infty} \times {}^1 1  {}^{\bar{1}} 1$, in which all the orbital E and ET multipoles are active, are omitted.
 \label{table:active_MP_nonmagSPG_2}}
\vspace{2mm}
\begin{tabular}{l@{\hspace{5mm}}l@{\hspace{3mm}}ccccccccccccc}
\hline \hline
&
& \multicolumn{6}{c}{Even-parity} && \multicolumn{6}{c}{Odd-parity} 
\\
\cline{3-8} \cline{10-15}
&
& \multicolumn{3}{c}{Orbital E} & & \multicolumn{2}{c}{Orbital ET} & & \multicolumn{2}{c}{Orbital E} & & \multicolumn{3}{c}{Orbital ET} 
\\
\cline{3-5} \cline{7-8} \cline{10-11} \cline{13-15}

Nonmagnetic SPG
&
Collinear SPG
& $l=0$ & $2$ & $4$ && $1$ & $3$ && 
$1$ & $3$ && $0$ & $2$ & $4$ 
\\
\hline

$\mathcal{G}_{\mathrm{SO}}^{\mathrm{NM}} \times {}^1 4 {}^1 3 {}^1 2$
&
$\mathcal{G}_{\mathrm{SO}}^{\infty} \times {}^1 4 {}^1 3 {}^1 2 {}^{\bar{1}} 1 $
& $Q_0^{\mathrm{(o)}}$ & & $Q_4^{\mathrm{(o)}}$ &&
& &&
& &&
$G_0^{\mathrm{(o)}}$ & & $G_4^{\mathrm{(o)}}$ 
\\

$\mathcal{G}_{\mathrm{SO}}^{\mathrm{NM}} \times {}^1 \bar{4} {}^1 3 {}^1 m$
&
$\mathcal{G}_{\mathrm{SO}}^{\infty} \times {}^1 \bar{4} {}^1 3 {}^1 m {}^{\bar{1}} 1 $
& $Q_0^{\mathrm{(o)}}$ & & $Q_4^{\mathrm{(o)}}$ &&
& &&
& $Q_{xyz}^{\mathrm{(o)}}$ &&
& &  
\\

$\mathcal{G}_{\mathrm{SO}}^{\mathrm{NM}} \times {}^1 2 {}^1 3$
&
$\mathcal{G}_{\mathrm{SO}}^{\infty} \times {}^1 2 {}^1 3 {}^{\bar{1}} 1 $
& $Q_0^{\mathrm{(o)}}$ & & $Q_4^{\mathrm{(o)}}$ &&
& $G_{xyz}^{\mathrm{(o)}}$ &&
& $Q_{xyz}^{\mathrm{(o)}}$ &&
$G_0^{\mathrm{(o)}}$ & & $G_4^{\mathrm{(o)}}$ 
\\

$\mathcal{G}_{\mathrm{SO}}^{\mathrm{NM}} \times {}^1 4 {}^1 2 {}^1 2$
&
$\mathcal{G}_{\mathrm{SO}}^{\infty} \times {}^1 4 {}^1 2 {}^1 2 {}^{\bar{1}} 1 $
& $Q_0^{\mathrm{(o)}}$ & $Q_u^{\mathrm{(o)}}$ & $Q_4^{\mathrm{(o)}}$, $Q_{4u}^{\mathrm{(o)}}$ &&
& &&
& &&
$G_0^{\mathrm{(o)}}$ & $G_u^{\mathrm{(o)}}$ & $G_4^{\mathrm{(o)}}$, $G_{4u}^{\mathrm{(o)}}$
\\

$\mathcal{G}_{\mathrm{SO}}^{\mathrm{NM}} \times {}^1 \bar{4} {}^1 2 {}^1 m$
&
$\mathcal{G}_{\mathrm{SO}}^{\infty} \times {}^1 \bar{4} {}^1 2 {}^1 m {}^{\bar{1}} 1 $
& $Q_0^{\mathrm{(o)}}$ & $Q_u^{\mathrm{(o)}}$ & $Q_4^{\mathrm{(o)}}$, $Q_{4u}^{\mathrm{(o)}}$ &&
& &&
& $Q_{xyz}^{\mathrm{(o)}}$ &&
& $G_v^{\mathrm{(o)}}$ & $G_{4v}^{\mathrm{(o)}}$ 
\\
\hline\hline
\end{tabular}
\end{table*}

\renewcommand{\arraystretch}{1.2}
\begin{table*}[htb!]
\centering
\caption{
Active orbital and spin electric (E) and electric-toroidal (ET) multipoles for noncentrosymmetric collinear SPGs with unitary nontrivial groups.
The triclinic point group $\mathcal{G}_{\mathrm{SO}}^{\infty} \times {}^1 1$, in which all the E and ET multipoles are active, is omitted.
 \label{table:active_MP_collinearSPG_unitary_2}}
\vspace{2mm}
\begin{tabular}{lccccccccccccc}
\hline \hline
& \multicolumn{6}{c}{Even-parity} && \multicolumn{6}{c}{Odd-parity} 
\\
\cline{2-7} \cline{9-14}

& \multicolumn{3}{c}{E} & & \multicolumn{2}{c}{ET} & & \multicolumn{2}{c}{E} & & \multicolumn{3}{c}{ET} 
\\
\cline{2-4} \cline{6-7} \cline{9-10} \cline{12-14}

Collinear SPG
& $l=0$ & $2$ & $4$ && $1$ & $3$ && 
$1$ & $3$ && $0$ & $2$ & $4$ 
\\
\hline

$\mathcal{G}_{\mathrm{SO}}^{\infty} \times {}^1 4 {}^1 3 {}^1 2$ 
& $Q_0^{\mathrm{(o)}}$, $Q_0^{\mathrm{(s)}}$ & & $Q_4^{\mathrm{(o)}}$, $Q_4^{\mathrm{(s)}}$ &&
& &&
& &&
$G_0^{\mathrm{(o)}}$, $G_0^{\mathrm{(s)}}$ & & $G_4^{\mathrm{(o)}}$, $G_4^{\mathrm{(s)}}$ 
\\
   
$\mathcal{G}_{\mathrm{SO}}^{\infty} \times {}^1 \bar{4} {}^1 3 {}^1 m$ 
& $Q_0^{\mathrm{(o)}}$, $Q_0^{\mathrm{(s)}}$ & & $Q_4^{\mathrm{(o)}}$, $Q_4^{\mathrm{(s)}}$ &&
& &&
& $Q_{xyz}^{\mathrm{(o)}}$, $Q_{xyz}^{\mathrm{(s)}}$ &&
& &  
\\

$\mathcal{G}_{\mathrm{SO}}^{\infty} \times {}^1 2 {}^1 3$ 
& $Q_0^{\mathrm{(o)}}$, $Q_0^{\mathrm{(s)}}$ & & $Q_4^{\mathrm{(o)}}$, $Q_4^{\mathrm{(s)}}$ &&
& $G_{xyz}^{\mathrm{(o)}}$, $G_{xyz}^{\mathrm{(s)}}$ &&
& $Q_{xyz}^{\mathrm{(o)}}$, $Q_{xyz}^{\mathrm{(s)}}$ &&
$G_0^{\mathrm{(o)}}$, $G_0^{\mathrm{(s)}}$ & & $G_4^{\mathrm{(o)}}$, $G_4^{\mathrm{(s)}}$ \\

$\mathcal{G}_{\mathrm{SO}}^{\infty} \times {}^1 4 {}^1 2 {}^1 2$ 
& $Q_0^{\mathrm{(o)}}$, $Q_0^{\mathrm{(s)}}$ & $Q_u^{\mathrm{(o)}}$, $Q_u^{\mathrm{(s)}}$ & $Q_4^{\mathrm{(o)}}$, $Q_4^{\mathrm{(s)}}$ &&
& &&
& &&
$G_0^{\mathrm{(o)}}$, $G_0^{\mathrm{(s)}}$ & $G_u^{\mathrm{(o)}}$, $G_u^{\mathrm{(s)}}$ & $G_4^{\mathrm{(o)}}$, $G_4^{\mathrm{(s)}}$
\\
&  &  & $Q_{4u}^{\mathrm{(o)}}$, $Q_{4u}^{\mathrm{(s)}}$ &&
& &&
& &&
&  & $G_{4u}^{\mathrm{(o)}}$, $G_{4u}^{\mathrm{(s)}}$  
\\
\hline\hline
\end{tabular}
\end{table*}

\subsection{Noncentrosymmetric spin point groups}

We next discuss noncentrosymmetric SPGs, for which spatial inversion symmetry is absent.
The active multipoles for the corresponding nonmagnetic SPGs and collinear SPGs with $\mathcal{T}$-symmetric nontrivial groups are summarized in Table~\ref{table:active_MP_nonmagSPG_2}, and those for collinear SPGs with unitary nontrivial groups are summarized in Table~\ref{table:active_MP_collinearSPG_unitary_2}.
As in the centrosymmetric case discussed above, the comparison between the two tables separates multipoles originating from the underlying crystal symmetry from those induced by collinear magnetic ordering.
The essential difference is that both even- and odd-parity multipoles are allowed once inversion symmetry is broken.

Table~\ref{table:active_MP_nonmagSPG_2} lists the active orbital electric and electric-toroidal multipoles for noncentrosymmetric nonmagnetic SPGs and $\mathcal{T}$-symmetric collinear SPGs.
Since time-reversal symmetry is preserved, spin multipoles are absent, as in Table~\ref{table:active_MP_nonmagSPG_1}.
However, in contrast to centrosymmetric SPGs, odd-parity orbital multipoles become active in addition to even-parity ones.
These odd-parity multipoles characterize structural inversion-symmetry breaking: odd-parity electric multipoles describe polar degrees of freedom, whereas odd-parity electric-toroidal multipoles provide descriptors of chiral structural degrees of freedom, depending on the crystallographic symmetry.

Table~\ref{table:active_MP_collinearSPG_unitary_2} presents the corresponding classification for noncentrosymmetric collinear SPGs with unitary nontrivial groups.
Compared with Table~\ref{table:active_MP_nonmagSPG_2}, the orbital multipoles remain unchanged, while the corresponding spin electric and electric-toroidal multipoles become active.
Thus, the newly activated spin multipoles again provide fingerprints of magnetic symmetry breaking.
This leads to a richer multipolar structure than in centrosymmetric systems, because spin multipoles can appear in both even- and odd-parity sectors.

As a representative polar example, consider the tetragonal point group $4mm$.
In the nonmagnetic SPG, the odd-parity orbital electric dipole $Q_z^{\mathrm{(o)}}$ is active, reflecting the polar axis along the $z$ direction.
When collinear ferromagnetic ordering develops, the corresponding spin electric dipole $Q_z^{\mathrm{(s)}}$ becomes active.
Thus, $Q_z^{\mathrm{(o)}}$ characterizes the structural polarity, whereas $Q_z^{\mathrm{(s)}}$ characterizes the magnetic contribution with the same spatial symmetry.

As a representative chiral example, consider the cubic point group $432$.
Although this point group has no polar axis, the odd-parity electric-toroidal multipoles such as $G_0^{\mathrm{(o)}}$ and $G_4^{\mathrm{(o)}}$ are active, reflecting the chiral nature of the crystal structure.
In the corresponding collinear SPG with unitary nontrivial group, their spin counterparts, such as $G_0^{\mathrm{(s)}}$ and $G_4^{\mathrm{(s)}}$, also become active.
This example demonstrates that polar and chiral symmetry breaking can be classified within the same framework while separating orbital and spin origins.

The activation of odd-parity multipoles is particularly important for physical responses.
Odd-parity electric multipoles are associated with polar responses, whereas odd-parity electric-toroidal multipoles are closely related to chiral responses.
Therefore, Tables~\ref{table:active_MP_nonmagSPG_2} and~\ref{table:active_MP_collinearSPG_unitary_2} provide a useful guide for identifying candidate noncentrosymmetric materials exhibiting nonreciprocal transport, current-induced spin polarization, and chirality-dependent spin phenomena~\cite{Hayami_PhysRevB.106.014420, Matsuda2025}.
As discussed in the next section, these response properties can be systematically inferred from the active multipoles listed in the tables.

\renewcommand{\arraystretch}{1.3}
\begin{table*}[htb!]
\centering
\caption{
Active even-parity electric (E) and electric-toroidal (ET) multipoles for centrosymmetric collinear SPG.
\label{table:active_MP_collinearSPG_nonunitary_1}}
\vspace{2mm}
\begin{tabular}{lccccccccccccc}\hline \hline
& \multicolumn{6}{c}{Even-parity} & & \multicolumn{6}{c}{Even-parity} \\ \cline{2-7} \cline{9-14}
&  \multicolumn{3}{c}{Orbital E} && \multicolumn{2}{c}{Orbital ET} &&  \multicolumn{3}{c}{Spin E} &&  \multicolumn{2}{c}{Spin ET} \\
\cline{2-4}
\cline{6-7}
\cline{9-11}
\cline{13-14}

$\mathcal{G}_{\mathrm{NT}}$
& $l=0$ & $2$ & $4$ && 
$1$ & $3$ && 
$0$ & $2$ & $4$ && 
$1$ & $3$  \\
\hline
 
${}^{\bar{1}} 4/{}^1 m {}^1 \bar{3} {}^{\bar{1}} 2/{}^{\bar{1}} m$  
& $Q_0^{\mathrm{(o)}}$ &  & $Q_4^{\mathrm{(o)}}$ &&
& && 
& & &&
&  $G_{xyz}^{\mathrm{(s)}}$ \rule[0pt]{0pt}{8pt}
\\

${}^1 4/{}^1 m {}^{\bar{1}} m {}^{\bar{1}} m$  
& $Q_0^{\mathrm{(o)}}$ & $Q_u^{\mathrm{(o)}}$ & $Q_4^{\mathrm{(o)}}$, $Q_{4u}^{\mathrm{(o)}}$ &&
& && 
& & $Q_{4z}^{\alpha \mathrm{(s)}}$ && 
$G_z^{\mathrm{(s)}}$ & $G_z^{\alpha \mathrm{(s)}}$ 
\\

${}^{\bar{1}} 4/{}^1 m {}^1 m {}^{\bar{1}} m$  
& $Q_0^{\mathrm{(o)}}$ & $Q_u^{\mathrm{(o)}}$ & $Q_4^{\mathrm{(o)}}$, $Q_{4u}^{\mathrm{(o)}}$ &&
& && 
& $Q_v^{\mathrm{(s)}}$ & $Q_{4v}^{\mathrm{(s)}}$ &&
& $G_{xyz}^{\mathrm{(s)}}$
\\

${}^{\bar{1}} 4/{}^1 m$  
& $Q_0^{\mathrm{(o)}}$ & $Q_u^{\mathrm{(o)}}$ & $Q_4^{\mathrm{(o)}}$, $Q_{4u}^{\mathrm{(o)}}$ &&
$G_z^{\mathrm{(o)}}$ & $G_z^{\alpha \mathrm{(o)}}$ &&
& $Q_v^{\mathrm{(s)}}$ & $Q_{4v}^{\mathrm{(s)}}$ &&
& $G_{xyz}^{\mathrm{(s)}}$
\\
& & & $Q_{4z}^{\alpha \mathrm{(o)}}$ &&
& &&
& $Q_{xy}^{\mathrm{(s)}}$ & $Q_{4z}^{\beta \mathrm{(s)}}$ &&
& $G_z^{\beta \mathrm{(s)}}$ 
\\
\hline\hline
\end{tabular}
\end{table*}

\renewcommand{\arraystretch}{1.3}
\begin{table*}[htb!]
\centering
\caption{
Active even-parity orbital and spin electric (E) and electric-toroidal (ET) multipoles for noncentrosymmetric collinear SPGs with non-unitary nontrivial group with the $\mathcal{PT}$ symmetry.
The triclinic point group $\mathcal{G}_{\mathrm{SO}}^{\infty} \times {}^{\bar{1}} \bar{1}$ is omitted, where all the even-parity E and ET multipoles are active.
\label{table:active_MP_collinearSPG_nonunitary_2}}
\vspace{2mm}
\begin{tabular}{lccccccccccccc}
\hline\hline
& \multicolumn{6}{c}{Even-parity} && \multicolumn{6}{c}{Odd-parity}
\\
\cline{2-7} \cline{9-14}

& \multicolumn{3}{c}{Orbital E} &&  \multicolumn{2}{c}{Orbital ET} && \multicolumn{2}{c}{Spin E} && \multicolumn{3}{c}{Spin ET} 
\\
\cline{2-4} \cline{6-7} \cline{9-10} \cline{12-14}

$\mathcal{G}_{\mathrm{NT}}$
& $l=0$ & $2$ & $4$ && $1$ & $3$ && $1$ & $3$ && $0$ & $2$ & $4$  
\\
\hline
 
${}^1 4/ {}^{\bar{1}} m {}^{\bar{1}} \bar{3} {}^1 2/ {}^{\bar{1}} m$ 
& $Q_0^{\mathrm{(o)}}$ & & $Q_4^{\mathrm{(o)}}$ && 
& &&
& && 
$G_0^{\mathrm{(s)}}$ & & $G_4^{\mathrm{(s)}}$ \rule[0pt]{0pt}{8pt} 
\\

${}^{\bar{1}} 4/ {}^{\bar{1}} m {}^{\bar{1}} \bar{3} {}^{\bar{1}} 2 /{}^1 m$ 
& $Q_0^{\mathrm{(o)}}$ & & $Q_4^{\mathrm{(o)}}$ && 
& &&
& $Q_{xyz}^{\mathrm{(s)}}$ && 
& & 
\\
  
${}^1 2/ {}^{\bar{1}} m {}^{\bar{1}} \bar{3}$ 
& $Q_0^{\mathrm{(o)}}$ & & $Q_4^{\mathrm{(o)}}$ && 
& $G_{zyz}^{\mathrm{(o)}}$ &&
& $Q_{xyz}^{\mathrm{(s)}}$ && 
$G_0^{\mathrm{(s)}}$ & & $G_4^{\mathrm{(s)}}$ 
\\

${}^1 4/{}^{\bar{1}} m {}^{\bar{1}} m {}^{\bar{1}} m$ 
& $Q_0^{\mathrm{(o)}}$ & $Q_u^{\mathrm{(o)}}$ & $Q_4^{\mathrm{(o)}}$, $Q_{4u}^{\mathrm{(o)}}$ && 
& &&
& && 
$G_0^{\mathrm{(s)}}$ & $G_u^{\mathrm{(s)}}$ & $G_{4}^{\mathrm{(s)}}$, $G_{4u}^{\mathrm{(s)}}$ 
\\

${}^{\bar{1}} 4/{}^{\bar{1}} m {}^{\bar{1}} m {}^1 m$ 
& $Q_0^{\mathrm{(o)}}$ & $Q_u^{\mathrm{(o)}}$ & $Q_4^{\mathrm{(o)}}$, $Q_{4u}^{\mathrm{(o)}}$  && 
& &&
& $Q_{xyz}^{\mathrm{(s)}}$ && 
& $G_{v}^{\mathrm{(s)}}$ & $G_{4v}^{\mathrm{(s)}}$  
\\
\hline\hline
\end{tabular}
\end{table*}

\renewcommand{\arraystretch}{1.3}
\begin{table*}[htb!]
\centering
\caption{
Active orbital and spin electric (E) and electric-toroidal (ET) multipoles for noncentrosymmetric collinear SPGs with non-unitary nontrivial group without the $\mathcal{PT}$ symmetry.
\label{table:active_MP_collinearSPG_nonunitary_3}}
\vspace{2mm}
\begin{tabular}{lccccccccccccccccccccccccccc}\hline \hline
& \multicolumn{6}{c}{Even-parity} && \multicolumn{6}{c}{Odd-parity} && \multicolumn{6}{c}{Even-parity} && \multicolumn{6}{c}{Odd-parity} \\
\cline{2-7}
\cline{9-14}
\cline{16-21}
\cline{23-28}

 & \multicolumn{3}{c}{Orbital E} && \multicolumn{2}{c}{Orbital ET} && \multicolumn{2}{c}{Orbital E} && \multicolumn{3}{c}{Orbital ET} 
 && \multicolumn{3}{c}{Spin E} && \multicolumn{2}{c}{Spin ET} && \multicolumn{2}{c}{Spin E} && \multicolumn{3}{c}{Spin ET}\\
 \cline{2-4} \cline{6-7}
 \cline{9-10} \cline{12-14}
 \cline{16-18} \cline{20-21}
 \cline{23-24} \cline{26-28}
 
 $\mathcal{G}_{\mathrm{NT}}$ 
 & $l=0$ & $2$ & $4$ && 
 $1$ & $3$ && 
 $1$ & $3$ && 
 $0$ & $2$ & $4$ && 
 $0$ & $2$ & $4$ && 
 $1$ & $3$ && 
 $1$ & $3$ && 
 $0$ & $2$ & $4$ \\
 \hline
 
${}^{\bar{1}} \bar{4} {}^1 3 {}^{\bar{1}} m$ 
& $Q_0^{\mathrm{(o)}}$ & & $Q_4^{\mathrm{(o)}}$ && 
& &&
& $Q_{xyz}^{\mathrm{(o)}}$ && 
& & &&
& & && 
& $G_{xyz}^{\mathrm{(s)}}$ &&
& && 
$G_0^{\mathrm{(s)}}$ & & $G_4^{\mathrm{(s)}}$ \rule[0pt]{0pt}{8pt}
\\
  
${}^{\bar{1}} 4 {}^1 3 {}^{\bar{1}} 2$ 
& $Q_0^{\mathrm{(o)}}$ & & $Q_4^{\mathrm{(o)}}$ && 
& &&
& && 
$G_0^{\mathrm{(o)}}$ & & $G_4^{\mathrm{(o)}}$ && 
& & && 
& $G_{xyz}^{\mathrm{(s)}}$ &&
& $Q_{xyz}^{\mathrm{(s)}}$ && 
& & \\

${}^1 4 {}^{\bar{1}} 2 {}^{\bar{1}} 2$ 
& $Q_0^{\mathrm{(o)}}$ & $Q_u^{\mathrm{(o)}}$  & $Q_4^{\mathrm{(o)}}$, $Q_{4u}^{\mathrm{(o)}}$ &&
& &&
& && 
$G_0^{\mathrm{(o)}}$ & $G_u^{\mathrm{(o)}}$ & $G_{4}^{\mathrm{(o)}}$, $G_{4u}^{\mathrm{(o)}}$ && 
& & $Q_{4z}^{\alpha \mathrm{(s)}}$ && 
$G_z^{\mathrm{(s)}}$ & $G_z^{\alpha \mathrm{(s)}}$ &&
$Q_z^{\mathrm{(s)}}$ & $Q_z^{\alpha \mathrm{(s)}}$ && 
& & $G_{4z}^{\alpha \mathrm{(s)}}$ 
\\
\hline\hline
\end{tabular}
\end{table*}

\subsection{Non-unitary spin point groups}

We finally discuss the 58 non-unitary collinear SPGs, which describe collinear magnetic structures possessing antiunitary symmetry operations.
This class includes a wide variety of collinear antiferromagnets and plays a central role in identifying nonrelativistic responses originating from antiferromagnetic order.
In contrast to the unitary SPGs discussed above, antiunitary operations constrain orbital and spin multipoles in a nontrivial manner, leading to characteristic patterns of active multipoles.

The complete classification is given in Tables~\ref{table:active_MP_collinearSPG_nonunitary_1}--\ref{table:active_MP_collinearSPG_nonunitary_3}.
According to the presence or absence of spatial inversion symmetry and combined $\mathcal{PT}$ symmetry, the non-unitary SPGs are divided into three cases:
centrosymmetric groups, noncentrosymmetric groups preserving $\mathcal{PT}$ symmetry, and noncentrosymmetric groups lacking $\mathcal{PT}$ symmetry.
This classification clarifies how antiunitary symmetry controls the allowed combinations of orbital and spin electric and electric-toroidal multipoles in collinear antiferromagnets.

Table~\ref{table:active_MP_collinearSPG_nonunitary_1} summarizes the centrosymmetric non-unitary SPGs.
Because spatial inversion symmetry is preserved, only even-parity multipoles are active.
Compared with the corresponding unitary SPGs, the antiunitary operations reduce the allowed spin multipoles and select only those compatible with the antiferromagnetic symmetry.
Thus, the active spin multipoles listed in this table serve as symmetry fingerprints of centrosymmetric collinear antiferromagnetic order.

Table~\ref{table:active_MP_collinearSPG_nonunitary_2} shows the noncentrosymmetric non-unitary SPGs that preserve combined $\mathcal{PT}$ symmetry.
Once inversion symmetry is broken, odd-parity multipoles can also become symmetry allowed in principle.
However, the remaining $\mathcal{PT}$ symmetry imposes a strong constraint on the allowed multipoles.
Since only $\mathcal{PT}$-even multipoles can be active, even-parity orbital electric and electric-toroidal multipoles and odd-parity spin electric and electric-toroidal multipoles are allowed, whereas odd-parity orbital multipoles and even-parity spin multipoles are forbidden.
This class therefore corresponds to antiferromagnets that break inversion symmetry but retain the combined $\mathcal{PT}$ symmetry.

The most general case is shown in Table~\ref{table:active_MP_collinearSPG_nonunitary_3}, where both inversion and $\mathcal{PT}$ symmetries are absent.
In this case, even- and odd-parity orbital and spin multipoles can be simultaneously active.
These SPGs therefore possess the richest multipolar degrees of freedom among the collinear SPGs considered here.
Such systems are expected to provide particularly broad platforms for nonrelativistic electromagnetic, transport, and spin responses.

The physical meaning of these tables is most clearly understood by comparison with the nonmagnetic and unitary cases.
Orbital multipoles common to the corresponding nonmagnetic SPG originate from the underlying crystallographic symmetry.
In contrast, spin multipoles appearing only in the non-unitary collinear SPG are induced by antiferromagnetic ordering.
Thus, the present classification separates structural and antiferromagnetic contributions to the active multipoles in a systematic way.

An important consequence is that different antiferromagnetic structures with the same crystallographic point group can be distinguished by their active spin multipoles.
This information is not accessible from the crystallographic point group alone and is obscured in conventional magnetic-point-group classifications where spin and orbital sectors are tied together by SOC.
The SPG classification therefore provides a direct symmetry fingerprint of antiferromagnetic order in the nonrelativistic limit~\cite{Liu2022, Watanabe2024, Xiao2024, Chen2024,Jiang2024, Chen2025, SciPostPhys.18.3.109,Etxebarria2025, Elcoro2026Automatic,Elcoro2026_arXiv:2606.19254}. 

From a practical viewpoint, Tables~\ref{table:active_MP_collinearSPG_nonunitary_1}--\ref{table:active_MP_collinearSPG_nonunitary_3} constitute a database of active multipoles for collinear antiferromagnets.
Once the SPG of a target material is identified, the corresponding table immediately gives the symmetry-allowed orbital and spin multipoles.
Together with the nonmagnetic and unitary classifications, these tables enable a hierarchical comparison among structural, ferromagnetic, and antiferromagnetic symmetry breaking.
This comparison is the central idea of the present work and provides the basis for identifying physical responses that remain active in the absence of SOC, as discussed in the next section.


\renewcommand{\arraystretch}{1.8}
\begin{table*}[htb!]
\begin{center}
\caption{
All nontrivial groups $\mathcal{G}_{\mathrm{NT}}$ 
supporting $d$- and $g$-wave altermagnetic (AM) states, classified by the 
lowest-rank active even-parity spin electric (E) multipoles in $\mathcal{PT}$-breaking collinear SPGs with non-unitary nontrivial groups.
For each SPG, the corresponding character table number, active multipoles, and representative candidate materials for altermagnets are listed.
SPGs connected by ``or” are equivalent, with the distinction arising solely from the choice of crystallographic axes.
\label{table:altermag}
}
\begin{tabular}{|c|c|c|c|c|} 
\hline\hline
AM type
& $\mathcal{G}_{\mathrm{NT}}$  
& Table \#
& Multipole 
& Candidate
\\ 
\hline
(Planar) $d$-wave 
& ${}^{\bar{1}} 4 / {}^1 m {}^1 m {}^{\bar{1}} m $ or ${}^{\bar{1}} 4 / {}^1 m {}^{\bar{1}} m {}^1 m$ 
& \ref{table:Litvin195collinear} 
& $Q_{v}^{\mathrm{(s)}}$ or $Q_{xy}^{\mathrm{(s)}}$
& $\mathrm{RuO_{2}}$~\cite{Ahn2019,Smejkal2020_Crystal, Fedchenko2024_Observation}, 
$\beta\mathrm{-MnO_{2}}$~\cite{Noda2016}, 
$\mathrm{V_{2}Se_{2}O}$~\cite{Ma2021}, 
$\mathrm{KV_{2}Se_{2}O}$~\cite{Jiang2025}
\\ 
& ${}^{\bar{1}} 4 {}^1 m {}^{\bar{1}} m$ or ${}^{\bar{1}} 4 {}^{\bar{1}} m {}^1 m$  
& \ref{table:Litvin152collinear}
& $Q_{v}^{\mathrm{(s)}}$ or $Q_{xy}^{\mathrm{(s)}}$ 
& $\mathrm{SrMn_{2}V_{2}O_{8}}$~\cite{Bera2014},
$\mathrm{GdAlSi}$~\cite{Nag2024}
\\ 
& ${}^{\bar{1}} \bar{4} {}^1 2 {}^{\bar{1}} m$ or ${}^{\bar{1}} \bar{4} {}^{\bar{1}} m {}^1 2$ 
& \ref{table:Litvin171collinear}
& $Q_{v}^{\mathrm{(s)}}$ or $Q_{xy}^{\mathrm{(s)}}$  
& $\mathrm{Fe_{2}WTe_{4}}$~\cite{Tan2026}
\\ 
& ${}^{\bar{1}} \bar{4} {}^1 m {}^{\bar{1}} 2$ or ${}^{\bar{1}} \bar{4} {}^{\bar{1}} 2 {}^1 m$ 
& \ref{table:Litvin168collinear}
& $Q_{v}^{\mathrm{(s)}}$ or $Q_{xy}^{\mathrm{(s)}}$
& $\mathrm{Ca(CoN)_{2}}$~\cite{Zhang2024_PhysRevLett.133.056401}
\\ 
& ${}^{\bar{1}} 4 {}^1 2 {}^{\bar{1}} 2$ or ${}^{\bar{1}} 4 {}^{\bar{1}} 2 {}^1 2$ 
& \ref{table:Litvin136collinear}
& $Q_{v}^{\mathrm{(s)}}$ or $Q_{xy}^{\mathrm{(s)}}$ 
& 
\\  
& ${}^{\bar{1}} 4 / {}^1 m$  
& \ref{table:Litvin105collinear}
& $Q_{v}^{\mathrm{(s)}}$, $Q_{xy}^{\mathrm{(s)}}$ 
& $\alpha\mathrm{-MnO_{2}}$~\cite{Noda2016}  $\mathrm{KRu_{4}O_{8}}$~\cite{Smejkal2022PRX3}
\\  
& ${}^{\bar{1}} \bar{4}$  
& \ref{table:Litvin99collinear}
& $Q_{v}^{\mathrm{(s)}}$, $Q_{xy}^{\mathrm{(s)}}$ 
& 
\\
\hline \hline 
\end{tabular}
\end{center}
\end{table*}

\renewcommand{\arraystretch}{1.8}
\begin{table*}[htb!]
\begin{center}
\caption{
All nontrivial groups $\mathcal{G}_{\mathrm{NT}}$ supporting $i$-wave altermagnetic (AM) states, classified by the 
lowest-rank active even-parity spin electric (E) multipoles in $\mathcal{PT}$-breaking collinear SPGs with non-unitary nontrivial groups.
For each SPG, the corresponding character table number, active multipoles, and representative candidate materials for altermagnets are listed.
\label{table:altermag_iwave}
}
\begin{tabular}{|c|c|c|c|c|} 
\hline\hline
AM type
& $\mathcal{G}_{\mathrm{NT}}$  
& Table \#
& Multipole 
& Candidate
\\ 
\hline
(Planar) $i$-wave
& ${}^1 6/{}^1 m {}^{\bar{1}} m {}^{\bar{1}} m$
& \ref{table:Litvin449collinear}
& $Q_{6s}^{\mathrm{(s)}}$
& 
\\  
& ${}^1 6 {}^{\bar{1}} m {}^{\bar{1}} m$
& \ref{table:Litvin396collinear}
& $Q_{6s}^{\mathrm{(s)}}$ 
&
\\ 
& ${}^1 \bar{6} {}^{\bar{1}} m {}^{\bar{1}} 2$
& \ref{table:Litvin415collinear}
& $Q_{6s}^{\mathrm{(s)}}$
& $\mathrm{FeBr_{3}}$~\cite{APL2024_10.1063/5.0198285}
\\ 
& ${}^1 6 {}^{\bar{1}} 2 {}^{\bar{1}} 2$
& \ref{table:Litvin341collinear}
& $Q_{6s}^{\mathrm{(s)}}$
& 
\\  
(Bulk) $i$-wave
& ${}^{\bar{1}} 4/{}^1 m {}^1 \bar{3} {}^{\bar{1}} 2/{}^{\bar{1}} m$
& \ref{table:Litvin570collinear}
& $Q_{6t}^{\mathrm{(s)}}$
& $\mathrm{LiTi_{2}O_{4}}$~\cite{He2021_PhysRevB.104.045143}, 
$\mathrm{Er_{2}Ru_{2}O_{7}}$~\cite{Taira2003}
\\ 
& ${}^{\bar{1}} \bar{4} {}^1 3 {}^{\bar{1}} m$
& \ref{table:Litvin554collinear}
& $Q_{6t}^{\mathrm{(s)}}$
& $\mathrm{Ce_{4}Sb_{3}}$~\cite{Nirmala2009}
\\ 
& ${}^{\bar{1}} 4 {}^1 3 {}^{\bar{1}} 2$
& \ref{table:Litvin562collinear}
& $Q_{6t}^{\mathrm{(s)}}$
& 
\\  
\hline \hline 
\end{tabular}
\end{center}
\end{table*}

\renewcommand{\arraystretch}{1.8}
\begin{table*}[htb!]
\begin{center}
\caption{
Momentum dependence of spin-split bands in $d$-, $g$-, and $i$-wave altermagnetic (AM) states.
For each nontrivial group $\mathcal{G}_{\mathrm{NT}}$, the corresponding active multipoles and functional forms of the spin-split bands in altermagnets are listed.
$C_1$ and $C_2$ denote constants. 
\label{table:altermag_band}
}
\begin{tabular}{|c|c|@{\hspace{3mm}}c@{\hspace{3mm}}|c|} 
\hline\hline
AM type
& $\mathcal{G}_{\mathrm{NT}}$  
& Multipole 
& Spin-split band form
\\ 
\hline
(Planar) $d$-wave 
& ${}^{\bar{1}} 4 / {}^1 m {}^1 m {}^{\bar{1}} m $ or ${}^{\bar{1}} 4 / {}^1 m {}^{\bar{1}} m {}^1 m$,
${}^{\bar{1}} 4 {}^1 m {}^{\bar{1}} m$ or ${}^{\bar{1}} 4 {}^{\bar{1}} m {}^1 m$,  
& $Q_{v}^{\mathrm{(s)}}$ or $Q_{xy}^{\mathrm{(s)}}$
& $(k_x^2 - k_y^2)\sigma$ \, or \,  $k_x k_y \sigma$
\\ 
& ${}^{\bar{1}} \bar{4} {}^1 2 {}^{\bar{1}} m$ or ${}^{\bar{1}} \bar{4} {}^{\bar{1}} m {}^1 2$, 
${}^{\bar{1}} \bar{4} {}^1 m {}^{\bar{1}} 2$ or ${}^{\bar{1}} \bar{4} {}^{\bar{1}} 2 {}^1 m$, 
${}^{\bar{1}} 4 {}^1 2 {}^{\bar{1}} 2$ or ${}^{\bar{1}} 4 {}^{\bar{1}} 2 {}^1 2$ 
& 
&  
\\ 
& ${}^{\bar{1}} 4 / {}^1 m$,
${}^{\bar{1}} \bar{4}$ ,
${}^{\bar{1}} 4$ 
& $Q_{v}^{\mathrm{(s)}}$, $Q_{xy}^{\mathrm{(s)}}$
& $[ C_1 (k_x^2 - k_y^2) + C_2 k_x k_y ] \sigma$ 
\\  
& ${}^{\bar{1}} m {}^{\bar{1}} m {}^1 m$,
${}^{\bar{1}} 2 {}^{\bar{1}} 2 {}^1 2$,
${}^{\bar{1}} m {}^{\bar{1}} m {}^1 2$ 
& $Q_{xy}^{\mathrm{(s)}}$
& $k_x k_y \sigma$
\\ 

(Bulk) $d$-wave 
& ${}^{\bar{1}} m {}^1 m {}^{\bar{1}} 2$ or ${}^1 m {}^{\bar{1}} m {}^{\bar{1}} 2$
& $Q_{zx}^{\mathrm{(s)}}$ or $Q_{yz}^{\mathrm{(s)}}$
& $k_z k_x \sigma$ \, or \, $k_y k_z \sigma$
\\ 
& ${}^{\bar{1}} 2 / {}^{\bar{1}} m $,
${}^{\bar{1}} m $, 
${}^{\bar{1}} 2 $
& $Q_{yz}^{\mathrm{(s)}}$, $Q_{xy}^{\mathrm{(s)}}$
& $[ C_1  k_y k_z + C_2 k_x k_y ] \sigma$ 
\\ 
\hline 
(Planar) $g$-wave 
& ${}^1 4 / {}^1 m {}^{\bar{1}} m {}^{\bar{1}} m$,
${}^1 4 {}^{\bar{1}} m {}^{\bar{1}} m$,
${}^1 \bar{4} {}^{\bar{1}} 2 {}^{\bar{1}} m$,
${}^1 4 {}^{\bar{1}} 2 {}^{\bar{1}} 2$,
& $Q_{4z}^{\alpha \mathrm{(s)}}$ 
& $k_x k_y(k_x^2 - k_y^2)\sigma$
\\
\hline \hline 
\end{tabular}
\end{center}
\end{table*}

\subsection{Application to the classification of altermagnets}

The present classification can be applied directly to identify SPGs hosting altermagnetic electronic structures.
The spin splitting characteristic of $d$- and $g$-wave altermagnets is described by even-parity spin electric multipoles with $l=2$ and $4$, respectively.
By inspecting the active multipoles in the above tables, one can systematically extract all $\mathcal{PT}$-breaking collinear non-unitary SPGs that allow these multipoles.
In addition, among the $\mathcal{PT}$-breaking collinear SPGs with non-unitary nontrivial groups, the seven groups that are not associated with either $d$- or $g$-wave altermagnets correspond to $i$-wave altermagnets, which are characterized by rank-6 spin electric multipoles. 
We further identify the active rank-6 spin electric multipoles for these seven SPGs.

The resulting classification is summarized in Tables~\ref{table:altermag} and ~\ref{table:altermag_iwave}, which lists the corresponding nontrivial groups, the relevant character table numbers, the active spin electric multipoles, and representative candidate materials.
These tables provide a practical guide for identifying not only candidate altermagnetic materials but also the symmetry and multipolar origin of their spin splitting.
Once the SPG is specified, the active spin electric multipoles immediately determine whether the corresponding altermagnetic spin splitting is of the $d$-, $g$-, or $i$-wave type as shown in Table~\ref{table:altermag_band}.

\subsection{Overview and practical use of the classification}

The classifications presented in Tables~\ref{table:active_MP_nonmagSPG_1}--\ref{table:active_MP_collinearSPG_nonunitary_3} constitute a complete database of active orbital and spin multipoles for all nonmagnetic and collinear SPGs.
Taken together, they provide a unified symmetry framework for describing structural, ferromagnetic, and antiferromagnetic order parameters in the absence of SOC.

A central advantage of the present classification is that it enables a hierarchical comparison among the corresponding nonmagnetic, $\mathcal{T}$-symmetric, unitary, and non-unitary SPGs sharing the same crystallographic symmetry.
The orbital multipoles common to all four groups originate from the underlying crystal symmetry, whereas the spin multipoles appearing only in the collinear SPGs represent symmetry breaking induced by magnetic ordering.
Furthermore, comparison between unitary and non-unitary SPGs distinguishes multipoles associated with ferromagnetic and antiferromagnetic order, respectively.
The present framework therefore separates structural, ferromagnetic, and antiferromagnetic contributions to the active multipoles in a systematic manner.

The tables also provide a practical route for identifying microscopic order parameters.
After determining the SPG from the crystal and magnetic structures of a material, one can immediately identify all symmetry-allowed orbital and spin multipoles from the corresponding table.
Comparing the collinear SPG with its nonmagnetic counterpart then isolates the multipoles induced solely by magnetic ordering.
Since this procedure relies only on symmetry, it is independent of any specific microscopic Hamiltonian or first-principles calculation.

The present classification further establishes a direct connection between magnetic symmetry and physical responses.
Each active multipole identified in the tables corresponds to a set of symmetry-allowed response tensors, while the observation of a particular response constrains the possible active multipoles and hence the symmetry of the underlying magnetic state.
The active-multipole tables therefore provide a practical starting point for both identifying microscopic order parameters and predicting symmetry-allowed physical responses.

In the next section, we make this correspondence explicit by classifying linear response tensors under 154 SPGs. 
Combined with the active-multipole tables presented here, the response classification provides a unified framework for identifying and predicting nonrelativistic electromagnetic, transport, and spin phenomena.
Unlike the conventional magnetic-point-group approach, the present framework not only identifies the relevant order parameters but also clarifies whether the corresponding physical responses survive in the absence of SOC.

\section{Responses under active multipoles \label{sec:res}} 

In this section, we show the correspondence between the active multipoles classified in the preceding section and symmetry-allowed physical responses. 
An active multipole determines the tensor components of macroscopic responses that can appear in the nonrelativistic limit. 
Thus, the present formulation provides a direct route from the SPG of a collinear magnetic structure to experimentally observable responses without SOC.
First, we discuss the relation between response tensors and multipoles based on the point-group symmetry in Sec.~\ref{sec:res_symmetry}.
We then illustrate the usefulness of this correspondence for two representative collinear antiferromagnetic structures: a $\mathcal{PT}$-symmetric antiferromagnet and a $d$-wave altermagnet that breaks both $\mathcal{T}$ and $\mathcal{PT}$. 
These examples demonstrate that active spin multipoles directly determine nonrelativistic spin-dependent responses, such as nonlinear spin-current generation, spin conductivity, and spin piezomagnetic effects.

\subsection{Correspondence between tensor component and multipole \label{sec:res_symmetry}}

We consider a general response tensor defined by
\begin{align}
B^{[n_B]} = \chi^{[n_{B} \times n_{F}]}_{\mathrm{(o/s)}} F^{[n_F]}, 
\end{align}
where $B^{[n_B]}$ and $F^{[n_F]}$ are the rank-$n_{B}$ (output) response and the rank-$n_{F}$ external (input) field, respectively. 
We distinguish between the response tensors $\chi^{[n_{B} \times n_{F}]}_{\mathrm{(o)}}$ and $\chi^{[n_{B} \times n_{F}]}_{\mathrm{(s)}}$; the former/latter one corresponds to active orbital/spin multipoles.

When both the input and output quantities belong to the orbital sector, the corresponding response tensor is denoted by 
$\chi^{[n_B\times n_F]}_{\mathrm{(o)}}$ and is described by active orbital multipoles. 
The same correspondence applies when both quantities belong to the spin sector.
Typical examples are the dielectric susceptibility, electric conductivity, orbital magnetic susceptibility, elastic responses, and orbital-current responses.
In contrast, when the response couples an orbital quantity to a spin quantity, the tensor is described by active spin multipoles. 
Examples include the spin magnetoelectric effect, spin conductivity, nonlinear spin conductivity, and spin piezomagnetic effect.
This distinction is a central advantage of the SPG formulation, because it separates responses originating from the orbital/crystal structure from those induced by collinear magnetic order itself.

\renewcommand{\arraystretch}{1.5}
\begin{table*}[htb!]
\centering
\caption{Correspondence of external fields and responses to multipoles. 
The spatial-inversion parity of the external field or the response is shown in the column of $\mathcal{P}$.
In the column of multipole, $X_{lm}^{\mathrm{(o/s)}}$ ($l=0,1,2$) means the rank-$l$ orbital/spin multipole with $X=Q,G,M,T$.
 \label{table:field_response_MP}}
\begin{tabular}{ccccc}
\hline \hline
$n_F$ & $\mathcal{P}$ & External field & Multipole \\ \hline
0
& $+$ & Isotropic stress tr[$\tau_{ij}$] & Orbital E monopole ($Q_{0}^{\mathrm{(o)}}$)\\
& & Spin magnetic field $H_{S}$($=\bm{n}\cdot\bm{H}_{S}$) &  Spin E monopole ($Q_{0}^{\mathrm{(s)}}$)\\
1 
& $+$ & Orbital magnetic field ${\bm H}_{L}$ &  Orbital M dipole ($M_{1m}^{\mathrm{(o)}}$)\\
& & Rotational distortion ${\bm \theta}$ & Orbital ET dipole ($G_{1m}^{\mathrm{(o)}}$)\\
& $-$ & Electric field ${\bm E}$ & Orbital E dipole ($Q_{1m}^{\mathrm{(o)}}$)\\
& & Thermal gradient $\nabla T$ & Orbital E dipole ($Q_{1m}^{\mathrm{(o)}}$)\\
2
& $+$ & (Symmetric) stress $\bm{\tau}$($\tau_{ij}$) & Orbital E quadrupole ($Q_{2m}^{\mathrm{(o)}}$)\\
\hline
$n_B$  & & Response &  \\ \hline
0
& $+$ & Isotropic strain tr[$\varepsilon_{ij}$] & Orbital E monopole ($Q_{0}^{\mathrm{(o)}}$)\\
& & Spin magnetization $S(= \bm{n}\cdot\bm{S})$ &  Spin E monopole ($Q_{0}^{\mathrm{(s)}}$)\\
& $-$ & Orbital current monopole tr[$J_i^{L_\alpha}$] & Orbital ET monopole ($G_0^{\mathrm{(o)}}$)\\
1 
& $+$ & Orbital magnetization ${\bm L}$ & Orbital M dipole ($M_{1m}^{\mathrm{(o)}}$)\\
& & Rotation ${\bm \omega}$& Orbital ET dipole ($G_{1m}^{\mathrm{(o)}}$)\\
& $-$ & Electric polarization ${\bm P}$ & Orbital E dipole ($Q_{1m}^{\mathrm{(o)}}$)\\
& & Orbital current pseudovector $\bm{J}^L$($\epsilon_{ki\alpha} J_i^{L_\alpha}$) & Orbital E dipole ($Q_{1m}^{\mathrm{(o)}}$)\\
& & Electric current ${\bm J}$ & Orbital MT dipole ($T_{1m}^{\mathrm{(o)}}$)\\
& & Thermal current ${\bm J}^{\rm Q}$ & Orbital MT dipole ($T_{1m}^{\mathrm{(o)}}$)\\
& & Spin current ${\bm J}^S$ & Spin E dipole ($Q_{1m}^{\mathrm{(s)}}$)\\
2 
& $+$ & Symmetric strain $\bm{\varepsilon}$($\varepsilon_{ij}$) & Orbital E quadrupole ($Q_{2m}^{\mathrm{(o)}}$)\\
& $-$ & (Symmetric) orbital current $\bm{J}^L$($J_i^{L_\alpha} + J_{\alpha}^{L_i}$) & Orbital ET quadrupole ($G_{2m}^{\mathrm{(o)}}$)\\
\hline\hline
\end{tabular}
\end{table*}

The quantities considered in this work are assigned to these two sectors as follows.
In the orbital sector, representative external fields include the electric field $\bm E$, orbital magnetic field $\bm{H}_L$, symmetric stress tensor $\tau_{ij}$, rotational distortion $\bm\theta=(\bm\nabla\times\bm u)/2$, and their products, where $\bm u$ is the displacement vector. 
Representative orbital responses include the electric polarization $\bm P$, orbital magnetization $\bm L$, electric and thermal currents $\bm J$ and $\bm J^{\rm Q}$, orbital current $J_i^{L_\alpha}$, symmetric strain $\varepsilon_{ij}=(\partial_i u_j+\partial_j u_i)/2$, and rotational response $\bm\omega$. 
In the spin sector, typical quantities are the spin magnetic field $H_S=\bm H_S\cdot\bm n$, spin magnetization $S=\bm S\cdot\bm n$, and spin current $J_i^S=\sigma J_i$, where $\bm n$ denotes the spin-quantization axis of the collinear magnetic structure.
Each of these external fields and responses can be assigned to a multipole with the same transformation property. 
For example, the electric field $\bm E$ transforms as an electric dipole, whereas the symmetric strain tensor $\varepsilon_{ij}$ decomposes into an electric monopole and electric quadrupoles. 
The representative correspondences are summarized in Table~\ref{table:field_response_MP}; the upper and lower panels show the multipole assignments of external fields and responses, respectively.
Hereafter, we focus on the response tensor in cubic, tetragonal, orthorhombic, monoclinic, and triclinic systems, and those in hexagonal and trigonal systems are summarized in Appendix~\ref{sec:ap_hexagonal}.

\subsubsection{Rank-1 tensor \label{sec:res_symmetry_rank1}}
The rank-1 response tensor $\chi^{[0\times 1]}$ for a scalar response $B^{[0]}=(B)$ with $n_{B}=0$ and a vector field $F^{[1]}=(F_x, F_y, F_z)$ with $n_{F}=1$ is related to the dipole $(X_x^{\mathrm{(o/s)}}, X_y^{\mathrm{(o/s)}}, X_z^{\mathrm{(o/s)}})$ as
\begin{align}
\label{eq:chi01}
\chi^{[0\times 1]}_{\mathrm{(o/s)}}
=
\begin{pmatrix}
X_x^{\mathrm{(o/s)}} \quad
X_y^{\mathrm{(o/s)}} \quad
X_z^{\mathrm{(o/s)}} 
\end{pmatrix}, 
\end{align}
where $X^{\mathrm{(o/s)}}$ stands for the polar multipoles ($Q^{\mathrm{(o/s)}}$) [axial multipoles ($G^{\mathrm{(o/s)}}$)] when $\chi^{[0\times 1]}_{\mathrm{(o/s)}}$ is the polar (axial) tensor.
The (spin) magnetoelectric effect where the spin magnetization $S$ is induced by the electric field as $S = \sum_i \alpha_i E_i$, is described by such a rank-1 response tensor. 
As $S$ corresponds to the spin electric monopole ($Q_0^{\mathrm{(s)}}$), the tensor component 
$\alpha_i$ is described by the spin electric dipole $(Q_x^{\mathrm{(s)}}, Q_y^{\mathrm{(s)}}, Q_z^{\mathrm{(s)}})$.

\subsubsection{Rank-2 tensor \label{sec:res_symmetry_rank2}}
We consider two types of rank-2 tensors, $\chi^{[1\times1]}_{\mathrm{(o/s)}}$ and $\chi^{[0\times2]}_{\mathrm{(o/s)}}$.
The tensor $\chi^{[1\times1]}_{\mathrm{(o/s)}}$ connects a vector response $B^{[1]}=(B_x, B_y, B_z)$ to a vector field 
$F^{[1]}=(F_x, F_y, F_z)$, which is related with the rank-0 to 2 multipoles as
monopole $X_0^{\mathrm{(o/s)}}$, dipole $(Y_x^{\mathrm{(o/s)}}, Y_y^{\mathrm{(o/s)}}, Y_z^{\mathrm{(o/s)}})$, and quadrupole $(X_u^{\mathrm{(o/s)}}, X_v^{\mathrm{(o/s)}}, X_{yz}^{\mathrm{(o/s)}}, X_{zx}^{\mathrm{(o/s)}}, X_{xy}^{\mathrm{(o/s)}})$.
The tensor component of $\chi^{[1\times1]}_{\mathrm{(o/s)}}$ is given by
\begin{widetext}
\begin{align}
\label{eq:rank11}
\chi^{[1\times1]}_{\mathrm{(o/s)}} 
&= 
\begin{pmatrix}
X_0^{\mathrm{(o/s)}} - X_u^{\mathrm{(o/s)}} + X_v^{\mathrm{(o/s)}} & 
X_{xy}^{\mathrm{(o/s)}} + Y_z^{\mathrm{(o/s)}} & 
X_{zx}^{\mathrm{(o/s)}} - Y_y^{\mathrm{(o/s)}} 
\\
X_{xy}^{\mathrm{(o/s)}} - Y_z^{\mathrm{(o/s)}} &
X_0^{\mathrm{(o/s)}} - X_u^{\mathrm{(o/s)}} - X_v^{\mathrm{(o/s)}} & 
X_{yz}^{\mathrm{(o/s)}} + Y_x^{\mathrm{(o/s)}} 
\\
X_{zx}^{\mathrm{(o/s)}} + Y_y^{\mathrm{(o/s)}} & 
X_{yz}^{\mathrm{(o/s)}} - Y_x^{\mathrm{(o/s)}} & 
X_0^{\mathrm{(o/s)}} + 2X_u^{\mathrm{(o/s)}} \\
\end{pmatrix},
\end{align}
\end{widetext}
where $X^{\mathrm{(o/s)}}=Q^{\mathrm{(o/s)}}$ ($G^{\mathrm{(o/s)}}$) and $Y^{\mathrm{(o/s)}}=G^{\mathrm{(o/s)}}$ ($Q^{\mathrm{(o/s)}}$) for the polar (axial) tensor.
The symmetric part of a polar (axial) rank-2 tensor is described by electric (electric-toroidal) monopole and quadrupole components, while its antisymmetric part is described by electric-toroidal (electric) dipole components. 
Typical examples are the dielectric susceptibility $(F^{[1]}=\bm E,B^{[1]}=\bm P)$ and the spin conductivity $(F^{[1]}=\bm E,B^{[1]}=\bm J^S)$. 
For the former orbital response, the relevant multipoles are $Q^{\mathrm{(o)}}$ and $G^{\mathrm{(o)}}$, whereas, for the latter spin conductivity, they are $Q^{\mathrm{(s)}}$ and $G^{\mathrm{(s)}}$.
Axial rank-2 tensors, such as the orbital magnetoelectric tensor relating $\bm E$ to $\bm L$ or $\bm H_L$ to $\bm P$, are obtained by interchanging electric and electric-toroidal multipoles.
Here, it should be noted that these tensor decompositions are general and apply to arbitrary input and output quantities; the tensor components may be further constrained by additional properties. 
For example, the $\mathcal{T}$-even electric-toroidal dipoles $G_i^{\mathrm{(o)}}$ ($i = x, y, z$), which correspond to the antisymmetric part of a polar tensor $\chi^{[1\times1]}_{\mathrm{(o)}}$, do not contribute to the electric conductivity, even when they are active.

The tensor $\chi^{[0\times2]}_{\mathrm{(o/s)}}$ is another rank-2 tensor for $B^{[0]}=(B)$ and $F^{[2]}=(F_{xx}, F_{yy}, F_{zz}, F_{yz}, F_{zx}, F_{xy})$ where $F_{ij}=F_{ji}$.  
As $F^{[2]}$ is decomposed into the monopole and quadrupole components, the tensor component of $\chi^{[0\times2]}_{\mathrm{(o/s)}}$ is given by
\begin{align}
\label{eq:rank20}
\chi^{[0\times2]}_{\mathrm{(o/s)}}
=
\begin{pmatrix}
X_0^{\mathrm{(o/s)}} - X_{u}^{\mathrm{(o/s)}} + X_{v}^{\mathrm{(o/s)}} \\ 
X_0^{\mathrm{(o/s)}} - X_{u}^{\mathrm{(o/s)}} - X_{v}^{\mathrm{(o/s)}} \\ 
X_0^{\mathrm{(o/s)}} + 2X_{u}^{\mathrm{(o/s)}} \\ 
X_{yz}^{\mathrm{(o/s)}} \\
X_{zx}^{\mathrm{(o/s)}} \\ 
X_{xy}^{\mathrm{(o/s)}} \\
\end{pmatrix}^{\rm T}. 
\end{align}
Thus, the active monopole and quadrupole contribute to $\chi^{[0\times2]}_{\mathrm{(o/s)}}$.
For example, the spin piezomagnetic tensor for $F^{[2]}=\bm{\tau}$ and $B^{[0]}= S$ corresponds to $\chi^{[0\times2]}$, where the spin electric monopole and quadrupole are relevant.
The expression of $\chi^{[2\times0]}_{\mathrm{(o/s)}}$ is obtained by transposing $\chi^{[0\times2]}_{\mathrm{(o/s)}}$.

\subsubsection{Rank-3 tensor \label{sec:res_symmetry_rank3}}
Rank-3 tensors include $\chi^{[1\times2]}_{\mathrm{(o/s)}}$ and $\chi^{[0\times3]}_{\mathrm{(o/s)}}$. 
The tensor $\chi^{[1\times2]}_{\mathrm{(o/s)}}$ is the rank-3 tensor for $B^{[1]}=(B_x, B_y, B_z)$ and $F^{[2]}=(F_{xx}, F_{yy}, F_{zz}, F_{yz}, F_{zx}, F_{xy})$, which is expressed by dipole $(X_x^{\mathrm{(o/s)}}, X_y^{\mathrm{(o/s)}}, X_z^{\mathrm{(o/s)}})$, quadrupole $(Y_{u}^{\mathrm{(o/s)}}, Y_{v}^{\mathrm{(o/s)}}, Y_{yz}^{\mathrm{(o/s)}}, Y_{zx}^{\mathrm{(o/s)}}, Y_{xy}^{\mathrm{(o/s)}})$, and octupole $(X_{xyz}^{\mathrm{(o/s)}}, X_x^{\alpha \mathrm{(o/s)}}, X_y^{\alpha \mathrm{(o/s)}}, X_z^{\alpha \mathrm{(o/s)}}, X_x^{\beta \mathrm{(o/s)}}, X_y^{\beta \mathrm{(o/s)}}, X_z^{\beta \mathrm{(o/s)}})$ as
\begin{widetext}
\begin{align}
\label{eq:rank12}
\chi^{[1\times2]}_{\mathrm{(o/s)}}
&=
\begin{pmatrix}
\tilde{X}_x^{\mathrm{(o/s)}} + 4 X_{x}^{\alpha \mathrm{(o/s)}} & 
\tilde{X}_y^{\prime \mathrm{(o/s)}} - 2Y_{zx}^{\mathrm{(o/s)}} - 2X_{y}^{\alpha \mathrm{(o/s)}} - 2X_{y}^{\beta \mathrm{(o/s)}} & 
\tilde{X}_z^{\prime \mathrm{(o/s)}} + 2Y_{xy}^{\mathrm{(o/s)}} - 2X_{z}^{\alpha \mathrm{(o/s)}} + 2X_{z}^{\beta \mathrm{(o/s)}} 
\\
\tilde{X}_x^{\prime \mathrm{(o/s)}} + 2Y_{yz}^{\mathrm{(o/s)}} - 2X_{x}^{\alpha \mathrm{(o/s)}} + 2X_{x}^{\beta \mathrm{(o/s)}} &
\tilde{X}_y^{\mathrm{(o/s)}} + 4 X_{y}^{\alpha \mathrm{(o/s)}} &
\tilde{X}_z^{\prime \mathrm{(o/s)}} - 2Y_{xy}^{\mathrm{(o/s)}} - 2X_{z}^{\alpha \mathrm{(o/s)}} - 2X_{z}^{\beta \mathrm{(o/s)}} 
\\
\tilde{X}_x^{\prime \mathrm{(o/s)}} - 2Y_{yz}^{\mathrm{(o/s)}} - 2X_{x}^{\alpha \mathrm{(o/s)}} - 2X_{x}^{\beta \mathrm{(o/s)}} & 
\tilde{X}_y^{\prime \mathrm{(o/s)}} + 2Y_{zx}^{\mathrm{(o/s)}} - 2X_{y}^{\alpha \mathrm{(o/s)}} + 2X_{y}^{\beta \mathrm{(o/s)}} &
\tilde{X}_z^{\mathrm{(o/s)}} + 4 X_{z}^{\alpha \mathrm{(o/s)}} 
\\
Y_{u}^{\mathrm{(o/s)}} + Y_{v}^{\mathrm{(o/s)}} + X_{xyz}^{\mathrm{(o/s)}} & 
- 3X_{z}^{\mathrm{(o/s)}} + Y_{xy}^{\mathrm{(o/s)}} - 2X_{z}^{\alpha \mathrm{(o/s)}} - 2X_{z}^{\beta \mathrm{(o/s)}} & 
- 3X_{y}^{\mathrm{(o/s)}} - Y_{zx}^{\mathrm{(o/s)}} - 2X_{y}^{\alpha \mathrm{(o/s)}} + 2X_{y}^{\beta \mathrm{(o/s)}} 
\\
- 3X_{z}^{\mathrm{(o/s)}} - Y_{xy}^{\mathrm{(o/s)}} - 2X_{z}^{\alpha \mathrm{(o/s)}} + 2X_{z}^{\beta \mathrm{(o/s)}} &
- Y_{u}^{\mathrm{(o/s)}} + Y_{v}^{\mathrm{(o/s)}} + X_{xyz}^{\mathrm{(o/s)}} & 
- 3X_{x}^{\mathrm{(o/s)}} + Y_{yz}^{\mathrm{(o/s)}} - 2X_{x}^{\alpha \mathrm{(o/s)}} - 2X_{x}^{\beta \mathrm{(o/s)}} 
\\
-3 X_{y}^{\mathrm{(o/s)}} + Y_{zx}^{\mathrm{(o/s)}} - 2 X_{y}^{\alpha \mathrm{(o/s)}} - 2X_{y}^{\beta \mathrm{(o/s)}} & 
- 3X_{x}^{\mathrm{(o/s)}} - Y_{yz}^{\mathrm{(o/s)}} - 2X_{x}^{\alpha \mathrm{(o/s)}} + 2X_{x}^{\beta \mathrm{(o/s)}} & 
- 2Y_{v}^{\mathrm{(o/s)}} + X_{xyz}^{\mathrm{(o/s)}} 
\\
\end{pmatrix}^{\rm T}.
\end{align}
\end{widetext}
It is noted that both $\tilde{X}_i^{\mathrm{(o/s)}}$ and $\tilde{X}_i^{\prime \mathrm{(o/s)}}$ ($i=x,y,z$) 
stand for the dipole but are independent with each other~\cite{Yatsushiro_PhysRevB.104.054412}: 
$(\tilde{X}_i^{\mathrm{(o/s)}}, \tilde{X}_i^{\prime \mathrm{(o/s)}})
\equiv
( X_i^{\prime \mathrm{(o/s)}} - 4X_i^{\mathrm{(o/s)}},  X_i^{\prime \mathrm{(o/s)}} + 2X_i^{\mathrm{(o/s)}} )
$
where
$X_i^{\mathrm{(o/s)}} = \frac{1}{10} \sum_{j}( \frac{1}{3}\chi^{[1\times2]}_{\mathrm{(o/s)} i;jj} - \chi^{[1\times2]}_{\mathrm{(o/s)} j;ij} )$
and
$X_i^{\prime \mathrm{(o/s)}} =  \frac{1}{3} \sum_{j} \chi^{[1\times2]}_{\mathrm{(o/s)} i;jj}$. 
For polar tensors, representative examples are the piezoelectric response $(F^{[2]}=\bm\tau,B^{[1]}=\bm P)$, the second-order nonlinear electric conductivity $(F_{ij}^{[2]}=E_iE_j,B^{[1]}=\bm J)$, and the second-order nonlinear spin conductivity $(F_{ij}^{[2]}=E_iE_j,B^{[1]}=\bm J^S)$. 
These responses are governed by electric dipoles/octupoles and electric-toroidal quadrupoles. 
For orbital responses, $X=Q^{\mathrm{(o)}}$ and $Y=G^{\mathrm{(o)}}$, whereas for spin responses, $X=Q^{\mathrm{(s)}}$ and $Y=G^{\mathrm{(s)}}$. 
Axial rank-3 tensors, such as the orbital piezomagnetic response $(F^{[2]}=\bm\tau, B^{[1]}=\bm L)$, are obtained by exchanging $Q$ and $G$.
The expression of the tensor $\chi^{[2\times1]}_{\mathrm{(o/s)}}$ is obtained by transposing $\chi^{[1\times2]}_{\mathrm{(o/s)}}$.

The tensor $\chi^{[0\times3]}_{\mathrm{(o/s)}}$ is the rank-3 response tensor for $B^{[0]}=(B)$ and $F^{[3]}=(F_{xxx}, F_{yyy}, F_{zzz}, F_{yyz}, F_{zzx}, F_{xxy}, F_{yzz}, F_{zxx}, F_{xyy}, F_{xyz})$ where $F_{ijk}=F_{jik}=F_{ikj}$. 
As $F^{[3]}$ is decomposed into the dipole and octupole components, $\chi^{[0\times3]}_{\mathrm{(o/s)}}$ is also related to them, which is shown as
\begin{align}
\label{eq:rank30}
\chi^{[0\times3]}=
\begin{pmatrix}
3X_x^{\mathrm{(o/s)}} + 2X_x^{\alpha \mathrm{(o/s)}} \\
3X_y^{\mathrm{(o/s)}} + 2X_y^{\alpha \mathrm{(o/s)}} \\
3X_z^{\mathrm{(o/s)}} + 2X_z^{\alpha \mathrm{(o/s)}} \\ 
X_z^{\mathrm{(o/s)}} - X_z^{\alpha \mathrm{(o/s)}} - X_z^{\beta \mathrm{(o/s)}} \\
X_x^{\mathrm{(o/s)}} - X_x^{\alpha \mathrm{(o/s)}} - X_x^{\beta \mathrm{(o/s)}} \\ 
X_y^{\mathrm{(o/s)}} - X_y^{\alpha \mathrm{(o/s)}} - X_y^{\beta \mathrm{(o/s)}} \\
X_y^{\mathrm{(o/s)}} - X_y^{\alpha \mathrm{(o/s)}} + X_y^{\beta \mathrm{(o/s)}} \\ 
X_z^{\mathrm{(o/s)}} - X_z^{\alpha \mathrm{(o/s)}} + X_z^{\beta \mathrm{(o/s)}} \\
X_x^{\mathrm{(o/s)}} - X_x^{\alpha \mathrm{(o/s)}} + X_x^{\beta \mathrm{(o/s)}} \\ 
X_{xyz}^{\mathrm{(o/s)}} \\
\end{pmatrix}^{\rm T}.
\end{align}
Examples include the third-order electrocaloric effect and the third-order spin piezoelectric effect, which is relevant with $X=Q$ ($G$) for the polar (axial) tensor. 
The expression of $\chi^{[3\times0]}_{\mathrm{(o/s)}}$ is obtained by transposing $\chi^{[0\times3]}_{\mathrm{(o/s)}}$.

\subsubsection{Rank-4 tensor \label{sec:res_symmetry_rank4}}
Rank-4 tensors include $\chi^{[1\times3]}_{\mathrm{(o/s)}}$ and $\chi^{[2\times2]}_{\mathrm{(o/s)}}$. 
The tensor $\chi^{[1\times3]}_{\mathrm{(o/s)}}$ is the rank-4 response tensor for  
$B^{[1]}=(B_x, B_y, B_z)$ and 
$F^{[3]}=(F_{xxx}, F_{yyy}, F_{zzz}, F_{yyz}, F_{zzx}, F_{xxy}, F_{yzz}, F_{zxx}, F_{xyy}, F_{xyz})$ where $F_{ijk}=F_{jik}=F_{ikj}$.
The relevant multipoles are from rank 0 to 4; 
monopole $X_0$, dipole $(Y_x, Y_y, Y_z)$, quadrupole $(X_{u}, X_{v}, X_{yz}, X_{zx}, X_{xy})$, octupole $(Y_{xyz}, Y_x^{\alpha \mathrm{(o/s)}} , Y_y^{\alpha \mathrm{(o/s)}} , Y_z^{\alpha \mathrm{(o/s)}} , Y_x^{\beta \mathrm{(o/s)}} , Y_y^{\beta \mathrm{(o/s)}} , Y_z^{\beta \mathrm{(o/s)}} )$, and hexadecapole $(X_4, X_{4u}, X_{4v}, X_{4x}^{\alpha \mathrm{(o/s)}} , X_{4y}^{\alpha \mathrm{(o/s)}} , X_{4z}^{\alpha \mathrm{(o/s)}} , X_{4x}^{\beta \mathrm{(o/s)}} , X_{4y}^{\beta \mathrm{(o/s)}} , X_{4z}^{\beta \mathrm{(o/s)}} )$
The tensor component of $\chi^{[1\times3]}_{\mathrm{(o/s)}}$ is given by as
\begin{widetext}
\begin{equation}
\resizebox{\textwidth}{!}{$
\begin{aligned}
\label{eq:rank13}
&
\chi^{[1\times3]}_{\mathrm{(o/s)}} 
=
^{\rm T}.
\end{aligned}
$}
\end{equation}
Note that $(\tilde{X}_u^{\mathrm{(o/s)}}, \tilde{X}_u^{\prime \mathrm{(o/s)}} , \tilde{X}_u^{\prime\prime \mathrm{(o/s)}} )$, $(\tilde{X}_v^{\mathrm{(o/s)}}, \tilde{X}_v^{\prime \mathrm{(o/s)}} , \tilde{X}_v^{\prime\prime \mathrm{(o/s)}} )$, and $(\tilde{X}_{yz}^{\mathrm{(o/s)}}, \tilde{X}_{yz}^{\prime \mathrm{(o/s)}} )$ (cyclic) are introduced to express the two independent quadrupoles~\cite{Yatsushiro_PhysRevB.104.054412}: 
$(\tilde{X}_u^{\mathrm{(o/s)}}, \tilde{X}_u^{\prime \mathrm{(o/s)}} , \tilde{X}_u^{\prime\prime \mathrm{(o/s)}} )
\equiv
(X_u^{\mathrm{(o/s)}}+X_u^{\prime \mathrm{(o/s)}}, 4X_u^{\mathrm{(o/s)}}-X_u^{\prime \mathrm{(o/s)}}, -3X_u^{\mathrm{(o/s)}}+2X_u^{\prime \mathrm{(o/s)}} )$, 
$(\tilde{X}_v^{\mathrm{(o/s)}}, \tilde{X}_v^{\prime \mathrm{(o/s)}} , \tilde{X}_v^{\prime\prime \mathrm{(o/s)}} )
\equiv
(3X_v^{\mathrm{(o/s)}}+X_v^{\prime \mathrm{(o/s)}}, 2X_v^{\mathrm{(o/s)}}-X_v^{\prime \mathrm{(o/s)}}, 7X_v^{\mathrm{(o/s)}}-X_v^{\prime \mathrm{(o/s)}} )$,
and 
$(\tilde{X}_{yz}^{\mathrm{(o/s)}}, \tilde{X}_{yz}^{\prime \mathrm{(o/s)}} )
\equiv
(2X_{yz}^{\mathrm{(o/s)}}-X_{yz}^{\prime \mathrm{(o/s)}}, 8X_{yz}^{\mathrm{(o/s)}}+X_{yz}^{\prime \mathrm{(o/s)}} )$, 
where
$X_u^{\mathrm{(o/s)}} = \frac{1}{42}[ 3\chi^{{\rm Q}(1\times3)}_{zz}-\sum_{i}\chi^{{\rm Q}(1\times3)}_{ii}]$, 
$X_v^{\mathrm{(o/s)}} = \frac{1}{42}[\chi^{{\rm Q}(1\times3)}_{xx}-\chi^{{\rm Q}(1\times3)}_{yy}]$, 
$X_{yz}^{\mathrm{(o/s)}} =\frac{1}{21}\chi^{{\rm Q}(1\times3)}_{yz}$,  
$X_u^{\prime \mathrm{(o/s)}} = \frac{1}{10}[3\chi^{{\rm Q}(1\times1)}_{zz}-\sum_{i}\chi^{{\rm Q}(1\times1)}_{ii}]$, 
$X_v^{\prime \mathrm{(o/s)}} = \frac{3}{10}[\chi^{{\rm Q}(1\times1)}_{xx}-\chi^{{\rm Q}(1\times1)}_{yy}]$, 
and
$X_{yz}^{\prime \mathrm{(o/s)}} = \frac{3}{5}\chi^{{\rm Q}(1\times1)}_{yz}$, 
with 
$\chi^{{\rm Q}(1\times1)}_{ij} = \frac{1}{6} \sum_{k}\left(\chi_{\mathrm{(o/s)} i;jkk}^{[1\times3]}+\chi_{\mathrm{(o/s)} j;ikk}^{[1\times3]} \right)=\chi^{{\rm Q}(1\times1)}_{ji}$  
and 
$\chi^{{\rm Q}(1\times 3)}_{ij} = \frac{1}{2}\sum_{k}\left[\left(\chi_{\mathrm{(o/s)} k;ijk}^{[1\times3]}+\chi_{\mathrm{(o/s)} k;jik}^{[1\times3]} \right)-\frac{2}{5}\left(\chi_{\mathrm{(o/s)} i;jkk}^{[1\times3]}+\chi_{\mathrm{(o/s)} j;ikk}^{[1\times3]}\right)\right] =\chi^{{\rm Q}(1\times 3)}_{ji}$.
This tensor includes, for example, the third-order nonlinear electric conductivity. 
The relevant multipoles are $X=Q$ and $Y=G$ for the polar tensor,
while those are $X=G$ and $Y=Q$ for the axial tensor.
The expression of $\chi^{[3\times1]}_{\mathrm{(o/s)}}$ is obtained by transposing $\chi^{[1\times3]}_{\mathrm{(o/s)}}$.

The tensor $\chi^{[2\times2]}_{\mathrm{(o/s)}}$ is another rank-4 tensor for 
$B^{[2]}=(B_{xx}, B_{yy}, B_{zz}, B_{yz}, B_{zx}, B_{xy})$ where $B_{ij}=B_{ji}$ and 
$F^{[2]}=(F_{xx}, F_{yy}, F_{zz}, F_{yz}, F_{zx}, F_{xy})$ where $F_{ij}=F_{ji}$. 
The tensor component of $\chi^{[2\times2]}_{\mathrm{(o/s)}}$ is related to the rank 0--4 multipoles, which is given by 
\begin{align}
\chi^{[2\times 2]}_{\mathrm{(o/s)}}
=
,
 \label{eq:rank22tt}
\end{aligned}
$}
\end{equation}
\end{widetext}
where $X_{4uv\pm}^{\mathrm{(o/s)}} =X_{4}^{\mathrm{(o/s)}} +X_{4u}^{\mathrm{(o/s)}} \pm X_{4v}^{\mathrm{(o/s)}} $.
We also introduce $(\tilde{X}_0^{\mathrm{(o/s)}} , \tilde{X}_0^{\prime \mathrm{(o/s)}} )$, $(\tilde{X}_u^{\mathrm{(o/s)}} , \tilde{X}_u^{(\pm)}, \tilde{X}_u^{\prime \mathrm{(o/s)}} )$, $(\tilde{X}_v^{\mathrm{(o/s)}} , \tilde{X}_v^{(\pm) \mathrm{(o/s)}})$, and $(\tilde{X}_{yz}^{(\pm) \mathrm{(o/s)}}, \tilde{X}_{yz}^{\prime(\pm) \mathrm{(o/s)}})$ (cyclic) 
for notational simplicity~\cite{Yatsushiro_PhysRevB.104.054412}: 
$(\tilde{X}_0^{\mathrm{(o/s)}}, \tilde{X}_0^{\prime \mathrm{(o/s)}}) \equiv (4X_0^{\mathrm{(o/s)}} +X_0^{\prime \mathrm{(o/s)}}, -2X_0^{\mathrm{(o/s)}} +X_0^{\prime \mathrm{(o/s)}})$, 
$(\tilde{X}_u^{\mathrm{(o/s)}}, \tilde{X}_u^{(\pm) \mathrm{(o/s)}}, \tilde{X}_u^{\prime \mathrm{(o/s)}}) \equiv 
(-4 {X}_{u}^{\mathrm{(o/s)}}-2 {X}_{u}^{(+) \mathrm{(o/s)}}, 
-4 {X}_{u}^{\mathrm{(o/s)}}+{X}_{u}^{(+) \mathrm{(o/s)}} \pm 3 {X}_{u}^{(-) \mathrm{(o/s)}}, 
-\tilde{X}_u^{(+) \mathrm{(o/s)}}-\tilde{X}_u^{(-) \mathrm{(o/s)}}
)$, 
$(\tilde{X}_v^{\mathrm{(o/s)}}, \tilde{X}_v^{(\pm) \mathrm{(o/s)}}) \equiv 
(4 {X}_{v}^{\mathrm{(o/s)}} +2 {X}^{(+) \mathrm{(o/s)}}_{v}, -4 {X}_{v}^{\mathrm{(o/s)}} +{X}^{(+) \mathrm{(o/s)}}_{v} \pm {X}^{(-) \mathrm{(o/s)}}_{v})$,  
and 
$(\tilde{X}_{yz}^{(\pm) \mathrm{(o/s)}}, \tilde{X}_{yz}^{\prime(\pm) \mathrm{(o/s)}}) \equiv (-4X_{yz}^{\mathrm{(o/s)}} +X^{(+) \mathrm{(o/s)}}_{yz}\pm X^{(-) \mathrm{(o/s)}}_{yz}, 2X_{yz}^{\mathrm{(o/s)}} +X^{(+) \mathrm{(o/s)}}_{yz}\pm X^{(-) \mathrm{(o/s)}}_{yz})$, 
where
$X_0^{\mathrm{(o/s)}} = \frac{1}{10}\chi^{{\rm M}(2\times2)}$, 
$X_0^{\prime \mathrm{(o/s)}} = \frac{1}{3}\chi^{{\rm M}(0\times0)}$, 
$X_u^{\mathrm{(o/s)}} = \frac{1}{42}[ 3\chi^{{\rm Q}(2\times2)}_{zz} - \sum_i \chi^{{\rm Q}(2\times2)}_{ii} ]$, 
$X_v^{\mathrm{(o/s)}} = \frac{1}{14}[ \chi^{{\rm Q}(2\times2)}_{xx}-\chi^{{\rm Q}(2\times2)}_{yy} ]$, 
$X_{yz}^{\mathrm{(o/s)}} = \frac{1}{7} \chi^{{\rm Q}(2\times2)}_{yz}$, 
$X^{(\pm) \mathrm{(o/s)}}_u = \frac{1}{6}[ 3 \chi^{{\rm Q}(0\times2,\pm)}_{zz} - \sum_i \chi^{{\rm Q}(0\times2,\pm)}_{ii} ]$, 
$X^{(\pm) \mathrm{(o/s)}}_v = \frac{1}{2}[ \chi^{{\rm Q}(0\times2,\pm)}_{xx}-\chi^{{\rm Q}(0\times2,\pm)}_{yy} ]$, 
and
$X^{(\pm) \mathrm{(o/s)}}_{yz} = \chi^{{\rm Q}(0\times2,\pm)}_{yz}$,
with 
$\chi^{{\rm M}(0\times0)} = \frac{1}{3} \sum_{ij} \chi_{\mathrm{(o/s)} ii;jj}^{[2\times 2]}$, 
$\chi^{{\rm M}(2\times2)} = \frac{1}{3} \sum_{ij}\left(\chi_{\mathrm{(o/s)} ij;ji}^{[2\times 2]}-\frac{1}{3}\chi_{\mathrm{(o/s)} ii;jj}^{[2\times2]}\right)$, 
$\chi_{ij}^{{\rm Q}(0\times2, \pm)}= \frac{1}{6}\sum_{k}\left(\chi_{\mathrm{(o/s)} kk;ij}^{[2\times2]}\pm\chi_{\mathrm{(o/s)} ij;kk}^{[2\times2]}\right)$,  
and 
$\chi_{ij}^{{\rm Q}(2\times2)}
= \frac{1}{2}\sum_{k}\left[
\left(\chi_{\mathrm{(o/s)} ik;kj}^{[2\times 2]}+\chi_{\mathrm{(o/s)} jk;ki}^{[2\times 2]}\right)
-\frac{2}{3}\left(\chi_{\mathrm{(o/s)} ij;kk}^{[2\times2]}+\chi_{\mathrm{(o/s)} kk;ij}^{[2\times2]}\right)
\right]
=\chi_{ji}^{{\rm Q}(2\times2)} $.
Examples include the elastic stiffness tensor and magneto-Seebeck-type tensors, which is related to the multipoles $X=Q$ and $Y =G$ for the polar tensor and to $X=G$ and $Y =Q$ for the axial tensor.

\subsection{Examples of responses \label{sec:res_linear}}

We now demonstrate how the above tensor--multipole correspondence can be used to identify nonrelativistic responses in representative collinear antiferromagnets. 
The following examples highlight the main advantage of the present approach: once the active spin multipoles are known from the SPG, the allowed spin-dependent responses can be read off immediately.

Focusing on systems with crystallographic point group $\bm{G} = 4 / mmm$, 
time-reversal-odd IRREPs of multipoles in the $\mathcal{T}$-symmetric collinear SPG 
$\mathcal{G}_{\mathrm{SO}}^{\infty} \times {}^1 4 / {}^1 m {}^1 m {}^1 m {}^{\bar{1}} 1$
are shown in Table~\ref{table:IREEP_D4h}.
One can see which nonrelativistic spin responses are induced by $\mathcal{T}$-breaking magnetic order.

\begin{table*}
\caption{
Time-reversal-odd IRREPs of multipoles in the $\mathcal{T}$-symmetric collinear SPG 
$\mathcal{G}_{\mathrm{SO}}^{\infty} \times {}^1 4 / {}^1 m {}^1 m {}^1 m {}^{\bar{1}} 1 
= {\rm SO(2)}\times ( {}^1 4 / {}^1 m {}^1 m {}^1 m + [\theta \,\|\, E] {}^1 4 / {}^1 m {}^1 m {}^1 m )$, 
together with the lowest-rank spin electric (E) and spin electric-toroidal (ET) multipoles belonging to each IRREP, their functional forms, the nonrelativistic phenomena expected to arise from these multipoles, and collinear SPG where these multipoles become active. 
It is defined that $\bm{t} = \bm{r} \times \bm{l}$ ($\bm{l}$: orbital angular momentum), and PZM stands for spin piezomagnetic effect.
\label{table:IREEP_D4h}
}
\begin{tabular}{cccccc@{\hspace{3mm}}c}
\hline\hline
IRREP
& Spin E & Function 
& Spin ET & Function 
& Phenomena & SPG  \\
\hline
${\rm A_{1g}^-}$
& $Q_{0}^{\rm (s)}$
& $\sigma$
& 
& 
& 
& $\mathcal{G}_{\mathrm{SO}}^{\infty} \times {}^1 4 / {}^1 m {}^1 m {}^1 m$ 
\\
${\rm A_{2g}^-}$
& $Q_{4z}^{\alpha {\rm (s) } }$
& $xy (x^2 - y^2)\sigma$
& $G_{z}^{\rm (s)}$
& $(\bm{r} \times \bm{r}')_z \sigma$
& Nonlinear PZM
& $\mathcal{G}_{\mathrm{SO}}^{\infty} \times {}^1 4/ {}^1 m {}^{\bar{1}} m {}^{\bar{1}} m$
\\
${\rm B_{1g}^-}$
& $Q_{v}^{\rm (s)}$
& $(x^2 - y^2)\sigma$
& $G_{xyz}^{\rm (s)}$
& $x(l_y t_z + l_z t_y)\sigma +  ({\rm cyclic})$
& Linear PZM, $\sigma^{\rm (s)}_{xx}=-\sigma^{\rm (s)}_{yy}$
& $\mathcal{G}_{\mathrm{SO}}^{\infty} \times {}^{\bar{1}} 4/ {}^1 m {}^1 m {}^{\bar{1}} m$
\\
${\rm B_{2g}^-}$
& $Q_{xy}^{\rm (s)}$
& $xy \sigma$
& $G_{z}^{\beta {\rm (s) } }$
& $[ x(l_zt_x + l_xt_z) - y(l_yt_z + l_zt_y)+z(l_xt_x - l_yt_y)]\sigma$
& Linear PZM, $\sigma^{\rm (s)}_{xy}=\sigma^{\rm (s)}_{yx}$
& $\mathcal{G}_{\mathrm{SO}}^{\infty} \times {}^{\bar{1}} 4/ {}^1 m {}^{\bar{1}} m {}^1 m $
\\
${\rm E_{g}^-}$
& \{$Q_{yz}^{\rm (s)}$, $Q_{zx}^{\rm (s)}$\} 
& \{$yz \sigma$, $zx \sigma$\} 
& \{$G_{x}^{\rm (s)}$, $G_{y}^{\rm (s)}$\} 
& \{$(\bm{r} \times \bm{r}')_x \sigma$, $(\bm{r} \times \bm{r}')_y \sigma$\}
& 
& 
\\
\hline
${\rm A_{1u}^-}$
& 
&  
& $G_{0}^{\rm (s)}$
& $\bm{r} \cdot (\bm{r}' \times \bm{r}'') \sigma$
& 
& $\mathcal{G}_{\mathrm{SO}}^{\infty} \times {}^1 4/ {}^{\bar{1}} m {}^{\bar{1}} m {}^{\bar{1}} m$
\\
${\rm A_{2u}^-}$
& $Q_{z}^{\rm (s)}$
& $z\sigma$
& $G_{4z}^{\alpha {\rm (s) } }$
& $[xy(l_xt_x-l_yt_y)+(x^2-y^2)(l_xt_y+l_yt_x)/2]\sigma$
& Magnetoelectric effect
& $\mathcal{G}_{\mathrm{SO}}^{\infty} \times {}^1 4/ {}^{\bar{1}} m {}^1 m {}^1 m$ 
\\
\hline\hline
\end{tabular}
\end{table*}

\begin{figure}[htb!]
\centering
\includegraphics[width=85mm]{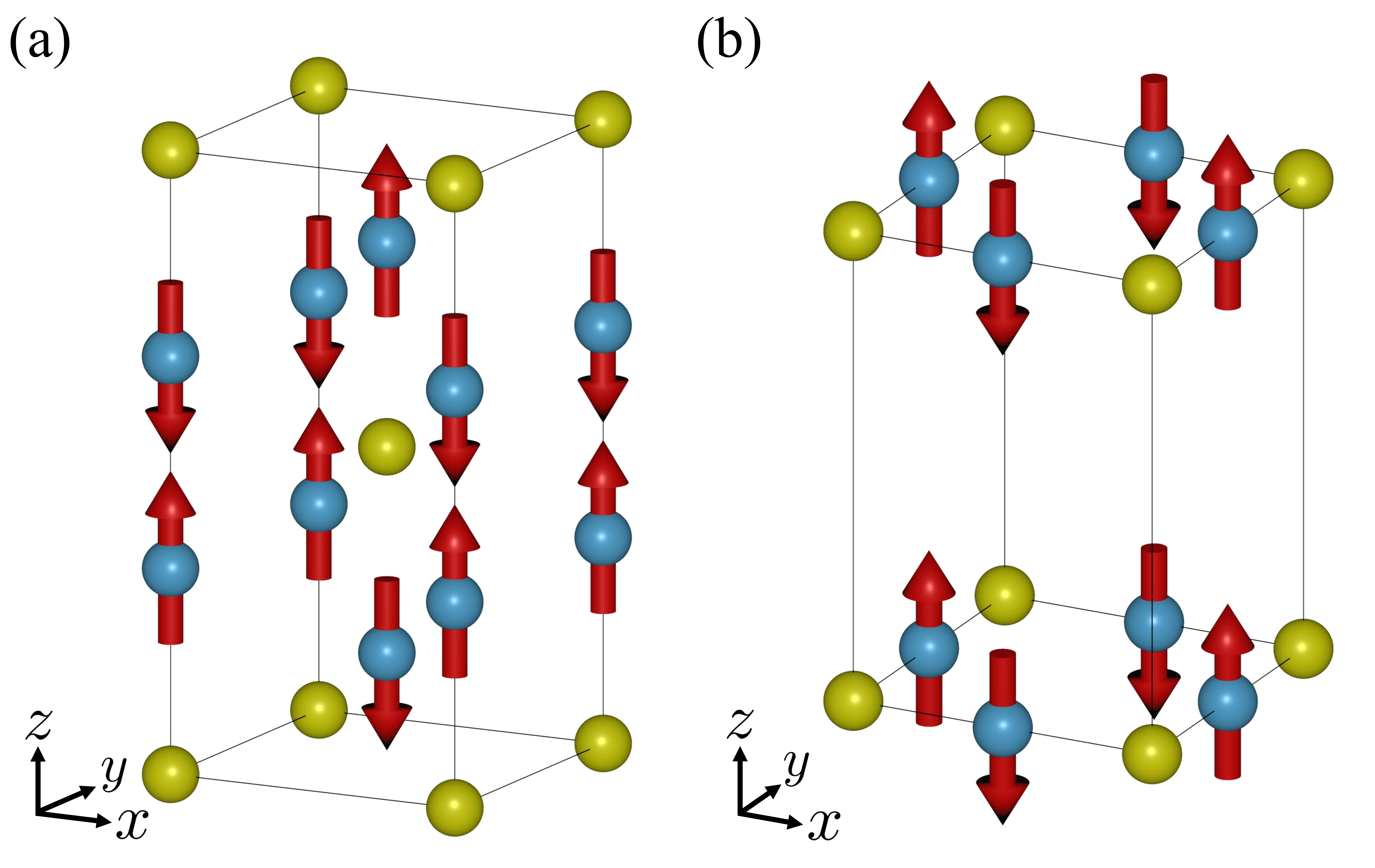}
\caption{
(a) Collinear antiferromagnetic structure with the $\mathcal{PT}$ symmetry.
(b) Collinear antiferromagnetic ($d$-wave altermagnetic) structure without the $\mathcal{PT}$ symmetry.
\label{fig:AFMstructure}
}
\end{figure}

\subsubsection{$\mathcal{PT}$-symmetric magnetic structure \label{sec:res_linear_staggered}}
The collinear antiferromagnetic structure shown in Fig.~\ref{fig:AFMstructure}(a) belongs to the $\mathcal{PT}$-symmetric SPG ${}^1 4/{}^{\bar{1}} m {}^1 m {}^1 m$.
According to Table~\ref{table:active_MP_collinearSPG_nonunitary_2}, the active orbital multipoles up to rank 4 are the even-parity orbital electric multipoles 
$Q_0^{\mathrm{(o)}}$ (monopole), $Q_u^{\mathrm{(o)}}$ (quadrupole), and 
($Q_{4}^{\mathrm{(o)}}$, $Q_{4u}^{\mathrm{(o)}}$) (hexadecapole). 
In addition, magnetic ordering activates the odd-parity spin electric multipoles 
$Q_z^{\mathrm{(s)}}$ (dipole) and $Q_z^{\alpha \mathrm{(s)}}$ (octupole), together with the even-parity spin electric-toroidal multipole $G_{4z}^{\alpha \mathrm{(s)}}$ (hexadecapole). 
These spin multipoles are absent in the corresponding nonmagnetic SPG and therefore represent symmetry breaking induced purely by the collinear antiferromagnetic order.

For example, the electric conductivity tensor $J_i=\sum_{j} \sigma_{ij}E_j$ is governed by the active orbital electric monopole and quadrupole:
\begin{align}
\sigma_{ij} \leftrightarrow 
\begin{pmatrix}
Q_0^{\mathrm{(o)}} -Q_u^{\mathrm{(o)}} & 0 &  0 \\
0 & Q_0^{\mathrm{(o)}} -Q_u^{\mathrm{(o)}} & 0 \\
0 &0 & Q_0^{\mathrm{(o)}} +2Q_u^{\mathrm{(o)}} \\
\end{pmatrix}. 
\end{align} 
Thus, even without SOC, the system can exhibit anisotropic longitudinal conductivity between the $xy$ plane and the $z$ direction. 
By contrast, the antisymmetric Hall component is absent in the nonrelativistic collinear antiferromagnet because no time-reversal-odd orbital magnetic multipole is active.

More characteristically, the active spin electric dipole and octupole generate nonlinear spin-current responses. 
For the second-order spin conductivity $J_i^{S} =\sum_{jk} \sigma_{i;jk}^{\mathrm{(s)}} E_jE_k$, 
the allowed tensor components are determined by $Q_z^{\mathrm{(s)}}$ and $Q_z^{\alpha\mathrm{(s)}}$ as
\begin{align}
\sigma_{i;jk}^{\mathrm{(s)}}
&=
\begin{pmatrix}
0 & 0 & \sigma_{z;xx}^{\mathrm{(s)}} \\
0 & 0 & \sigma_{z;yy}^{\mathrm{(s)}} (=\sigma_{z;xx}^{\mathrm{(s)}} )\\
0 & 0 & \sigma_{z;zz}^{\mathrm{(s)}} \\
0 & \sigma_{y;yz}^{\mathrm{(s)}} & 0 \\
\sigma_{x;zx}^{\mathrm{(s)}} (= \sigma_{y;yz}^{\mathrm{(s)}}) & 0 & 0 \\
0 & 0 & 0 \\
\end{pmatrix}
\notag\\
&\leftrightarrow
\begin{pmatrix}
0 & 0 & \tilde{Q}_{z}^{\prime \mathrm{(s)}} -2Q_z^{\alpha \mathrm{(s)}} \\
0 & 0 & \tilde{Q}_{z}^{\prime \mathrm{(s)}} -2Q_z^{\alpha \mathrm{(s)}} \\
0 & 0 & \tilde{Q}_{z}^{\mathrm{(s)}} +4Q_z^{\alpha \mathrm{(s)}} \\
0 & -3Q_{z}^{\mathrm{(s)}} -2Q_z^{\alpha \mathrm{(s)}}  & 0 \\
-3Q_{z}^{\mathrm{(s)}} -2Q_z^{\alpha \mathrm{(s)}} & 0 & 0 \\
 0 & 0 & 0 \\
\end{pmatrix}.
\end{align}
This result shows that $\mathcal{PT}$-symmetric collinear antiferromagnets can support nonlinear spin-current generation even in the absence of SOC~\cite{Hayami_PhysRevB.106.024405}. 
Such an SOC-free response is difficult to identify within the conventional magnetic-point-group approach, because that framework does not explicitly separate orbital and spin sectors in the nonrelativistic limit.

\subsubsection{$d$-wave altermagnetic structure \label{sec:res_linear_altermag}}
We next consider the spin point group ${}^{\bar{1}} 4 / {}^1 m {}^{\bar{1}} m {}^1 m$, which corresponds to a $d$-wave altermagnetic structure shown in Fig.~\ref{fig:AFMstructure} (b). 
In this case, both $\mathcal{T}$ and $\mathcal{PT}$ symmetries are broken.
Accordingly, the even-parity spin electric quadrupole $Q_{v}^{\mathrm{(s)}}$ becomes active, together with the higher-rank spin electric hexadecapole $Q_{4v}^{\mathrm{(s)}}$ and the odd-parity spin electric-toroidal octupole $G_{xyz}^{\mathrm{(s)}}$.  
The active spin electric quadrupole $Q_{v}^{\mathrm{(s)}}$ is the multipolar order parameter of the $d$-wave altermagnetic spin splitting, as shown in Table~\ref{table:altermag}.

This immediately leads to a nonrelativistic spin conductivity $J_i^{S}=\sum_{j} \sigma_{ij}^{\mathrm{(s)}} E_j$, whose tensor component is given by 
\begin{align}
\sigma_{ij}^{\mathrm{(s)}}  
\leftrightarrow 
\begin{pmatrix}
Q_v^{\mathrm{(s)}} & 0 &  0 \\
0 & - Q_v^{\mathrm{(s)}} & 0 \\
0 & 0 & 0 
\end{pmatrix}. 
\end{align} 
Thus, the longitudinal spin current generated along the $x$ direction has the opposite sign to that generated along the $y$ direction, reflecting the $d_{x^2-y^2}$-type spin splitting encoded in $Q_v^{\mathrm{(s)}}$~\cite{Ahn2019, Naka2019}. 
This provides a direct multipolar interpretation of anisotropic spin-current generation in altermagnets.

The same active spin quadrupole also allows spin piezomagnetic and nonlinear spin magnetoelectric responses.
For the spin piezomagnetic effect $S = \sum_{ij}\Lambda_{ij} \tau_{ij}$, one obtains 
\begin{align}
\Lambda_{ij}
&=
\begin{pmatrix}
\Lambda_{xx} \\
\Lambda_{yy} (= - \Lambda_{xx}) \\
0 \\
0 \\
0 \\
0 \\
\end{pmatrix}^{\rm T} 
\leftrightarrow
\begin{pmatrix}
Q_v^{\mathrm{(s)}} \\
- Q_v^{\mathrm{(s)}} \\
0 \\
0 \\
0 \\
0 \\
\end{pmatrix}^{\rm T}. 
\end{align}
Similarly, the nonlinear spin magnetoelectric effect $S = \sum_{ij} \alpha_{ij}^{(2)} E_iE_j$ has the same form:
\begin{align}
\alpha_{ij}^{(2)}
&=
\begin{pmatrix}
\alpha_{xx}^{(2)} \\
\alpha_{yy}^{(2)} (= -\alpha_{xx}^{(2)} ) \\
0 \\
0 \\
0 \\
0 \\
\end{pmatrix}^{\rm T} 
\leftrightarrow
\begin{pmatrix}
Q_v^{\mathrm{(s)}} \\
- Q_v^{\mathrm{(s)}} \\
0 \\
0 \\
0 \\
0 \\
\end{pmatrix}^{\rm T}.
\end{align}
These examples show that the active spin electric quadrupole is not only a descriptor of the altermagnetic band splitting but also a unifying symmetry origin of several experimentally relevant nonrelativistic responses. 
This demonstrates the advantage of the present multipole classification: it connects the symmetry of altermagnetic spin splitting directly to observable spintronic functionalities.

\section{Summary \label{sec:summary}}
In this work, we completed a comprehensive multipole classification for all 32 nonmagnetic SPGs and all 122 collinear SPGs. 
By treating orbital and spin degrees of freedom independently within the SPG framework, we established a unified classification of orbital and spin multipoles that is directly applicable to collinaer magnetic systems in the absence of SOC. 
Comparing the active multipoles in a collinear SPG with those in the corresponding nonmagnetic SPG provides a simple and general criterion for identifying multipoles induced solely by magnetic ordering.

Based on this classification, we systematically established the correspondence between active multipoles and symmetry-allowed physical responses. 
Unlike the conventional magnetic-point-group approach, the present framework distinguishes responses that intrinsically survive in the nonrelativistic limit from those that require SOC. 
It thereby provides a unified symmetry-based understanding of a broad range of nonrelativistic spin-dependent phenomena in collinear magnetic systems, including spin magnetization, spin-current generation, magnetoelectric effects, and nonlinear spin responses.

The present classification provides a general platform for exploring nonrelativistic functionalities in collinear magnetic materials. 
Beyond serving as a symmetry database for identifying active multipoles and response tensors, it offers practical guidelines for discovering unconventional spin-dependent phenomena in antiferromagnets, altermagnets, and related materials without relying on SOC. 
Building on the present framework, a natural next step is to extend the classification to noncollinear magnetic systems, in view of recent generalizations of the concept of altermagnetism to noncollinear magnetic structures both with and without broken spatial inversion symmetry~\cite{Yuan2021PRMat, Hayami_PhysRevB.101.220403, Liu2022, Hayami_PhysRevB.105.024413, Hayami_PhysRevB.106.014420, Brekke_PhysRevLett.133.236703, cheong2024altermagnetism, zhu2024observation, cheong2025altermagnetism, uchino2026analysis}.
Such an extension would provide a systematic basis for identifying multipoles and nonrelativistic responses unique to noncollinear magnetic orders.
We expect that the present framework will facilitate both theoretical studies and experimental searches for emergent nonrelativistic responses in magnetic materials.

\begin{acknowledgments}
This research was supported by JSPS KAKENHI (Grant Nos.~JP22H00101, JP23H04869, and JP26K22277), JST CREST (Grant No.~JPMJCR23O4), and JST FOREST (Grant No.~JPMJFR2366).
\end{acknowledgments}

\appendix

\section{Multipole notation under cubic and hexagonal point groups \label{sec:ap_real}}

We show the cubic expressions of $O_{lm}({\bm r})$ up to rank 4 as follows: 
the rank 0 is 
\begin{align}
O_0 = 1,
\end{align}
the rank 1 is 
\begin{align}
(O_x, O_y, O_z)= (x,y,z), 
\end{align}
the rank 2 is 
\begin{align}
O_u &= \frac{1}{2}(3z^2-r^2), \\
O_v &= \frac{\sqrt{3}}{2}(x^2-y^2), \\
(O_{yz}, O_{zx}, O_{xy}) &= \sqrt{3}(yz,zx,xy),
\end{align}
the rank 3 is 
\begin{align}
\label{eq:rank3_xyz}
O_{xyz} &= \sqrt{15}xyz, \\
\label{eq:rank3_alpha}
 \left(O_x^\alpha, O_y^\alpha, O_z^\alpha \right)  &= \frac{1}{2}\left( x(5x^2-3r^2), y(5y^2-3r^2), z(5z^2-3r^2) \right),  \\
\label{eq:rank3_beta}
 \left(O_x^\beta, O_y^\beta, O_z^\beta \right)   &= \frac{\sqrt{15}}{2}\left( x(y^2-z^2), y(z^2-x^2), z(x^2-y^2)\right), 
\end{align}
and the rank 4 is 
\begin{align}
\label{eq:rank4_4}
O_4 &= \frac{5\sqrt{21}}{12}\left( x^4+y^4+z^4-\frac{3}{5}r^4 \right), \\
\label{eq:rank4_4u}
O_{4u} &= \frac{7\sqrt{15}}{6}\left[ z^4-\frac{x^4+y^4}{2}-\frac{3}{7}r^2(3z^2-r^2)\right], \\
\label{eq:rank4_4v}
O_{4v} &= \frac{7\sqrt{5}}{4}\left[ x^4-y^4-\frac{6}{7}r^2(x^2-y^2) \right], \\
\label{eq:rank4_alpha}
\left(O_{4x}^\alpha, O_{4y}^\alpha, O_{4z}^\alpha \right) &= \frac{\sqrt{35}}{2}\left( yz(y^2-z^2), zx(z^2-x^2), xy(x^2-y^2) \right), \\
\label{eq:rank4_beta}
\left(O_{4x}^\beta, O_{4y}^\beta, O_{4z}^\beta \right) &= \frac{\sqrt{5}}{2} \left( yz(7x^2-r^2), zx(7y^2-r^2), xy(7z^2-r^2) \right), 
\end{align}
where we denote $O_{lm}({\bm r})\to O_{lm}$ for notation simplicity. 
In the case of the hexagonal and trigonal point groups, it is useful to replace 
$O_x^\alpha$, $O_y^\alpha$, $O_x^\beta$, $O_y^\beta$ with rank 3 in Eqs.~\eqref{eq:rank3_alpha} and \eqref{eq:rank3_beta} and $O_{lm}$ with rank 4 in Eqs.~\eqref{eq:rank4_4}--\eqref{eq:rank4_beta} by 
\begin{align}
O_{3a} &= \frac{\sqrt{10}}{4}x(x^2-3y^2), \\
O_{3b} &= \frac{\sqrt{10}}{4}y(3x^2-y^2), \\
(O_{3u}, O_{3v}) &=\frac{\sqrt{6}}{4} \left(x(5z^2-r^2),y(5z^2-r^2)\right), 
\end{align}
and 
\begin{align}
O_{40} &= \frac{1}{8}(35z^4-30z^2r^2+3r^4), \\
O_{4a} &=\frac{\sqrt{70}}{4} yz (3x^2-y^2), \\
O_{4b} &=\frac{\sqrt{70}}{4} zx (x^2-3y^2), \\
\left(O_{4u}^\alpha, O_{4v}^\alpha \right) &= \frac{\sqrt{10}}{4}\left( zx(7z^2-3r^2), yz(7z^2-3r^2) \right), \\
\left( O_{4u}^{\beta1}, O_{4v}^{\beta1} \right) &= \frac{\sqrt{35}}{8} \left(x^4-6x^2y^2+y^4, 4xy(x^2-y^2) \right), \\
\left(O_{4u}^{\beta2}, O_{4v}^{\beta2} \right) &= \frac{\sqrt{5}}{4} \left( (x^2-y^2)(7z^2-r^2), 2xy(7z^2-r^2) \right), 
\end{align}
with the use of the tesseral harmonics. 
It is noted that three components $\{O_{xyz}, O_z^\alpha, O_z^\beta\}$ among the rank-3 functions are common to the cubic harmonics in Eqs.~\eqref{eq:rank3_xyz}--\eqref{eq:rank3_beta}.

\section{Tensor expression in hexagonal/trigonal system \label{sec:ap_hexagonal}}

The rank-3 and -4 tensors in hexagonal and trigonal system are represented by
\begin{align}
\chi^{[0\times3]}_{\mathrm{(o/s)}}
=
.
\end{aligned}
$}
\end{equation}
\end{widetext}
\section{Collinear spin point group with non-unitary nontrivial group and its unitary subgroup \label{sec:ap_SPG_subgroup}}

The non-unitary nontrivial group of collinear SPGs can be decomposed into its halving unitary subgroup $[E \,\|\, \bm{H}]$ and the antiunitary coset $[E \,\|\, \bm{H}] + [\theta \,\|\, \bm{G} - \bm{H}]$ in Eq.~\eqref{eq:nontrivial};
Their correspondence is summarized in Table~\ref{table:subgroup}
~\cite{tavger1956magnetic,dimmock1964symmetry,Hamermesh:100343,bradley2009mathematical,dimmock1962irreducible,cracknell1966corepresentations}.

\renewcommand{\arraystretch}{1.5}
\begin{table*}[htb!]
\centering
\caption{ 
List of 58 non-unitary nontrivial spin point groups $\mathcal{G}_{\rm NT} = [E \,\|\, \bm{H}] + [\theta \,\|\, \bm{G} - \bm{H}] $.
The parentheses represent the corresponding unitary subgroup $[E \,\|\, \bm{H}]$.
\label{table:subgroup}}
\begin{tabular}{l@{\hspace{3mm}}l@{\hspace{3mm}}l@{\hspace{3mm}}l@{\hspace{5mm}}l@{\hspace{5mm}}l@{\hspace{3mm}}l}
\hline\hline
Cubic & Tetragonal & Orthorhombic & Monoclinic & Triclinic & Hexagonal & Trigonal \\ \hline
${}^1 4/ {}^{\bar{1}} m {}^{\bar{1}} \bar{3} {}^1 2/{}^{\bar{1}} m$ (${}^1 4 {}^1 3 {}^1 2$) 
& ${}^1 4/{}^{\bar{1}} m {}^{\bar{1}} m {}^{\bar{1}} m$ (${}^1 4 {}^1 2 {}^1 2$) 
& ${}^{\bar{1}} m {}^{\bar{1}} m {}^{\bar{1}} m$ (${}^1 2 {}^1 2 {}^1 2$) 
& ${}^{\bar{1}} 2/{}^{\bar{1}} m$ (${}^1 \bar{1}$) 
& ${}^{\bar{1}} \bar{1}$ (${}^1 1$) 
& ${}^1 6/{}^{\bar{1}} m {}^{\bar{1}} m {}^{\bar{1}} m$ (${}^1 6 {}^1 2 {}^1 2$) 
& ${}^1 \bar{3} {}^{\bar{1}} m$ (${}^1 \bar{3}$) \rule[0pt]{0pt}{8pt} 
\\
${}^{\bar{1}} 4/{}^1 m {}^1 \bar{3} {}^{\bar{1}} 2/{}^{\bar{1}} m$ (${}^1 2/{}^1 m {}^1 \bar{3}$) 
& ${}^1 4/{}^1 m {}^{\bar{1}} m {}^{\bar{1}} m$ (${}^1 4/{}^1 m$) 
& ${}^{\bar{1}} m {}^{\bar{1}} m {}^1 m$ (${}^1 2/{}^1 m$)
& ${}^1 2/{}^{\bar{1}} m$ (${}^1 2$) 
& & ${}^1 6/{}^1 m {}^{\bar{1}} m {}^{\bar{1}} m$ (${}^1 6/{}^1 m$) & 
${}^{\bar{1}} \bar{3} {}^{\bar{1}} m$ (${}^1 3 {}^1 2$)  
\\
${}^{\bar{1}} 4/{}^{\bar{1}} m {}^{\bar{1}} \bar{3} {}^{\bar{1}} 2/{}^1 m$ (${}^1 \bar{4} {}^1 3 {}^1 m$) 
& ${}^{\bar{1}} 4/{}^{\bar{1}} m {}^{\bar{1}} m {}^1 m$ (${}^1 \bar{4} {}^1 2 {}^1 m$) 
& ${}^{\bar{1}} m {}^1 m {}^1 m$ (${}^1 2 {}^1 m {}^1 m$)
& ${}^{\bar{1}} 2/{}^1 m$ (${}^1 m$) 
& & ${}^{\bar{1}} 6/{}^{\bar{1}} m {}^1 m {}^{\bar{1}} m$ (${}^1 \bar{3} {}^1 m$) 
& ${}^{\bar{1}} \bar{3} {}^1 m$ (${}^1 3 {}^1 m$) 
\\
${}^{\bar{1}} \bar{4} {}^1 3 {}^{\bar{1}} m$ (${}^1 2 {}^1 3$) 
& ${}^{\bar{1}} 4/{}^1 m {}^{\bar{1}} m {}^1 m$ (${}^1 m {}^1 m {}^1 m$) 
& ${}^{\bar{1}} 2 {}^{\bar{1}} 2 {}^1 2$ (${}^1 2$) 
& ${}^{\bar{1}} m$ (${}^1 1$) 
& &${}^{\bar{1}} 6/{}^1 m {}^1 m {}^{\bar{1}} m$ (${}^1 \bar{6} {}^1 m {}^1 2$) 
& ${}^1 3 {}^{\bar{1}} 2$ (${}^1 3$)
\\
\hline \hline
\end{tabular}
\end{table*}

\clearpage

\section{Multipole classification \label{sec:ap_classification}}

We present the complete classification tables of active orbital and spin multipoles for all nonmagnetic and collinear SPGs. 
The tables are organized into four categories:
(i) 32 collinear SPGs with unitary nontrivial groups,
(ii) 58 collinear SPGs with non-unitary nontrivial groups, 
(iii) 32 collinear SPGs with $\mathcal{T}$-symmetric nontrivial groups, and
(iv) 32 nonmagnetic SPGs.
Table~\ref{table:Tablelist_tetragonal} provides a list of the classification tables for tetragonal, orthorhombic, monoclinic, and triclinic systems, Table~\ref{table:Tablelist_cubic} for cubic systems, and Table~\ref{table:Tablelist_hexagonal} for hexagonal and trigonal systems, with each table indicating the correspondence between the parent crystallographic point groups, the overall SPGs, and the corresponding classification table numbers in this appendix.
The notation in terms of primary, secondary, and tertiary axes with respect to the symmetry operations is shown in Table~\ref{table:op_axis}.

\begin{table*}
\caption{
List of the classification tables for tetragonal, orthorhombic, monoclinic, and triclinic systems.
\label{table:Tablelist_tetragonal}}
\begingroup
\renewcommand{\arraystretch}{1.5} 
\begin{tabular}{l@{\hspace{5mm}}l@{\hspace{5mm}}l@{\hspace{8mm}}|l@{\hspace{5mm}}l@{\hspace{5mm}}l@{\hspace{5mm}}}
\hline\hline
Parent group 
& Overall SPG 
& Table \#
& Parent group 
& Overall SPG 
& Table \#
\\
\hline
$1$ $(C_1)$
& $\mathcal{G}_{\mathrm{SO}}^{\mathrm{NM}} \times {}^1 1$
& 
&
$4$ $(C_4)$
& $\mathcal{G}_{\mathrm{SO}}^{\mathrm{NM}} \times {}^1 4$
& \ref{table:Litvin90nonmag}
\\
& $\mathcal{G}_{\mathrm{SO}}^{\infty} \times {}^1 1 {}^{\bar{1}} 1 $
& \ref{table:Tsymmetric1collinear}
&
& $\mathcal{G}_{\mathrm{SO}}^{\infty} \times {}^1 4 {}^{\bar{1}} 1 $
& \ref{table:Tsymmetric90collinear}
\\
& $\mathcal{G}_{\mathrm{SO}}^{\infty} \times {}^1 1$
& 
&
& $\mathcal{G}_{\mathrm{SO}}^{\infty} \times {}^{\bar{1}} 4$
& \ref{table:Litvin93collinear}
\\
$\bar{1}$ $(C_{\mathrm{i}})$
& $\mathcal{G}_{\mathrm{SO}}^{\mathrm{NM}} \times {}^1 \bar{1}$
& \ref{table:Litvin2nonmag}
&
& $\mathcal{G}_{\mathrm{SO}}^{\infty} \times {}^1 4$
& \ref{table:Litvin90collinear}
\\
& $\mathcal{G}_{\mathrm{SO}}^{\infty} \times {}^1 \bar{1} {}^{\bar{1}} 1 $
& \ref{table:Tsymmetric2collinear}
&
$\bar{4}$ $(S_4)$
& $\mathcal{G}_{\mathrm{SO}}^{\mathrm{NM}} \times {}^1 \bar{4}$
& \ref{table:Litvin96nonmag}
\\
& $\mathcal{G}_{\mathrm{SO}}^{\infty} \times {}^{\bar{1}} \bar{1}$
& \ref{table:Litvin5collinear}
&
& $\mathcal{G}_{\mathrm{SO}}^{\infty} \times {}^1 \bar{4} {}^{\bar{1}} 1 $
& \ref{table:Tsymmetric96collinear}
\\
& $\mathcal{G}_{\mathrm{SO}}^{\infty} \times {}^1 \bar{1}$
& \ref{table:Litvin2collinear}
&
& $\mathcal{G}_{\mathrm{SO}}^{\infty} \times {}^{\bar{1}} \bar{4}$
& \ref{table:Litvin99collinear}
\\
\hline\hline
\end{tabular}
\endgroup
\end{table*}

\begin{table*}
\caption{
List of the classification tables for cubic systems.
\label{table:Tablelist_cubic}}
\begingroup
\renewcommand{\arraystretch}{1.5} 
\begin{tabular}{l@{\hspace{5mm}}l@{\hspace{5mm}}l@{\hspace{5mm}}}
\hline\hline
Parent group 
& Overall SPG 
& Table \#
\\
\hline
$23$ ($T$)
& $\mathcal{G}_{\mathrm{SO}}^{\mathrm{NM}} \times {}^1 2 {}^1 3$
& \ref{table:Litvin538nonmag}
\\
& $\mathcal{G}_{\mathrm{SO}}^{\infty} \times {}^1 2 {}^1 3 {}^{\bar{1}} 1 $
& \ref{table:Tsymmetric538collinear}
\\
& $\mathcal{G}_{\mathrm{SO}}^{\infty} \times {}^1 2 {}^1 3$
& \ref{table:Litvin538collinear}
\\
$2/m\bar{3}$ ($T_{\rm h}$)
& $\mathcal{G}_{\mathrm{SO}}^{\mathrm{NM}} \times {}^1 2 / {}^1 m {}^1 \bar{3}$
& \ref{table:Litvin541nonmag}
\\
& $\mathcal{G}_{\mathrm{SO}}^{\infty} \times {}^1 2 / {}^1 m {}^1 \bar{3} {}^{\bar{1}} 1 $
& \ref{table:Tsymmetric541collinear}
\\
& $\mathcal{G}_{\mathrm{SO}}^{\infty} \times {}^1 2 / {}^{\bar{1}} m {}^{\bar{1}} \bar{3}$
& \ref{table:Litvin544collinear}
\\
& $\mathcal{G}_{\mathrm{SO}}^{\infty} \times {}^1 2 / {}^1 m {}^1 \bar{3}$
& \ref{table:Litvin541collinear}
\\
\hline\hline
\end{tabular}
\endgroup
\end{table*}

\begin{table*}
\caption{
List of the classification tables for hexagonal and trigonal systems.
\label{table:Tablelist_hexagonal}}
\begingroup
\renewcommand{\arraystretch}{1.5} 
\begin{tabular}{l@{\hspace{5mm}}l@{\hspace{5mm}}l@{\hspace{8mm}}|l@{\hspace{5mm}}l@{\hspace{5mm}}l@{\hspace{5mm}}}
\hline\hline
Parent group 
& Overall SPG 
& Table \#
& Parent group 
& Overall SPG 
& Table \#
\\
\hline
$3$ $(C_3)$
& $\mathcal{G}_{\mathrm{SO}}^{\mathrm{NM}} \times {}^1 3$
& \ref{table:Litvin272nonmag}
&
$6/m$ $(C_{\rm 6h})$
& $\mathcal{G}_{\mathrm{SO}}^{\mathrm{NM}} \times {}^1 6 / {}^1 m$
& \ref{table:Litvin357nonmag}
\\
& $\mathcal{G}_{\mathrm{SO}}^{\infty} \times {}^1 3 {}^{\bar{1}} 1 $
& \ref{table:Tsymmetric272collinear}
&
& $\mathcal{G}_{\mathrm{SO}}^{\infty} \times {}^1 6 / {}^1 m {}^{\bar{1}} 1$
& \ref{table:Tsymmetric357collinear}
\\
& $\mathcal{G}_{\mathrm{SO}}^{\infty} \times {}^1 3$
& \ref{table:Litvin272collinear}
&
& $\mathcal{G}_{\mathrm{SO}}^{\infty} \times {}^{\bar{1}} 6 / {}^{\bar{1}} m$
& \ref{table:Litvin360collinear}
\\
$\bar{3}$ $(C_{\mathrm{3i}})$
& $\mathcal{G}_{\mathrm{SO}}^{\mathrm{NM}} \times {}^1 \bar{3}$
& \ref{table:Litvin274nonmag}
&
& $\mathcal{G}_{\mathrm{SO}}^{\infty} \times {}^{\bar{1}} 6 / {}^1 m$
& \ref{table:Litvin363collinear}
\\
& $\mathcal{G}_{\mathrm{SO}}^{\infty} \times {}^1 \bar{3} {}^{\bar{1}} 1 $
& \ref{table:Tsymmetric274collinear}
&
& $\mathcal{G}_{\mathrm{SO}}^{\infty} \times {}^1 6 / {}^{\bar{1}} m$
& \ref{table:Litvin366collinear}
\\
& $\mathcal{G}_{\mathrm{SO}}^{\infty} \times {}^{\bar{1}} \bar{3}$
& \ref{table:Litvin277collinear}
&
& $\mathcal{G}_{\mathrm{SO}}^{\infty} \times {}^1 6 / {}^1 m$
& \ref{table:Litvin357collinear}
\\
& $\mathcal{G}_{\mathrm{SO}}^{\infty} \times {}^1 \bar{3}$
& \ref{table:Litvin274collinear}
&
$622$ $(D_6)$
& $\mathcal{G}_{\mathrm{SO}}^{\mathrm{NM}} \times {}^1 6 {}^1 2 {}^1 2$
& \ref{table:Litvin338nonmag}
\\
\hline\hline
\end{tabular}
\endgroup
\end{table*}

\begin{table}[htb!]
\centering
\caption{
Primary, secondary, and tertiary axes with respect to the symmetry operations in the Cartesian coordinates. 
 \label{table:op_axis}}
\begin{tabular}{cccc}
\hline\hline
& Primary & Secondary & Tertiary \\ \hline
Cubic & $\left<100\right>$ & $\left<111\right>$ & $\left<110\right>$  \\
Tetragonal & $[001]$ & $[100]$ & $[110]$ \\
Orthorhombic & $[100]$ & $[010]$ & $[001]$ \\
Monoclinic & $[010]$ &  & \\
Triclinic &  &  & \\
Hexagonal & $[001]$ & $[100]$ & $[010]$ \\
Trigonal & $[001]$ & $[010]$ & \\
\hline \hline
\end{tabular}
\end{table}

\clearpage 

\subsection{Collinear spin point groups with unitary nontrivial groups \label{sec:ap_classification_unitaryCSPG}}

The complete classification of orbital and spin multipoles for the 32 collinear SPGs with unitary nontrivial groups is summarized in Tables~\ref{table:Litvin2collinear} -- ~\ref{table:Litvin440collinear}.

\renewcommand{\arraystretch}{1.8}
\begin{table*}[t]
\begin{center}
\caption{
IRREPs of multipoles in the collinear SPG
$\mathcal{G}_{\mathrm{SO}}^\infty \times {}^1 \bar{1} 
= {\rm SO(2)}\times ( {}^1 \bar{1} + [\theta 2_{\perp\bm{n}} \,\|\, E] {}^1 \bar{1} )$.
The absence [presence] of a bar over the IRREPs indicates parity +1 [$-1$] under the antiunitary operation $[\theta 2_{\perp\bm{n}} \,\|\, E]$. 
}
\label{table:Litvin2collinear}
\begin{tabular}{|c|c|c|c|c|c|c|c|c|c|} 
\hline\hline 
IRREP  &  $[E \,\|\, E]$  &  $[E \,\|\, I]$ & $[\theta 2_{\perp\bm{n}} \,\|\, E]$ & $[\theta 2_{\perp\bm{n}} \,\|\, I]$ &  E  &  M  &  ET  &   MT       
\\ 
\hline
${\rm A_g}$
& $1$  & $1$ & $1$  & $1$ 
& $Q_{lm}^{\rm (o)}(l:{\rm even}),$  
& 
& $G_{lm}^{\rm (o)}(l:{\rm odd}),$ 
& 
\\  
& & & & 
& $Q_{lm}^{\rm (s)}(l:{\rm even})$
& 
& $G_{lm}^{\rm (s)}(l:{\rm odd})$
& 
\\ 
$\bar{{\rm A}}_{\rm g}$
& $1$  & $1$ & $-1$  & $-1$ 
& 
& $M_{lm}^{\rm (o)}(l:{\rm odd}),$ 
&   
& $T_{lm}^{\rm (o)}(l:{\rm even}),$
\\  
& & & & 
& 
& $M_{lm}^{\rm (s)}(l:{\rm odd})$
& 
& $T_{lm}^{\rm (s)}(l:{\rm even})$
\\ \hline 
${\rm A_u}$
& $1$  & $-1$ & $1$  & $-1$ 
& $Q_{lm}^{\rm (o)}(l:{\rm odd}),$  
&  
& $G_{lm}^{\rm (o)}(l:{\rm even}),$
& 
\\  
& & & & 
& $Q_{lm}^{\rm (s)}(l:{\rm odd})$
& 
& $G_{lm}^{\rm (s)}(l:{\rm even})$
&
\\ 
$\bar{{\rm A}}_{\rm u}$
& $1$  & $-1$ & $-1$  & $1$ 
& 
& $M_{lm}^{\rm (o)} (l:{\rm even}),$ 
&  
& $T_{lm}^{\rm (o)}(l:{\rm odd}),$
\\  
& & & & 
& 
& $M_{lm}^{\rm (s)}(l:{\rm even})$
& 
& $T_{lm}^{\rm (s)}(l:{\rm odd})$
\\ 
\hline\hline 
\end{tabular}
\end{center}
\end{table*}

\renewcommand{\arraystretch}{1.5}
\begin{table*}
\caption{
IRREPs of multipoles in the collinear SPG
$\mathcal{G}_{\mathrm{SO}}^\infty \times {}^1 2 
= \mathrm{SO(2)}\times ( {}^1 2 + [\theta 2_{\perp\bm{n}} \,\|\, E] {}^1 2 )$.
All the symmetry operations of ${}^1 2 + [\theta 2_{\perp\bm{n}} \,\|\, E] {}^1 2$ are listed.
The absence [presence] of a bar over the IRREPs indicates parity +1 [$-1$] under the antiunitary operation $[\theta 2_{\perp\bm{n}} \,\|\, E]$. 
}
\label{table:Litvin6collinear}
\begin{tabular}{|c|c|c|c|c|c|c|}
\hline\hline
IRREP & \rotatebox{90}{$[E \,\|\, E]$} & \rotatebox{90}{$[E \,\|\, C_{2y}]$} & \rotatebox{90}{$[\theta 2_{\perp \bm{n}} \,\|\, E]$} & \rotatebox{90}{$[\theta 2_{\perp \bm{n}} \,\|\, C_{2y}]$} & E, M & ET, MT 
\\
\hline
$\mathrm{A}$ 
& $1$ & $1$ & $1$ & $1$ 
& $Q_{0}^{\mathrm{(o)}}$, $Q_{y}^{\mathrm{(o)}}$, $Q_{u}^{\mathrm{(o)}}$, $Q_{v}^{\mathrm{(o)}}$, $Q_{zx}^{\mathrm{(o)}}$, $Q_{xyz}^{\mathrm{(o)}}$, 
& $G_{0}^{\mathrm{(o)}}$, $G_{y}^{\mathrm{(o)}}$, $G_{u}^{\mathrm{(o)}}$, $G_{v}^{\mathrm{(o)}}$, $G_{zx}^{\mathrm{(o)}}$, $G_{xyz}^{\mathrm{(o)}}$, 
\\  
& & & & 
& 
$Q_{y}^{\alpha \mathrm{(o)}}$, $Q_{y}^{\beta \mathrm{(o)}}$, $Q_{4}^{\mathrm{(o)}}$, $Q_{4u}^{\mathrm{(o)}}$, $Q_{4v}^{\mathrm{(o)}}$, $Q_{4y}^{\alpha \mathrm{(o)}}$, $Q_{4y}^{\beta \mathrm{(o)}}$, 
&
$G_{y}^{\alpha \mathrm{(o)}}$, $G_{y}^{\beta \mathrm{(o)}}$, $G_{4}^{\mathrm{(o)}}$, $G_{4u}^{\mathrm{(o)}}$, $G_{4v}^{\mathrm{(o)}}$, $G_{4y}^{\alpha \mathrm{(o)}}$, $G_{4y}^{\beta \mathrm{(o)}}$, 
\\  
& & & & 
& 
$Q_{0}^{\mathrm{(s)}}$, $Q_{y}^{\mathrm{(s)}}$, $Q_{u}^{\mathrm{(s)}}$, $Q_{v}^{\mathrm{(s)}}$, $Q_{zx}^{\mathrm{(s)}}$, $Q_{xyz}^{\mathrm{(s)}}$, 
&
$G_{0}^{\mathrm{(s)}}$, $G_{y}^{\mathrm{(s)}}$, $G_{u}^{\mathrm{(s)}}$, $G_{v}^{\mathrm{(s)}}$, $G_{zx}^{\mathrm{(s)}}$, $G_{xyz}^{\mathrm{(s)}}$, 
\\  
& & & & 
& 
$Q_{y}^{\alpha \mathrm{(s)}}$, $Q_{y}^{\beta \mathrm{(s)}}$, $Q_{4}^{\mathrm{(s)}}$, $Q_{4u}^{\mathrm{(s)}}$, $Q_{4v}^{\mathrm{(s)}}$, $Q_{4y}^{\alpha \mathrm{(s)}}$, $Q_{4y}^{\beta \mathrm{(s)}}$ 
&
$G_{y}^{\alpha \mathrm{(s)}}$, $G_{y}^{\beta \mathrm{(s)}}$, $G_{4}^{\mathrm{(s)}}$, $G_{4u}^{\mathrm{(s)}}$, $G_{4v}^{\mathrm{(s)}}$, $G_{4y}^{\alpha \mathrm{(s)}}$, $G_{4y}^{\beta \mathrm{(s)}}$ 
\\
\hline
\hline
\end{tabular}
\end{table*}

\begin{table*}
\caption{
IRREPs of multipoles in the collinear SPG
$\mathcal{G}_{\mathrm{SO}}^\infty \times {}^1 m 
= \mathrm{SO(2)}\times ( {}^1 m + [\theta 2_{\perp\bm{n}} \,\|\, E] {}^1 m )$.
All the symmetry operations of ${}^1 m + [\theta 2_{\perp\bm{n}} \,\|\, E] {}^1 m$ are listed.
The absence [presence] of a bar over the IRREPs indicates parity +1 [$-1$] under the antiunitary operation $[\theta 2_{\perp\bm{n}} \,\|\, E]$. 
}
\label{table:Litvin10collinear}
\begin{tabular}{|c|c|c|c|c|c|c|}
\hline\hline
IRREP & \rotatebox{90}{$[E \,\|\, E]$} & \rotatebox{90}{$[E \,\|\, IC_{2y}]$} & \rotatebox{90}{$[\theta 2_{\perp \bm{n}} \,\|\, E]$} & \rotatebox{90}{$[\theta 2_{\perp \bm{n}} \,\|\, IC_{2y}]$} & E, M & ET, MT 
\\
\hline
$\mathrm{A}^{\prime}$ 
& $1$ & $1$ & $1$ & $1$ 
& $Q_{0}^{\mathrm{(o)}}$, $Q_{x}^{\mathrm{(o)}}$, $Q_{z}^{\mathrm{(o)}}$, $Q_{u}^{\mathrm{(o)}}$, $Q_{v}^{\mathrm{(o)}}$, $Q_{zx}^{\mathrm{(o)}}$, $Q_{x}^{\alpha \mathrm{(o)}}$, $Q_{z}^{\alpha \mathrm{(o)}}$,
& $G_{y}^{\mathrm{(o)}}$, $G_{yz}^{\mathrm{(o)}}$, $G_{xy}^{\mathrm{(o)}}$, $G_{xyz}^{\mathrm{(o)}}$, $G_{y}^{\alpha \mathrm{(o)}}$, $G_{y}^{\beta \mathrm{(o)}}$, 
\\
& & & & 
& 
$Q_{x}^{\beta \mathrm{(o)}}$, $Q_{z}^{\beta \mathrm{(o)}}$, $Q_{4}^{\mathrm{(o)}}$, $Q_{4u}^{\mathrm{(o)}}$, $Q_{4v}^{\mathrm{(o)}}$, $Q_{4y}^{\alpha \mathrm{(o)}}$, $Q_{4y}^{\beta \mathrm{(o)}}$, 
&
$G_{4x}^{\alpha \mathrm{(o)}}$, $G_{4z}^{\alpha \mathrm{(o)}}$, $G_{4x}^{\beta \mathrm{(o)}}$, $G_{4z}^{\beta \mathrm{(o)}}$, 
\\  
& & & & 
& 
$Q_{0}^{\mathrm{(s)}}$, $Q_{x}^{\mathrm{(s)}}$, $Q_{z}^{\mathrm{(s)}}$, $Q_{u}^{\mathrm{(s)}}$, $Q_{v}^{\mathrm{(s)}}$, $Q_{zx}^{\mathrm{(s)}}$, $Q_{x}^{\alpha \mathrm{(s)}}$, $Q_{z}^{\alpha \mathrm{(s)}}$, 
&
$G_{y}^{\mathrm{(s)}}$, $G_{yz}^{\mathrm{(s)}}$, $G_{xy}^{\mathrm{(s)}}$, $G_{xyz}^{\mathrm{(s)}}$, $G_{y}^{\alpha \mathrm{(s)}}$, $G_{y}^{\beta \mathrm{(s)}}$, 
\\  
& & & & 
& 
$Q_{x}^{\beta \mathrm{(s)}}$, $Q_{z}^{\beta \mathrm{(s)}}$, $Q_{4}^{\mathrm{(s)}}$, $Q_{4u}^{\mathrm{(s)}}$, $Q_{4v}^{\mathrm{(s)}}$, $Q_{4y}^{\alpha \mathrm{(s)}}$, $Q_{4y}^{\beta \mathrm{(s)}}$ 
&
$G_{4x}^{\alpha \mathrm{(s)}}$, $G_{4z}^{\alpha \mathrm{(s)}}$, $G_{4x}^{\beta \mathrm{(s)}}$, $G_{4z}^{\beta \mathrm{(s)}}$
\\
\hline
\hline
\end{tabular}
\end{table*}

\begin{table*}
\caption{
IRREPs of multipoles in the collinear SPG
$\mathcal{G}_{\mathrm{SO}}^\infty \times {}^1 2 / {}^1 m 
= \mathrm{SO(2)}\times ( {}^1 2 / {}^1 m + [\theta 2_{\perp\bm{n}} \,\|\, E] {}^1 2 / {}^1 m )$.
All the symmetry operations of ${}^1 2 / {}^1 m + [\theta 2_{\perp\bm{n}} \,\|\, E] {}^1 2 / {}^1 m$ are listed.
The absence [presence] of a bar over the IRREPs indicates parity +1 [$-1$] under the antiunitary operation $[\theta 2_{\perp\bm{n}} \,\|\, E]$. 
}
\label{table:Litvin14collinear}
\begin{tabular}{|c|c|c|c|c|c|c|c|c|c|c|}
\hline\hline
IRREP & \rotatebox{90}{$[E \,\|\, E]$} & \rotatebox{90}{$[E \,\|\, C_{2y}]$} & \rotatebox{90}{$[E \,\|\, I]$} & \rotatebox{90}{$[E \,\|\, IC_{2y}]$} & \rotatebox{90}{$[\theta 2_{\perp \bm{n}} \,\|\, E]$} & \rotatebox{90}{$[\theta 2_{\perp \bm{n}} \,\|\, C_{2y}]$} & \rotatebox{90}{$[\theta 2_{\perp \bm{n}} \,\|\, I]$} & \rotatebox{90}{$[\theta 2_{\perp \bm{n}} \,\|\, IC_{2y}]$} & E, M & ET, MT 
\\
\hline
$\mathrm{A}_{\mathrm{g}}$ 
& $1$ & $1$ & $1$ & $1$ & $1$ & $1$ & $1$ & $1$ 
& $Q_{0}^{\mathrm{(o)}}$, $Q_{u}^{\mathrm{(o)}}$, $Q_{v}^{\mathrm{(o)}}$, $Q_{zx}^{\mathrm{(o)}}$, 
& $G_{y}^{\mathrm{(o)}}$, $G_{xyz}^{\mathrm{(o)}}$, $G_{y}^{\alpha \mathrm{(o)}}$, $G_{y}^{\beta \mathrm{(o)}}$, 
\\  
& & & & & & & & 
&
$Q_{4}^{\mathrm{(o)}}$, $Q_{4u}^{\mathrm{(o)}}$, $Q_{4v}^{\mathrm{(o)}}$, $Q_{4y}^{\alpha \mathrm{(o)}}$, $Q_{4y}^{\beta \mathrm{(o)}}$, 
&
$G_{y}^{\mathrm{(s)}}$, $G_{xyz}^{\mathrm{(s)}}$, $G_{y}^{\alpha \mathrm{(s)}}$, $G_{y}^{\beta \mathrm{(s)}}$ 
\\  
& & & & & & & & 
&
$Q_{0}^{\mathrm{(s)}}$, $Q_{u}^{\mathrm{(s)}}$, $Q_{v}^{\mathrm{(s)}}$, $Q_{zx}^{\mathrm{(s)}}$, 
&
\\  
& & & & & & & & 
&
$Q_{4}^{\mathrm{(s)}}$, $Q_{4u}^{\mathrm{(s)}}$, $Q_{4v}^{\mathrm{(s)}}$, $Q_{4y}^{\alpha \mathrm{(s)}}$, $Q_{4y}^{\beta \mathrm{(s)}}$ 
&
\\
$\bar{\mathrm{A}}_{\mathrm{g}}$ 
& $1$ & $1$ & $1$ & $1$ & $-1$ & $-1$ & $-1$ & $-1$ 
& $M_{y}^{\mathrm{(o)}}$, $M_{xyz}^{\mathrm{(o)}}$, $M_{y}^{\alpha \mathrm{(o)}}$, $M_{y}^{\beta \mathrm{(o)}}$, 
& $T_{0}^{\mathrm{(o)}}$, $T_{u}^{\mathrm{(o)}}$, $T_{v}^{\mathrm{(o)}}$, $T_{zx}^{\mathrm{(o)}}$, 
\\  
& & & & & & & & 
&
$M_{y}^{\mathrm{(s)}}$, $M_{xyz}^{\mathrm{(s)}}$, $M_{y}^{\alpha \mathrm{(s)}}$, $M_{y}^{\beta \mathrm{(s)}}$ 
&
$T_{4}^{\mathrm{(o)}}$, $T_{4u}^{\mathrm{(o)}}$, $T_{4v}^{\mathrm{(o)}}$, $T_{4y}^{\alpha \mathrm{(o)}}$, $T_{4y}^{\beta \mathrm{(o)}}$, 
\\
\hline
\hline
\end{tabular}
\end{table*}

\begin{table*}
\caption{
IRREPs of multipoles in the collinear SPG
$\mathcal{G}_{\mathrm{SO}}^\infty \times {}^1 m {}^1 m {}^1 2 
= \mathrm{SO(2)}\times ( {}^1 m {}^1 m {}^1 2 + [\theta 2_{\perp\bm{n}} \,\|\, E] {}^1 m {}^1 m {}^1 2 )$.
All the symmetry operations of ${}^1 m {}^1 m {}^1 2 + [\theta 2_{\perp\bm{n}} \,\|\, E] {}^1 m {}^1 m {}^1 2$ are listed.
The absence [presence] of a bar over the IRREPs indicates parity +1 [$-1$] under the antiunitary operation $[\theta 2_{\perp\bm{n}} \,\|\, E]$. 
}
\label{table:Litvin34collinear}
\begin{tabular}{|c|c|c|c|c|c|c|c|c|c|c|}
\hline\hline
IRREP & \rotatebox{90}{$[E \,\|\, E]$} & \rotatebox{90}{$[E \,\|\, C_{2z}]$} & \rotatebox{90}{$[E \,\|\, IC_{2y}]$} & \rotatebox{90}{$[E \,\|\, IC_{2x}]$} & \rotatebox{90}{$[\theta 2_{\perp \bm{n}} \,\|\, E]$} & \rotatebox{90}{$[\theta 2_{\perp \bm{n}} \,\|\, C_{2z}]$} & \rotatebox{90}{$[\theta 2_{\perp \bm{n}} \,\|\, IC_{2y}]$} & \rotatebox{90}{$[\theta 2_{\perp \bm{n}} \,\|\, IC_{2x}]$} & E, M & ET, MT 
\\
\hline
$\mathrm{A}_{1}$ 
& $1$ & $1$ & $1$ & $1$ & $1$ & $1$ & $1$ & $1$ 
& $Q_{0}^{\mathrm{(o)}}$, $Q_{z}^{\mathrm{(o)}}$, $Q_{u}^{\mathrm{(o)}}$, $Q_{v}^{\mathrm{(o)}}$, $Q_{z}^{\alpha \mathrm{(o)}}$, $Q_{z}^{\beta \mathrm{(o)}}$, 
& $G_{xy}^{\mathrm{(o)}}$, $G_{xyz}^{\mathrm{(o)}}$, $G_{4z}^{\alpha \mathrm{(o)}}$, $G_{4z}^{\beta \mathrm{(o)}}$, 
\\  
& & & & & & & & 
&
$Q_{4}^{\mathrm{(o)}}$, $Q_{4u}^{\mathrm{(o)}}$, $Q_{4v}^{\mathrm{(o)}}$, 
&
$G_{xy}^{\mathrm{(s)}}$, $G_{xyz}^{\mathrm{(s)}}$, $G_{4z}^{\alpha \mathrm{(s)}}$, $G_{4z}^{\beta \mathrm{(s)}}$ 
\\  
& & & & & & & & 
&
$Q_{0}^{\mathrm{(s)}}$, $Q_{z}^{\mathrm{(s)}}$, $Q_{u}^{\mathrm{(s)}}$, $Q_{v}^{\mathrm{(s)}}$, $Q_{z}^{\alpha \mathrm{(s)}}$, $Q_{z}^{\beta \mathrm{(s)}}$, 
&
\\  
& & & & & & & & 
&
$Q_{4}^{\mathrm{(s)}}$, $Q_{4u}^{\mathrm{(s)}}$, $Q_{4v}^{\mathrm{(s)}}$ 
&
\\
$\bar{\mathrm{A}}_{1}$ 
& $1$ & $1$ & $1$ & $1$ & $-1$ & $-1$ & $-1$ & $-1$ 
& $M_{xy}^{\mathrm{(o)}}$, $M_{xyz}^{\mathrm{(o)}}$, $M_{4z}^{\alpha \mathrm{(o)}}$, $M_{4z}^{\beta \mathrm{(o)}}$, 
& $T_{0}^{\mathrm{(o)}}$, $T_{z}^{\mathrm{(o)}}$, $T_{u}^{\mathrm{(o)}}$, $T_{v}^{\mathrm{(o)}}$, $T_{z}^{\alpha \mathrm{(o)}}$, $T_{z}^{\beta \mathrm{(o)}}$, 
\\  
& & & & & & & & 
&
$M_{xy}^{\mathrm{(s)}}$, $M_{xyz}^{\mathrm{(s)}}$, $M_{4z}^{\alpha \mathrm{(s)}}$, $M_{4z}^{\beta \mathrm{(s)}}$ 
&
$T_{4}^{\mathrm{(o)}}$, $T_{4u}^{\mathrm{(o)}}$, $T_{4v}^{\mathrm{(o)}}$, 
\\
\hline
\hline
\end{tabular}
\end{table*}

\begin{table*}
\caption{
IRREPs of multipoles in the collinear SPG
$\mathcal{G}_{\mathrm{SO}}^\infty \times {}^1 2 {}^1 2 {}^1 2 
= \mathrm{SO(2)}\times ( {}^1 2 {}^1 2 {}^1 2 + [\theta 2_{\perp\bm{n}} \,\|\, E] {}^1 2 {}^1 2 {}^1 2 )$.
All the symmetry operations of ${}^1 2 {}^1 2 {}^1 2 + [\theta 2_{\perp\bm{n}} \,\|\, E] {}^1 2 {}^1 2 {}^1 2$ are listed.
The absence [presence] of a bar over the IRREPs indicates parity +1 [$-1$] under the antiunitary operation $[\theta 2_{\perp\bm{n}} \,\|\, E]$. 
}
\label{table:Litvin47collinear}
\begin{tabular}{|c|c|c|c|c|c|c|c|c|c|c|}
\hline\hline
IRREP & \rotatebox{90}{$[E \,\|\, E]$} & \rotatebox{90}{$[E \,\|\, C_{2z}]$} & \rotatebox{90}{$[E \,\|\, C_{2x}]$} & \rotatebox{90}{$[E \,\|\, C_{2y}]$} & \rotatebox{90}{$[\theta 2_{\perp \bm{n}} \,\|\, E]$} & \rotatebox{90}{$[\theta 2_{\perp \bm{n}} \,\|\, C_{2z}]$} & \rotatebox{90}{$[\theta 2_{\perp \bm{n}} \,\|\, C_{2x}]$} & \rotatebox{90}{$[\theta 2_{\perp \bm{n}} \,\|\, C_{2y}]$} & E, M & ET, MT 
\\
\hline
$\mathrm{A}$ 
& $1$ & $1$ & $1$ & $1$ & $1$ & $1$ & $1$ & $1$ 
& $Q_{0}^{\mathrm{(o)}}$, $Q_{u}^{\mathrm{(o)}}$, $Q_{v}^{\mathrm{(o)}}$, $Q_{xyz}^{\mathrm{(o)}}$, $Q_{4}^{\mathrm{(o)}}$, $Q_{4u}^{\mathrm{(o)}}$, $Q_{4v}^{\mathrm{(o)}}$, 
& $G_{0}^{\mathrm{(o)}}$, $G_{u}^{\mathrm{(o)}}$, $G_{v}^{\mathrm{(o)}}$, $G_{xyz}^{\mathrm{(o)}}$, $G_{4}^{\mathrm{(o)}}$, $G_{4u}^{\mathrm{(o)}}$, $G_{4v}^{\mathrm{(o)}}$, 
\\  
& & & & & & & & 
&
$Q_{0}^{\mathrm{(s)}}$, $Q_{u}^{\mathrm{(s)}}$, $Q_{v}^{\mathrm{(s)}}$, $Q_{xyz}^{\mathrm{(s)}}$, $Q_{4}^{\mathrm{(s)}}$, $Q_{4u}^{\mathrm{(s)}}$, $Q_{4v}^{\mathrm{(s)}}$ 
&
$G_{0}^{\mathrm{(s)}}$, $G_{u}^{\mathrm{(s)}}$, $G_{v}^{\mathrm{(s)}}$, $G_{xyz}^{\mathrm{(s)}}$, $G_{4}^{\mathrm{(s)}}$, $G_{4u}^{\mathrm{(s)}}$, $G_{4v}^{\mathrm{(s)}}$ 
\\
$\bar{\mathrm{A}}$ 
& $1$ & $1$ & $1$ & $1$ & $-1$ & $-1$ & $-1$ & $-1$ 
& $M_{0}^{\mathrm{(o)}}$, $M_{u}^{\mathrm{(o)}}$, $M_{v}^{\mathrm{(o)}}$, $M_{xyz}^{\mathrm{(o)}}$, $M_{4}^{\mathrm{(o)}}$, $M_{4u}^{\mathrm{(o)}}$, $M_{4v}^{\mathrm{(o)}}$, 
& $T_{0}^{\mathrm{(o)}}$, $T_{u}^{\mathrm{(o)}}$, $T_{v}^{\mathrm{(o)}}$, $T_{xyz}^{\mathrm{(o)}}$, $T_{4}^{\mathrm{(o)}}$, $T_{4u}^{\mathrm{(o)}}$, $T_{4v}^{\mathrm{(o)}}$, 
\\
\hline
\hline
\end{tabular}
\end{table*}

\begin{table*}
\caption{
IRREPs of multipoles in the collinear SPG
$\mathcal{G}_{\mathrm{SO}}^\infty \times {}^1 m {}^1 m {}^1 m 
= \mathrm{SO(2)}\times ( {}^1 m {}^1 m {}^1 m + [\theta 2_{\perp\bm{n}} \,\|\, E] {}^1 m {}^1 m {}^1 m )$.
All the symmetry operations of ${}^1 m {}^1 m {}^1 m$ are listed.
The other operations are omitted except for $[\theta 2_{\perp \bm{n}} \,\|\, E]$.
The absence [presence] of a bar over the IRREPs indicates parity +1 [$-1$] under the antiunitary operation $[\theta 2_{\perp\bm{n}} \,\|\, E]$. 
}
\label{table:Litvin54collinear}
\begin{tabular}{|c|c|c|c|c|c|c|c|c|c|c|c|}
\hline\hline
IRREP & \rotatebox{90}{$[E \,\|\, E]$} & \rotatebox{90}{$[E \,\|\, C_{2z}]$} & \rotatebox{90}{$[E \,\|\, C_{2x}]$} & \rotatebox{90}{$[E \,\|\, C_{2y}]$} & \rotatebox{90}{$[E \,\|\, I]$} & \rotatebox{90}{$[E \,\|\, IC_{2z}]$} & \rotatebox{90}{$[E \,\|\, IC_{2x}]$} & \rotatebox{90}{$[E \,\|\, IC_{2y}]$} & \rotatebox{90}{$[\theta 2_{\perp\bm{n}} \,\|\, E]$} & E, M & ET, MT 
\\
\hline
$\mathrm{A}_{\mathrm{g}}$ 
& $1$ & $1$ & $1$ & $1$ & $1$ & $1$ & $1$ & $1$ & $1$ 
& $Q_{0}^{\mathrm{(o)}}$, $Q_{u}^{\mathrm{(o)}}$, $Q_{v}^{\mathrm{(o)}}$, $Q_{4}^{\mathrm{(o)}}$, $Q_{4u}^{\mathrm{(o)}}$, $Q_{4v}^{\mathrm{(o)}}$, 
& $G_{xyz}^{\mathrm{(o)}}$, $G_{xyz}^{\mathrm{(s)}}$ 
\\  
& & & & & & & & & 
&
$Q_{0}^{\mathrm{(s)}}$, $Q_{u}^{\mathrm{(s)}}$, $Q_{v}^{\mathrm{(s)}}$, $Q_{4}^{\mathrm{(s)}}$, $Q_{4u}^{\mathrm{(s)}}$, $Q_{4v}^{\mathrm{(s)}}$ 
&
\\
$\bar{\mathrm{A}}_{\mathrm{g}}$ 
& $1$ & $1$ & $1$ & $1$ & $1$ & $1$ & $1$ & $1$ & $-1$ 
& $M_{xyz}^{\mathrm{(o)}}$, $M_{xyz}^{\mathrm{(s)}}$ 
& $T_{0}^{\mathrm{(o)}}$, $T_{u}^{\mathrm{(o)}}$, $T_{v}^{\mathrm{(o)}}$, $T_{4}^{\mathrm{(o)}}$, $T_{4u}^{\mathrm{(o)}}$, $T_{4v}^{\mathrm{(o)}}$, 
\\  
& & & & & & & & & 
&
&
$T_{0}^{\mathrm{(s)}}$, $T_{u}^{\mathrm{(s)}}$, $T_{v}^{\mathrm{(s)}}$, $T_{4}^{\mathrm{(s)}}$, $T_{4u}^{\mathrm{(s)}}$, $T_{4v}^{\mathrm{(s)}}$ 
\\
\hline
$\mathrm{B}_{1\mathrm{g}}$ 
& $1$ & $1$ & $-1$ & $-1$ & $1$ & $1$ & $-1$ & $-1$ & $1$ 
& $Q_{xy}^{\mathrm{(o)}}$, $Q_{4z}^{\alpha \mathrm{(o)}}$, $Q_{4z}^{\beta \mathrm{(o)}}$, $Q_{xy}^{\mathrm{(s)}}$, $Q_{4z}^{\alpha \mathrm{(s)}}$, $Q_{4z}^{\beta \mathrm{(s)}}$ 
& $G_{z}^{\mathrm{(o)}}$, $G_{z}^{\alpha \mathrm{(o)}}$, $G_{z}^{\beta \mathrm{(o)}}$, $G_{z}^{\mathrm{(s)}}$, $G_{z}^{\alpha \mathrm{(s)}}$, $G_{z}^{\beta \mathrm{(s)}}$ 
\\
\hline
\hline
\end{tabular}
\end{table*}

\clearpage 

\begin{table*}
\caption{
IRREPs of multipoles in the collinear SPG
$\mathcal{G}_{\mathrm{SO}}^\infty \times {}^1 4 
= \mathrm{SO(2)}\times ( {}^1 4 + [\theta 2_{\perp\bm{n}} \,\|\, E] {}^1 4 )$.
All the symmetry operations of ${}^1 4 + [\theta 2_{\perp\bm{n}} \,\|\, E] {}^1 4$ are listed.
The absence [presence] of a bar over the IRREPs indicates parity +1 [$-1$] under the antiunitary operation $[\theta 2_{\perp\bm{n}} \,\|\, E]$. 
}
\label{table:Litvin90collinear}
\begin{tabular}{|c|c|c|c|c|c|c|c|c|}
\hline\hline
IRREP & \rotatebox{90}{$[E \,\|\, E]$} & \rotatebox{90}{$[E \,\|\, 2C_{4z}]$} & \rotatebox{90}{$[E \,\|\, C_{2z}]$} & \rotatebox{90}{$[\theta 2_{\perp \bm{n}} \,\|\, E]$} & \rotatebox{90}{$[\theta 2_{\perp \bm{n}} \,\|\, 2C_{4z}]$} & \rotatebox{90}{$[\theta 2_{\perp \bm{n}} \,\|\, C_{2z}]$} & E, M & ET, MT 
\\
\hline
$\mathrm{A}$ 
& $1$ & $1$ & $1$ & $1$ & $1$ & $1$ 
& $Q_{0}^{\mathrm{(o)}}$, $Q_{z}^{\mathrm{(o)}}$, $Q_{u}^{\mathrm{(o)}}$, $Q_{z}^{\alpha \mathrm{(o)}}$, $Q_{4}^{\mathrm{(o)}}$, $Q_{4u}^{\mathrm{(o)}}$, $Q_{4z}^{\alpha \mathrm{(o)}}$, 
& $G_{0}^{\mathrm{(o)}}$, $G_{z}^{\mathrm{(o)}}$, $G_{u}^{\mathrm{(o)}}$, $G_{z}^{\alpha \mathrm{(o)}}$, $G_{4}^{\mathrm{(o)}}$, $G_{4u}^{\mathrm{(o)}}$, $G_{4z}^{\alpha \mathrm{(o)}}$, 
\\  
& & & & & & 
&
$Q_{0}^{\mathrm{(s)}}$, $Q_{z}^{\mathrm{(s)}}$, $Q_{u}^{\mathrm{(s)}}$, $Q_{z}^{\alpha \mathrm{(s)}}$, $Q_{4}^{\mathrm{(s)}}$, $Q_{4u}^{\mathrm{(s)}}$, $Q_{4z}^{\alpha \mathrm{(s)}}$ 
&
$G_{0}^{\mathrm{(s)}}$, $G_{z}^{\mathrm{(s)}}$, $G_{u}^{\mathrm{(s)}}$, $G_{z}^{\alpha \mathrm{(s)}}$, $G_{4}^{\mathrm{(s)}}$, $G_{4u}^{\mathrm{(s)}}$, $G_{4z}^{\alpha \mathrm{(s)}}$
\\
$\bar{\mathrm{A}}$ 
& $1$ & $1$ & $1$ & $-1$ & $-1$ & $-1$ 
& $M_{0}^{\mathrm{(o)}}$, $M_{z}^{\mathrm{(o)}}$, $M_{u}^{\mathrm{(o)}}$, $M_{z}^{\alpha \mathrm{(o)}}$, $M_{4}^{\mathrm{(o)}}$, $M_{4u}^{\mathrm{(o)}}$, $M_{4z}^{\alpha \mathrm{(o)}}$, 
& $T_{0}^{\mathrm{(o)}}$, $T_{z}^{\mathrm{(o)}}$, $T_{u}^{\mathrm{(o)}}$, $T_{z}^{\alpha \mathrm{(o)}}$, $T_{4}^{\mathrm{(o)}}$, $T_{4u}^{\mathrm{(o)}}$, $T_{4z}^{\alpha \mathrm{(o)}}$, 
\\  
& & & & & & 
&
$M_{0}^{\mathrm{(s)}}$, $M_{z}^{\mathrm{(s)}}$, $M_{u}^{\mathrm{(s)}}$, $M_{z}^{\alpha \mathrm{(s)}}$, $M_{4}^{\mathrm{(s)}}$, $M_{4u}^{\mathrm{(s)}}$, $M_{4z}^{\alpha \mathrm{(s)}}$ 
&
$T_{0}^{\mathrm{(s)}}$, $T_{z}^{\mathrm{(s)}}$, $T_{u}^{\mathrm{(s)}}$, $T_{z}^{\alpha \mathrm{(s)}}$, $T_{4}^{\mathrm{(s)}}$, $T_{4u}^{\mathrm{(s)}}$, $T_{4z}^{\alpha \mathrm{(s)}}$
\\
\hline
\hline
\end{tabular}
\end{table*}

\begin{table*}
\caption{
IRREPs of multipoles in the collinear SPG
$\mathcal{G}_{\mathrm{SO}}^\infty \times {}^1 \bar{4} 
= \mathrm{SO(2)}\times ( {}^1 \bar{4} + [\theta 2_{\perp\bm{n}} \,\|\, E] {}^1 \bar{4} )$.
All the symmetry operations of ${}^1 \bar{4} + [\theta 2_{\perp\bm{n}} \,\|\, E] {}^1 \bar{4}$ are listed.
The absence [presence] of a bar over the IRREPs indicates parity +1 [$-1$] under the antiunitary operation $[\theta 2_{\perp\bm{n}} \,\|\, E]$. 
}
\label{table:Litvin96collinear}
\begin{tabular}{|c|c|c|c|c|c|c|c|c|}
\hline\hline
IRREP & \rotatebox{90}{$[E \,\|\, E]$} & \rotatebox{90}{$[E \,\|\, 2IC_{4z}]$} & \rotatebox{90}{$[E \,\|\, C_{2z}]$} & \rotatebox{90}{$[\theta 2_{\perp\bm{n}} \,\|\, E]$} & \rotatebox{90}{$[\theta 2_{\perp\bm{n}} \,\|\, 2IC_{4z}]$} & \rotatebox{90}{$[\theta 2_{\perp\bm{n}} \,\|\, C_{2z}]$} & E, M & ET, MT 
\\
\hline
$\mathrm{A}$ 
& $1$ & $1$ & $1$ & $1$ & $1$ & $1$ 
& $Q_{0}^{\mathrm{(o)}}$, $Q_{u}^{\mathrm{(o)}}$, $Q_{xyz}^{\mathrm{(o)}}$, $Q_{z}^{\beta \mathrm{(o)}}$, $Q_{4}^{\mathrm{(o)}}$, $Q_{4u}^{\mathrm{(o)}}$, $Q_{4z}^{\alpha \mathrm{(o)}}$,  
& $G_{z}^{\mathrm{(o)}}$, $G_{v}^{\mathrm{(o)}}$, $G_{xy}^{\mathrm{(o)}}$, $G_{z}^{\alpha \mathrm{(o)}}$, $G_{4v}^{\mathrm{(o)}}$, $G_{4z}^{\beta \mathrm{(o)}}$, 
\\
& & & & & & 
&
$Q_{0}^{\mathrm{(s)}}$, $Q_{u}^{\mathrm{(s)}}$, $Q_{xyz}^{\mathrm{(s)}}$, $Q_{z}^{\beta \mathrm{(s)}}$, $Q_{4}^{\mathrm{(s)}}$, $Q_{4u}^{\mathrm{(s)}}$, $Q_{4z}^{\alpha \mathrm{(s)}}$
&
$G_{z}^{\mathrm{(s)}}$, $G_{v}^{\mathrm{(s)}}$, $G_{xy}^{\mathrm{(s)}}$, $G_{z}^{\alpha \mathrm{(s)}}$, $G_{4v}^{\mathrm{(s)}}$, $G_{4z}^{\beta \mathrm{(s)}}$ 
\\
$\bar{\mathrm{A}}$ 
& $1$ & $1$ & $1$ & $-1$ & $-1$ & $-1$ 
& $M_{z}^{\mathrm{(o)}}$, $M_{v}^{\mathrm{(o)}}$, $M_{xy}^{\mathrm{(o)}}$, $M_{z}^{\alpha \mathrm{(o)}}$, $M_{4v}^{\mathrm{(o)}}$, $M_{4z}^{\beta \mathrm{(o)}}$, 
& $T_{0}^{\mathrm{(o)}}$, $T_{u}^{\mathrm{(o)}}$, $T_{xyz}^{\mathrm{(o)}}$, $T_{z}^{\beta \mathrm{(o)}}$, $T_{4}^{\mathrm{(o)}}$, $T_{4u}^{\mathrm{(o)}}$, $T_{4z}^{\alpha \mathrm{(o)}}$,  
\\
& & & & & & 
&
$M_{z}^{\mathrm{(s)}}$, $M_{v}^{\mathrm{(s)}}$, $M_{xy}^{\mathrm{(s)}}$, $M_{z}^{\alpha \mathrm{(s)}}$, $M_{4v}^{\mathrm{(s)}}$, $M_{4z}^{\beta \mathrm{(s)}}$ 
&
$T_{0}^{\mathrm{(s)}}$, $T_{u}^{\mathrm{(s)}}$, $T_{xyz}^{\mathrm{(s)}}$, $T_{z}^{\beta \mathrm{(s)}}$, $T_{4}^{\mathrm{(s)}}$, $T_{4u}^{\mathrm{(s)}}$, $T_{4z}^{\alpha \mathrm{(s)}}$
\\
\hline
\hline
\end{tabular}
\end{table*}

\begin{table*}
\caption{
IRREPs of multipoles in the collinear SPG
$\mathcal{G}_{\mathrm{SO}}^\infty \times {}^1 4 / {}^1 m 
= \mathrm{SO(2)}\times ( {}^1 4 / {}^1 m + [\theta 2_{\perp\bm{n}} \,\|\, E] {}^1 4 / {}^1 m )$.
All the symmetry operations of ${}^1 4 / {}^1 m + [\theta 2_{\perp\bm{n}} \,\|\, E] {}^1 4 / {}^1 m$ are listed.
The absence [presence] of a bar over the IRREPs indicates parity +1 [$-1$] under the antiunitary operation $[\theta 2_{\perp\bm{n}} \,\|\, E]$. 
}
\label{table:Litvin102collinear}
\begin{tabular}{|c|c|c|c|c|c|c|c|c|c|c|c|c|c|c|}
\hline\hline
IRREP & \rotatebox{90}{$[E \,\|\, E]$} & \rotatebox{90}{$[E \,\|\, 2C_{4z}]$} & \rotatebox{90}{$[E \,\|\, C_{2z}]$} & \rotatebox{90}{$[E \,\|\, I]$} & \rotatebox{90}{$[E \,\|\, 2IC_{4z}]$} & \rotatebox{90}{$[E \,\|\, IC_{2z}]$} & \rotatebox{90}{$[\theta 2_{\perp\bm{n}} \,\|\, E]$} & \rotatebox{90}{$[\theta 2_{\perp\bm{n}} \,\|\, 2C_{4z}]$} & \rotatebox{90}{$[\theta 2_{\perp\bm{n}} \,\|\, C_{2z}]$} & \rotatebox{90}{$[\theta 2_{\perp\bm{n}} \,\|\, I]$} & \rotatebox{90}{$[\theta 2_{\perp\bm{n}} \,\|\, 2IC_{4z}]$} & \rotatebox{90}{$[\theta 2_{\perp\bm{n}} \,\|\, IC_{2z}]$} & E, M & ET, MT 
\\
\hline
$\mathrm{A}_{\mathrm{g}}$ 
& $1$ & $1$ & $1$ & $1$ & $1$ & $1$ & $1$ & $1$ & $1$ & $1$ & $1$ & $1$ 
& $Q_{0}^{\mathrm{(o)}}$, $Q_{u}^{\mathrm{(o)}}$, $Q_{4}^{\mathrm{(o)}}$, $Q_{4u}^{\mathrm{(o)}}$, $Q_{4z}^{\alpha \mathrm{(o)}}$, 
& $G_{z}^{\mathrm{(o)}}$, $G_{z}^{\alpha \mathrm{(o)}}$, 
\\
& & & & & & & & & & & & 
&
$Q_{0}^{\mathrm{(s)}}$, $Q_{u}^{\mathrm{(s)}}$, $Q_{4}^{\mathrm{(s)}}$, $Q_{4u}^{\mathrm{(s)}}$, $Q_{4z}^{\alpha \mathrm{(s)}}$ 
&
$G_{z}^{\mathrm{(s)}}$, $G_{z}^{\alpha \mathrm{(s)}}$ 
\\
$\bar{\mathrm{A}}_{\mathrm{g}}$ 
& $1$ & $1$ & $1$ & $1$ & $1$ & $1$ & $-1$ & $-1$ & $-1$ & $-1$ & $-1$ & $-1$ 
& $M_{z}^{\mathrm{(o)}}$, $M_{z}^{\alpha \mathrm{(o)}}$, 
& $T_{0}^{\mathrm{(o)}}$, $T_{u}^{\mathrm{(o)}}$, $T_{4}^{\mathrm{(o)}}$, $T_{4u}^{\mathrm{(o)}}$, $T_{4z}^{\alpha \mathrm{(o)}}$, 
\\
& & & & & & & & & & & & 
&
$M_{z}^{\mathrm{(s)}}$, $M_{z}^{\alpha \mathrm{(s)}}$ 
&
$T_{0}^{\mathrm{(s)}}$, $T_{u}^{\mathrm{(s)}}$, $T_{4}^{\mathrm{(s)}}$, $T_{4u}^{\mathrm{(s)}}$, $T_{4z}^{\alpha \mathrm{(s)}}$ 
\\
\hline
$\mathrm{B}_{\mathrm{g}}$ 
& $1$ & $-1$ & $1$ & $1$ & $-1$ & $1$ & $1$ & $-1$ & $1$ & $1$ & $-1$ & $1$ 
& $Q_{v}^{\mathrm{(o)}}$, $Q_{xy}^{\mathrm{(o)}}$, $Q_{4v}^{\mathrm{(o)}}$, $Q_{4z}^{\beta \mathrm{(o)}}$,  
& $G_{xyz}^{\mathrm{(o)}}$, $G_{z}^{\beta \mathrm{(o)}}$, 
\\
\hline
\hline
\end{tabular}
\end{table*}

\begin{table*}
\caption{
IRREPs of multipoles in the collinear SPG
$\mathcal{G}_{\mathrm{SO}}^\infty \times {}^1 4 {}^1 2 {}^1 2 
= \mathrm{SO(2)}\times ( {}^1 4 {}^1 2 {}^1 2 + [\theta 2_{\perp\bm{n}} \,\|\, E] {}^1 4 {}^1 2 {}^1 2 )$.
All the symmetry operations of ${}^1 4 {}^1 2 {}^1 2 + [\theta 2_{\perp\bm{n}} \,\|\, E] {}^1 4 {}^1 2 {}^1 2$ are listed.
Among the listed operations, $2C^{\prime}_{2x}$ denotes the conjugacy class consisting of two twofold rotations, with $C_{2x}$ as a representative element ($2C^{\prime}_{2x} \equiv C_{2x}, C_{2y}$).
The absence [presence] of a bar over the IRREPs indicates parity +1 [$-1$] under the antiunitary operation $[\theta 2_{\perp\bm{n}} \,\|\, E]$. 
}
\label{table:Litvin130collinear}
\begin{tabular}{|c|c|c|c|c|c|c|c|c|c|c|c|c|}
\hline\hline
IRREP & \rotatebox{90}{$[E \,\|\, E]$} & \rotatebox{90}{$[E \,\|\, 2C_{4z}]$} & \rotatebox{90}{$[E \,\|\, C_{2z}]$} 
& \rotatebox{90}{$[E \,\|\, 2C^{\prime}_{2x}]$} 
& \rotatebox{90}{$[E \,\|\, 2C^{\prime\prime}_{2}]$} & \rotatebox{90}{$[\theta 2_{\perp\bm{n}} \,\|\, E]$} & \rotatebox{90}{$[\theta 2_{\perp\bm{n}} \,\|\, 2C_{4z}]$} & \rotatebox{90}{$[\theta 2_{\perp\bm{n}} \,\|\, C_{2z}]$} 
& \rotatebox{90}{$[\theta 2_{\perp\bm{n}} \,\|\, 2C^{\prime}_{2x}]$} & \rotatebox{90}{$[\theta 2_{\perp\bm{n}} \,\|\, 2C^{\prime\prime}_{2}]$} & E, M & ET, MT 
\\
\hline
$\mathrm{A}_{1}$ 
& $1$ & $1$ & $1$ & $1$ & $1$ & $1$ & $1$ & $1$ & $1$ & $1$ 
& $Q_{0}^{\mathrm{(o)}}$, $Q_{u}^{\mathrm{(o)}}$, $Q_{4}^{\mathrm{(o)}}$, $Q_{4u}^{\mathrm{(o)}}$, 
& $G_{0}^{\mathrm{(o)}}$, $G_{u}^{\mathrm{(o)}}$, $G_{4}^{\mathrm{(o)}}$, $G_{4u}^{\mathrm{(o)}}$, 
\\
& & & & & & & & & & 
&
$Q_{0}^{\mathrm{(s)}}$, $Q_{u}^{\mathrm{(s)}}$, $Q_{4}^{\mathrm{(s)}}$, $Q_{4u}^{\mathrm{(s)}}$ 
&
$G_{0}^{\mathrm{(s)}}$, $G_{u}^{\mathrm{(s)}}$, $G_{4}^{\mathrm{(s)}}$, $G_{4u}^{\mathrm{(s)}}$ 
\\
$\bar{\mathrm{A}}_{1}$ 
& $1$ & $1$ & $1$ & $1$ & $1$ & $-1$ & $-1$ & $-1$ & $-1$ & $-1$ 
& $M_{0}^{\mathrm{(o)}}$, $M_{u}^{\mathrm{(o)}}$, $M_{4}^{\mathrm{(o)}}$, $M_{4u}^{\mathrm{(o)}}$, 
& $T_{0}^{\mathrm{(o)}}$, $T_{u}^{\mathrm{(o)}}$, $T_{4}^{\mathrm{(o)}}$, $T_{4u}^{\mathrm{(o)}}$, 
\\
& & & & & & & & & & 
&
$M_{0}^{\mathrm{(s)}}$, $M_{u}^{\mathrm{(s)}}$, $M_{4}^{\mathrm{(s)}}$, $M_{4u}^{\mathrm{(s)}}$ 
&
$T_{0}^{\mathrm{(s)}}$, $T_{u}^{\mathrm{(s)}}$, $T_{4}^{\mathrm{(s)}}$, $T_{4u}^{\mathrm{(s)}}$ 
\\
\hline
$\mathrm{A}_{2}$ 
& $1$ & $1$ & $1$ & $-1$ & $-1$ & $1$ & $1$ & $1$ & $-1$ & $-1$ 
& $Q_{z}^{\mathrm{(o)}}$, $Q_{z}^{\alpha \mathrm{(o)}}$, $Q_{4z}^{\alpha \mathrm{(o)}}$, 
& $G_{z}^{\mathrm{(o)}}$, $G_{z}^{\alpha \mathrm{(o)}}$, $G_{4z}^{\alpha \mathrm{(o)}}$, 
\\
\hline
\hline
\end{tabular}
\end{table*}

\begin{table*}
\caption{
IRREPs of multipoles in the collinear SPG
$\mathcal{G}_{\mathrm{SO}}^\infty \times {}^1 4 {}^1 m {}^1 m 
= \mathrm{SO(2)}\times ( {}^1 4 {}^1 m {}^1 m + [\theta 2_{\perp\bm{n}} \,\|\, E] {}^1 4 {}^1 m {}^1 m )$.
All the symmetry operations of ${}^1 4 {}^1 m {}^1 m + [\theta 2_{\perp\bm{n}} \,\|\, E] {}^1 4 {}^1 m {}^1 m$ are listed.
Among the listed operations, $2C^{\prime}_{2x}$ denotes the conjugacy class consisting of two twofold rotations, with $C_{2x}$ as a representative element ($2C^{\prime}_{2x} \equiv C_{2x}, C_{2y}$).
The absence [presence] of a bar over the IRREPs indicates parity +1 [$-1$] under the antiunitary operation $[\theta 2_{\perp\bm{n}} \,\|\, E]$. 
}
\label{table:Litvin146collinear}
\begin{tabular}{|c|c|c|c|c|c|c|c|c|c|c|c|c|}
\hline\hline
IRREP & \rotatebox{90}{$[E \,\|\, E]$} & \rotatebox{90}{$[E \,\|\, 2C_{4z}]$} & \rotatebox{90}{$[E \,\|\, C_{2z}]$} 
& \rotatebox{90}{$[E \,\|\, 2IC^{\prime}_{2x}]$} 
& \rotatebox{90}{$[E \,\|\, 2IC^{\prime\prime}_{2}]$} & \rotatebox{90}{$[\theta 2_{\perp\bm{n}} \,\|\, E]$} & \rotatebox{90}{$[\theta 2_{\perp\bm{n}} \,\|\, 2C_{4z}]$} & \rotatebox{90}{$[\theta 2_{\perp\bm{n}} \,\|\, C_{2z}]$} 
& \rotatebox{90}{$[\theta 2_{\perp\bm{n}} \,\|\, 2IC^{\prime}_{2x}]$} 
& \rotatebox{90}{$[\theta 2_{\perp\bm{n}} \,\|\, 2IC^{\prime\prime}_{2}]$} & E, M & ET, MT 
\\
\hline
$\mathrm{A}_{1}$ 
& $1$ & $1$ & $1$ & $1$ & $1$ & $1$ & $1$ & $1$ & $1$ & $1$ 
& $Q_{0}^{\mathrm{(o)}}$, $Q_{z}^{\mathrm{(o)}}$, $Q_{u}^{\mathrm{(o)}}$, $Q_{z}^{\alpha \mathrm{(o)}}$, $Q_{4}^{\mathrm{(o)}}$, $Q_{4u}^{\mathrm{(o)}}$, 
& $G_{4z}^{\alpha \mathrm{(o)}}$, 
\\
& & & & & & & & & & 
&
$Q_{0}^{\mathrm{(s)}}$, $Q_{z}^{\mathrm{(s)}}$, $Q_{u}^{\mathrm{(s)}}$, $Q_{z}^{\alpha \mathrm{(s)}}$, $Q_{4}^{\mathrm{(s)}}$, $Q_{4u}^{\mathrm{(s)}}$ 
&
$G_{4z}^{\alpha \mathrm{(s)}}$ 
\\
$\bar{\mathrm{A}}_{1}$ 
& $1$ & $1$ & $1$ & $1$ & $1$ & $-1$ & $-1$ & $-1$ & $-1$ & $-1$ 
& $M_{4z}^{\alpha \mathrm{(o)}}$, 
& $T_{0}^{\mathrm{(o)}}$, $T_{z}^{\mathrm{(o)}}$, $T_{u}^{\mathrm{(o)}}$, $T_{z}^{\alpha \mathrm{(o)}}$, $T_{4}^{\mathrm{(o)}}$, $T_{4u}^{\mathrm{(o)}}$, 
\\
& & & & & & & & & & 
&
$M_{4z}^{\alpha \mathrm{(s)}}$ 
&
$T_{0}^{\mathrm{(s)}}$, $T_{z}^{\mathrm{(s)}}$, $T_{u}^{\mathrm{(s)}}$, $T_{z}^{\alpha \mathrm{(s)}}$, $T_{4}^{\mathrm{(s)}}$, $T_{4u}^{\mathrm{(s)}}$ 
\\
\hline
$\mathrm{A}_{2}$ 
& $1$ & $1$ & $1$ & $-1$ & $-1$ & $1$ & $1$ & $1$ & $-1$ & $-1$ 
& $Q_{4z}^{\alpha \mathrm{(o)}}$, 
& $G_{0}^{\mathrm{(o)}}$, $G_{z}^{\mathrm{(o)}}$, $G_{u}^{\mathrm{(o)}}$, $G_{z}^{\alpha \mathrm{(o)}}$, $G_{4}^{\mathrm{(o)}}$, $G_{4u}^{\mathrm{(o)}}$, 
\\
\hline
\hline
\end{tabular}
\end{table*}

\begin{table*}
\caption{
IRREPs of multipoles in the collinear SPG
$\mathcal{G}_{\mathrm{SO}}^\infty \times {}^1 \bar{4} {}^1 2 {}^1 m 
= \mathrm{SO(2)}\times ( {}^1 \bar{4} {}^1 2 {}^1 m + [\theta 2_{\perp\bm{n}} \,\|\, E] {}^1 \bar{4} {}^1 2 {}^1 m )$.
All the symmetry operations of ${}^1 \bar{4} {}^1 2 {}^1 m + [\theta 2_{\perp\bm{n}} \,\|\, E] {}^1 \bar{4} {}^1 2 {}^1 m$ are listed.
Among the listed operations, $2C^{\prime}_{2x}$ denotes the conjugacy class consisting of two twofold rotations, with $C_{2x}$ as a representative element ($2C^{\prime}_{2x} \equiv C_{2x}, C_{2y}$).
The absence [presence] of a bar over the IRREPs indicates parity +1 [$-1$] under the antiunitary operation $[\theta 2_{\perp\bm{n}} \,\|\, E]$. 
}
\label{table:Litvin162collinear}
\begin{tabular}{|c|c|c|c|c|c|c|c|c|c|c|c|c|}
\hline\hline
IRREP & \rotatebox{90}{$[E \,\|\, E]$} & \rotatebox{90}{$[E \,\|\, 2IC_{4z}]$} & \rotatebox{90}{$[E \,\|\, C_{2z}]$} 
& \rotatebox{90}{$[E \,\|\, 2C^{\prime}_{2x}]$} 
& \rotatebox{90}{$[E \,\|\, 2IC^{\prime\prime}_{2}]$} & \rotatebox{90}{$[\theta 2_{\perp\bm{n}} \,\|\, E]$} & \rotatebox{90}{$[\theta 2_{\perp\bm{n}} \,\|\, 2IC_{4z}]$} & \rotatebox{90}{$[\theta 2_{\perp\bm{n}} \,\|\, C_{2z}]$} 
& \rotatebox{90}{$[\theta 2_{\perp\bm{n}} \,\|\, 2C^{\prime}_{2x}]$} & \rotatebox{90}{$[\theta 2_{\perp\bm{n}} \,\|\, 2IC^{\prime\prime}_{2}]$} & E, M & ET, MT 
\\
\hline
$\mathrm{A}_{1}$ 
& $1$ & $1$ & $1$ & $1$ & $1$ & $1$ & $1$ & $1$ & $1$ & $1$ 
& $Q_{0}^{\mathrm{(o)}}$, $Q_{u}^{\mathrm{(o)}}$, $Q_{xyz}^{\mathrm{(o)}}$, $Q_{4}^{\mathrm{(o)}}$, $Q_{4u}^{\mathrm{(o)}}$, 
& $G_{v}^{\mathrm{(o)}}$, $G_{4v}^{\mathrm{(o)}}$, 
\\
& & & & & & & & & & 
&
$Q_{0}^{\mathrm{(s)}}$, $Q_{u}^{\mathrm{(s)}}$, $Q_{xyz}^{\mathrm{(s)}}$, $Q_{4}^{\mathrm{(s)}}$, $Q_{4u}^{\mathrm{(s)}}$ 
&
$G_{v}^{\mathrm{(s)}}$, $G_{4v}^{\mathrm{(s)}}$ 
\\
$\bar{\mathrm{A}}_{1}$ 
& $1$ & $1$ & $1$ & $1$ & $1$ & $-1$ & $-1$ & $-1$ & $-1$ & $-1$ 
& $M_{v}^{\mathrm{(o)}}$, $M_{4v}^{\mathrm{(o)}}$,  
& $T_{0}^{\mathrm{(o)}}$, $T_{u}^{\mathrm{(o)}}$, $T_{xyz}^{\mathrm{(o)}}$, $T_{4}^{\mathrm{(o)}}$, $T_{4u}^{\mathrm{(o)}}$,  
\\
& & & & & & & & & & 
&
$M_{v}^{\mathrm{(s)}}$, $M_{4v}^{\mathrm{(s)}}$
&
$T_{0}^{\mathrm{(s)}}$, $T_{u}^{\mathrm{(s)}}$, $T_{xyz}^{\mathrm{(s)}}$, $T_{4}^{\mathrm{(s)}}$, $T_{4u}^{\mathrm{(s)}}$
\\
\hline
$\mathrm{A}_{2}$ 
& $1$ & $1$ & $1$ & $-1$ & $-1$ & $1$ & $1$ & $1$ & $-1$ & $-1$ 
& $Q_{z}^{\beta \mathrm{(o)}}$, $Q_{4z}^{\alpha \mathrm{(o)}}$,
& $G_{z}^{\mathrm{(o)}}$, $G_{xy}^{\mathrm{(o)}}$, $G_{z}^{\alpha \mathrm{(o)}}$, $G_{4z}^{\beta \mathrm{(o)}}$, 
\\
\hline
\hline
\end{tabular}
\end{table*}

\begin{table*}
\caption{
IRREPs of multipoles in the collinear SPG
$\mathcal{G}_{\mathrm{SO}}^\infty \times {}^1 4 / {}^1 m {}^1 m {}^1 m 
= \mathrm{SO(2)}\times ( {}^1 4 / {}^1 m {}^1 m {}^1 m + [\theta 2_{\perp\bm{n}} \,\|\, E] {}^1 4 / {}^1 m {}^1 m {}^1 m )$.
All the symmetry operations of ${}^1 4 / {}^1 m {}^1 m {}^1 m$ are listed.
The other operations are omitted except for $[\theta 2_{\perp \bm{n}} \,\|\, E]$. 
Among the listed operations, $2C^{\prime}_{2x}$ denotes the conjugacy class consisting of two twofold rotations, with $C_{2x}$ as a representative element ($2C^{\prime}_{2x} \equiv C_{2x}, C_{2y}$).
The absence [presence] of a bar over the IRREPs indicates parity +1 [$-1$] under the antiunitary operation $[\theta 2_{\perp\bm{n}} \,\|\, E]$. 
}
\label{table:Litvin186collinear}

\end{table*}

\clearpage 

\begin{table*}
\caption{
IRREPs of multipoles in the collinear SPG
$\mathcal{G}_{\mathrm{SO}}^\infty \times {}^1 2 {}^1 3 
= \mathrm{SO(2)}\times ( {}^1 2 {}^1 3 + [\theta 2_{\perp\bm{n}} \,\|\, E] {}^1 2 {}^1 3 )$.
All the symmetry operations of ${}^1 2 {}^1 3 + [\theta 2_{\perp\bm{n}} \,\|\, E] {}^1 2 {}^1 3$ are listed.
The absence [presence] of a bar over the IRREPs indicates parity +1 [$-1$] under the antiunitary operation $[\theta 2_{\perp\bm{n}} \,\|\, E]$. 
}
\label{table:Litvin538collinear}
%
\end{table*}

\begin{table*}
\caption{
IRREPs of multipoles in the collinear SPG
$\mathcal{G}_{\mathrm{SO}}^\infty \times {}^1 2 / {}^1 m {}^1 \bar{3} 
= \mathrm{SO(2)}\times ( {}^1 2 / {}^1 m {}^1 \bar{3} + [\theta 2_{\perp\bm{n}} \,\|\, E] {}^1 2 / {}^1 m {}^1 \bar{3} )$.
All the symmetry operations of ${}^1 2 /{}^1 m {}^1 \bar{3}$ are listed.
The other operations are omitted except for $[\theta 2_{\perp \bm{n}} \,\|\, E]$.
The absence [presence] of a bar over the IRREPs indicates parity +1 [$-1$] under the antiunitary operation $[\theta 2_{\perp\bm{n}} \,\|\, E]$. 
}
\label{table:Litvin541collinear}

\end{table*}

\begin{table*}
\caption{
IRREPs of multipoles in the collinear SPG
$\mathcal{G}_{\mathrm{SO}}^\infty \times {}^1 \bar{4} {}^1 3 {}^1 m 
= \mathrm{SO(2)}\times ( {}^1 \bar{4} {}^1 3 {}^1 m + [\theta 2_{\perp\bm{n}} \,\|\, E] {}^1 \bar{4} {}^1 3 {}^1 m )$.
All the symmetry operations of ${}^1 \bar{4} {}^1 3 {}^1 m$ are listed.
The other operations are omitted except for $[\theta 2_{\perp \bm{n}} \,\|\, E]$.
The absence [presence] of a bar over the IRREPs indicates parity +1 [$-1$] under the antiunitary operation $[\theta 2_{\perp\bm{n}} \,\|\, E]$. 
}
\label{table:Litvin551collinear}

\end{table*}

\begin{table*}
\caption{
IRREPs of multipoles in the collinear SPG
$\mathcal{G}_{\mathrm{SO}}^\infty \times {}^1 4 {}^1 3 {}^1 2 
= \mathrm{SO(2)}\times ( {}^1 4 {}^1 3 {}^1 2 + [\theta 2_{\perp\bm{n}} \,\|\, E] {}^1 4 {}^1 3 {}^1 2 )$.
All the symmetry operations of ${}^1 4 {}^1 3 {}^1 2$ are listed.
The other operations are omitted except for $[\theta 2_{\perp \bm{n}} \,\|\, E]$.
The absence [presence] of a bar over the IRREPs indicates parity +1 [$-1$] under the antiunitary operation $[\theta 2_{\perp\bm{n}} \,\|\, E]$. 
}
\label{table:Litvin559collinear}
%
\end{table*}

\begin{table*}
\caption{
IRREPs of multipoles in the collinear SPG
$\mathcal{G}_{\mathrm{SO}}^\infty \times {}^1 4 / {}^1 m {}^1 \bar{3} {}^1 2 / {}^{1} m 
= \mathrm{SO(2)}\times ( {}^1 4 / {}^1 m {}^1 \bar{3} {}^1 2 / {}^1 m + [\theta 2_{\perp\bm{n}} \,\|\, E] {}^1 4 / {}^1 m {}^1 \bar{3} {}^1 2 / {}^1 m )$.
All the symmetry operations of ${}^1 4 / {}^1 m {}^1 \bar{3} {}^1 2 / {}^1 m$ are listed.
The other operations are omitted except for $[\theta 2_{\perp \bm{n}} \,\|\, E]$.
The absence [presence] of a bar over the IRREPs indicates parity +1 [$-1$] under the antiunitary operation $[\theta 2_{\perp\bm{n}} \,\|\, E]$. 
}
\label{table:Litvin567collinear}

\end{table*}

\clearpage 

\begin{table*}
\caption{
IRREPs of multipoles in the collinear SPG
$\mathcal{G}_{\mathrm{SO}}^\infty \times {}^1 3 
= \mathrm{SO(2)}\times ( {}^1 3 + [\theta 2_{\perp\bm{n}} \,\|\, E] {}^1 3 )$.
All the symmetry operations of ${}^1 3 + [\theta 2_{\perp\bm{n}} \,\|\, E] {}^1 3$ are listed.
The absence [presence] of a bar over the IRREPs indicates parity +1 [$-1$] under the antiunitary operation $[\theta 2_{\perp\bm{n}} \,\|\, E]$. 
}
\label{table:Litvin272collinear}
%
\end{table*}

\begin{table*}
\caption{
IRREPs of multipoles in the collinear SPG
$\mathcal{G}_{\mathrm{SO}}^\infty \times {}^1 \bar{3} 
= \mathrm{SO(2)}\times ( {}^1 \bar{3} + [\theta 2_{\perp\bm{n}} \,\|\, E] {}^1 \bar{3} )$.
All the symmetry operations of ${}^1 \bar{3} + [\theta 2_{\perp\bm{n}} \,\|\, E] {}^1 \bar{3}$ are listed.
The absence [presence] of a bar over the IRREPs indicates parity +1 [$-1$] under the antiunitary operation $[\theta 2_{\perp\bm{n}} \,\|\, E]$. 
}
\label{table:Litvin274collinear}
%
\end{table*}

\begin{table*}
\caption{
IRREPs of multipoles in the collinear SPG
$\mathcal{G}_{\mathrm{SO}}^\infty \times {}^1 3 {}^1 2 
= \mathrm{SO(2)}\times ( {}^1 3 {}^1 2 + [\theta 2_{\perp\bm{n}} \,\|\, E] {}^1 3 {}^1 2 )$.
All the symmetry operations of ${}^1 3 {}^1 2 + [\theta 2_{\perp\bm{n}} \,\|\, E] {}^1 3 {}^1 2$ are listed.
Among the listed operations, $3C^{\prime}_{2y}$ denotes the conjugacy class consisting of three twofold rotations, with $C_{2y}$ as a representative element.
The absence [presence] of a bar over the IRREPs indicates parity +1 [$-1$] under the antiunitary operation $[\theta 2_{\perp\bm{n}} \,\|\, E]$. 
}
\label{table:Litvin282collinear}

\end{table*}

\begin{table*}
\caption{
IRREPs of multipoles in the collinear SPG
$\mathcal{G}_{\mathrm{SO}}^\infty \times {}^1 3 {}^1 m 
= \mathrm{SO(2)}\times ( {}^1 3 {}^1 m + [\theta 2_{\perp\bm{n}} \,\|\, E] {}^1 3 {}^1 m )$.
All the symmetry operations of ${}^1 3 {}^1 m + [\theta 2_{\perp\bm{n}} \,\|\, E] {}^1 3 {}^1 m$ are listed.
Among the listed operations, $3C^{\prime}_{2y}$ denotes the conjugacy class consisting of three twofold rotations, with $C_{2y}$ as a representative element.
The absence [presence] of a bar over the IRREPs indicates parity +1 [$-1$] under the antiunitary operation $[\theta 2_{\perp\bm{n}} \,\|\, E]$. 
}
\label{table:Litvin288collinear}

\end{table*}

\begin{table*}
\caption{
IRREPs of multipoles in the collinear SPG
$\mathcal{G}_{\mathrm{SO}}^\infty \times {}^1 \bar{3} {}^1 m 
= \mathrm{SO(2)}\times ( {}^1 \bar{3} {}^1 m + [\theta 2_{\perp\bm{n}} \,\|\, E] {}^1 \bar{3} {}^1 m )$.
All the symmetry operations of ${}^1 \bar{3} {}^1 m$ are listed.
The other operations are omitted except for $[\theta 2_{\perp \bm{n}} \,\|\, E]$.
Among the listed operations, $3C^{\prime}_{2y}$ denotes the conjugacy class consisting of three twofold rotations, with $C_{2y}$ as a representative element.
The absence [presence] of a bar over the IRREPs indicates parity +1 [$-1$] under the antiunitary operation $[\theta 2_{\perp\bm{n}} \,\|\, E]$. 
}
\label{table:Litvin294collinear}

\end{table*}

\clearpage 

\begin{table*}
\caption{
IRREPs of multipoles in the collinear SPG
$\mathcal{G}_{\mathrm{SO}}^\infty \times {}^1 6 
= \mathrm{SO(2)}\times ( {}^1 6 + [\theta 2_{\perp\bm{n}} \,\|\, E] {}^1 6 )$.
All the symmetry operations of ${}^1 6 + [\theta 2_{\perp\bm{n}} \,\|\, E] {}^1 6$ are listed.
The absence [presence] of a bar over the IRREPs indicates parity +1 [$-1$] under the antiunitary operation $[\theta 2_{\perp\bm{n}} \,\|\, E]$. 
}
\label{table:Litvin330collinear}
%
\end{table*}

\begin{table*}
\caption{
IRREPs of multipoles in the collinear SPG
$\mathcal{G}_{\mathrm{SO}}^\infty \times {}^1 \bar{6} 
= \mathrm{SO(2)}\times ( {}^1 \bar{6} + [\theta 2_{\perp\bm{n}} \,\|\, E] {}^1 \bar{6} )$.
All the symmetry operations of ${}^1 \bar{6} + [\theta 2_{\perp\bm{n}} \,\|\, E] {}^1 \bar{6}$ are listed.
The absence [presence] of a bar over the IRREPs indicates parity +1 [$-1$] under the antiunitary operation $[\theta 2_{\perp\bm{n}} \,\|\, E]$. 
}
\label{table:Litvin322collinear}
%
\end{table*}

\begin{table*}
\caption{
IRREPs of multipoles in the collinear SPG
$\mathcal{G}_{\mathrm{SO}}^\infty \times {}^1 6 / {}^1 m 
= \mathrm{SO(2)}\times ( {}^1 6 / {}^1 m + [\theta 2_{\perp\bm{n}} \,\|\, E] {}^1 6 / {}^1 m )$.
All the symmetry operations of ${}^1 6 / {}^1 m$ are listed.
The other operations are omitted except for $[\theta 2_{\perp \bm{n}} \,\|\, E]$.
The absence [presence] of a bar over the IRREPs indicates parity +1 [$-1$] under the antiunitary operation $[\theta 2_{\perp\bm{n}} \,\|\, E]$. 
}
\label{table:Litvin357collinear}
%
\end{table*}

\begin{table*}
\caption{
IRREPs of multipoles in the collinear SPG
$\mathcal{G}_{\mathrm{SO}}^\infty \times {}^1 6 {}^1 2 {}^1 2 
= \mathrm{SO(2)}\times ( {}^1 6 {}^1 2 {}^1 2 + [\theta 2_{\perp\bm{n}} \,\|\, E] {}^1 6 {}^1 2 {}^1 2 )$.
All the symmetry operations of ${}^1 6 {}^1 2 {}^1 2$ are listed.
The other operations are omitted except for $[\theta 2_{\perp \bm{n}} \,\|\, E]$.
Among the listed operations, $3C^{\prime}_{2x} (3C^{\prime\prime}_{2y})$ denotes the conjugacy class consisting of three twofold rotations, with $C_{2x} (C_{2y})$ as a representative element.
The absence [presence] of a bar over the IRREPs indicates parity +1 [$-1$] under the antiunitary operation $[\theta 2_{\perp\bm{n}} \,\|\, E]$. 
}
\label{table:Litvin338collinear}

\end{table*}

\begin{table*}
\caption{
IRREPs of multipoles in the collinear SPG
$\mathcal{G}_{\mathrm{SO}}^\infty \times {}^1 6 {}^1 m {}^1 m 
= \mathrm{SO(2)}\times ( {}^1 6 {}^1 m {}^1 m + [\theta 2_{\perp\bm{n}} \,\|\, E] {}^1 6 {}^1 m {}^1 m )$.
All the symmetry operations of ${}^1 6 {}^1 m {}^1 m$ are listed.
The other operations are omitted except for $[\theta 2_{\perp \bm{n}} \,\|\, E]$.
Among the listed operations, $3C^{\prime}_{2x} (3C^{\prime\prime}_{2y})$ denotes the conjugacy class consisting of three twofold rotations, with $C_{2x} (C_{2y})$ as a representative element.
The absence [presence] of a bar over the IRREPs indicates parity +1 [$-1$] under the antiunitary operation $[\theta 2_{\perp\bm{n}} \,\|\, E]$. 
}
\label{table:Litvin393collinear}

\end{table*}

\begin{table*}
\caption{
IRREPs of multipoles in the collinear SPG
$\mathcal{G}_{\mathrm{SO}}^\infty \times {}^1 \bar{6} {}^1 m {}^1 2 
= \mathrm{SO(2)}\times ( {}^1 \bar{6} {}^1 m {}^1 2 + [\theta 2_{\perp\bm{n}} \,\|\, E] {}^1 \bar{6} {}^1 m {}^1 2 )$.
All the symmetry operations of ${}^1 \bar{6} {}^1 m {}^1 2$ are listed.
The other operations are omitted except for $[\theta 2_{\perp \bm{n}} \,\|\, E]$.
Among the listed operations, $3C^{\prime}_{2x} (3C^{\prime\prime}_{2y})$ denotes the conjugacy class consisting of three twofold rotations, with $C_{2x} (C_{2y})$ as a representative element.
The absence [presence] of a bar over the IRREPs indicates parity +1 [$-1$] under the antiunitary operation $[\theta 2_{\perp\bm{n}} \,\|\, E]$. 
}
\label{table:Litvin412collinear}

\end{table*}

\begin{table*}
\caption{
IRREPs of multipoles in the collinear SPG
$\mathcal{G}_{\mathrm{SO}}^\infty \times {}^1 6 / {}^1 m {}^1 m {}^1 m 
= \mathrm{SO(2)}\times ( {}^1 6 / {}^1 m {}^1 m {}^1 m + [\theta 2_{\perp\bm{n}} \,\|\, E] {}^1 6 / {}^1 m {}^1 m {}^1 m )$.
All the symmetry operations of ${}^1 6 / {}^1 m {}^1 m {}^1 m$ are listed.
The other operations are omitted except for $[\theta 2_{\perp \bm{n}} \,\|\, E]$.
Among the listed operations, $3C^{\prime}_{2x} (3C^{\prime\prime}_{2y})$ denotes the conjugacy class consisting of three twofold rotations, with $C_{2x} (C_{2y})$ as a representative element.
The absence [presence] of a bar over the IRREPs indicates parity +1 [$-1$] under the antiunitary operation $[\theta 2_{\perp\bm{n}} \,\|\, E]$. 
}
\label{table:Litvin440collinear}

\end{table*}

\clearpage 

\subsection{Collinear spin point groups with non-unitary nontrivial groups \label{sec:ap_classification_nonunitaryCSPG2}}

The complete classification of orbital and spin multipoles for the 58 collinear SPGs with non-unitary nontrivial groups is summarized in Tables~\ref{table:Litvin5collinear} -- ~\ref{table:Litvin455collinear}.

\renewcommand{\arraystretch}{1.8}
\begin{table*}[t]
\begin{center}
\caption{
IRREPs of multipoles in the collinear SPG
$\mathcal{G}_{\mathrm{SO}}^\infty \times {}^{\bar{1}} \bar{1}
= \mathrm{SO(2)}\times( {}^1 1 + [\theta \,\|\, I] {}^1 1 + [\theta 2_{\perp\bm{n}} \,\|\, E]( {}^1 1 + [\theta \,\|\, I] {}^1 1 ) )$.
All the symmetry operations of ${}^1 1 + [\theta \,\|\, I] {}^1 1 + [\theta 2_{\perp\bm{n}} \,\|\, E]( {}^1 1 + [\theta \,\|\, I] {}^1 1 )$ are listed.
The absence [presence] of a bar over the IRREPs indicates parity +1 [$-1$] under the antiunitary operation $[\theta 2_{\perp\bm{n}} \,\|\, E]$. 
}
\label{table:Litvin5collinear}
\begin{tabular}{|c|c|@{\hspace{2mm}}c@{\hspace{2mm}}|c|@{\hspace{1mm}}c@{\hspace{1mm}}|c|c|c|c|} 
\hline\hline
IRREP  &  $[E \,\|\, E]$  &  $[\theta \,\|\, I]$  
& $[\theta 2_{\perp \bm{n}} \,\|\, E]$  & $[2_{\perp \bm{n}} \,\|\, I]$ 
&  E  &  M  &  ET  &  MT         
\\ 
\hline
$\mathrm{A}^+$
& $1$  & $1$ & $1$  & $1$ 
& $Q_{lm}^{\rm (o)}(l:{\rm even}),$  
&  
& $G_{lm}^{\rm (o)}(l:{\rm odd}),$
& 
\\  
& & & & 
& $Q_{lm}^{\rm (s)}(l:{\rm odd})$
& 
& $G_{lm}^{\rm (s)}(l:{\rm even})$
& 
\\ 
$\bar{\mathrm{A}}^+$
& $1$  & $1$ & $-1$  & $-1$  
& 
& $M_{lm}^{\rm (o)}(l:{\rm even}),$
&  
& $T_{lm}^{\rm (o)}(l:{\rm odd}),$   
\\  
& & & &
& 
& $M_{lm}^{\rm (s)}(l:{\rm odd})$
& 
& $T_{lm}^{\rm (s)}(l:{\rm even})$
\\ \hline 
$\mathrm{A}^-$
& $1$  & $-1$ & $1$  & $-1$ 
& $Q_{lm}^{\rm (o)}(l:{\rm odd}),$  
& 
& $G_{lm}^{\rm (o)}(l:{\rm even}),$ 
& 
\\  
& & & & 
& $Q_{lm}^{\rm (s)}(l:{\rm even})$
& 
& $G_{lm}^{\rm (s)}(l:{\rm odd})$
&
\\ 
$\bar{\mathrm{A}}^-$
& $1$  & $-1$ & $-1$  & $1$ 
& 
& $M_{lm}^{\rm (o)} (l:{\rm odd}),$
& 
& $T_{lm}^{\rm (o)}(l:{\rm even}),$  
\\  
& & & &
& 
& $M_{lm}^{\rm (s)}(l:{\rm even})$
& 
& $T_{lm}^{\rm (s)}(l:{\rm odd})$
\\ 
\hline \hline 
\end{tabular}
\end{center}
\end{table*}

\renewcommand{\arraystretch}{1.5}
\begin{table*}
\caption{
IRREPs of multipoles in the collinear SPG
$\mathcal{G}_{\mathrm{SO}}^\infty \times {}^{\bar{1}} 2 
= \mathrm{SO(2)}\times( {}^1 1 + [\theta \,\|\, C_{2y}] {}^1 1 + [\theta 2_{\perp\bm{n}} \,\|\, E]( {}^1 1 + [\theta \,\|\, C_{2y}] {}^1 1 ) )$.
All the symmetry operations of ${}^1 1 + [\theta \,\|\, C_{2y}] {}^1 1 + [\theta 2_{\perp\bm{n}} \,\|\, E]( {}^1 1 + [\theta \,\|\, C_{2y}] {}^1 1 ) $ are listed.
The absence [presence] of a bar over the IRREPs indicates parity +1 [$-1$] under the antiunitary operation $[\theta 2_{\perp\bm{n}} \,\|\, E]$. 
}
\label{table:Litvin9collinear}

\end{table*}

\begin{table*}
\caption{
IRREPs of multipoles in the collinear SPG
$\mathcal{G}_{\mathrm{SO}}^\infty \times {}^{\bar{1}} m 
= \mathrm{SO(2)}\times( {}^1 1 + [\theta \,\|\, IC_{2y}] {}^1 1 + [\theta 2_{\perp\bm{n}} \,\|\, E]( {}^1 1 + [\theta \,\|\, IC_{2y}] {}^1 1 ) )$.
All the symmetry operations of ${}^1 1 + [\theta \,\|\, IC_{2y}] {}^1 1 + [\theta 2_{\perp\bm{n}} \,\|\, E]( {}^1 1 + [\theta \,\|\, IC_{2y}] {}^1 1 ) $ are listed.
The absence [presence] of a bar over the IRREPs indicates parity +1 [$-1$] under the antiunitary operation $[\theta 2_{\perp\bm{n}} \,\|\, E]$. 
}
\label{table:Litvin13collinear}

\end{table*}

\begin{table*}
\caption{
IRREPs of multipoles in the collinear SPG
$\mathcal{G}_{\mathrm{SO}}^\infty \times {}^1 2 / {}^{\bar{1}} m 
= \mathrm{SO(2)}\times( {}^1 2 + [\theta \,\|\, I] {}^1 2 + [\theta 2_{\perp\bm{n}} \,\|\, E]( {}^1 2 + [\theta \,\|\, I] {}^1 2 ) )$.
All the symmetry operations of ${}^1 2 + [\theta \,\|\, I] {}^1 2$ are listed.
The other operations are omitted except for $[\theta 2_{\perp \bm{n}} \,\|\, E]$.
The absence [presence] of a bar over the IRREPs indicates parity +1 [$-1$] under the antiunitary operation $[\theta 2_{\perp\bm{n}} \,\|\, E]$. 
}
\label{table:Litvin17collinear}

\end{table*}

\begin{table*}
\caption{
IRREPs of multipoles in the collinear SPG
$\mathcal{G}_{\mathrm{SO}}^\infty \times {}^{\bar{1}} 2 / {}^1 m 
= \mathrm{SO(2)}\times( {}^1 m + [\theta \,\|\, I] {}^1 m + [\theta 2_{\perp\bm{n}} \,\|\, E]( {}^1 m + [\theta \,\|\, I] {}^1 m ) )$.
All the symmetry operations of ${}^1 m + [\theta \,\|\, I] {}^1 m$ are listed.
The other operations are omitted except for $[\theta 2_{\perp \bm{n}} \,\|\, E]$.
The absence [presence] of a bar over the IRREPs indicates parity +1 [$-1$] under the antiunitary operation $[\theta 2_{\perp\bm{n}} \,\|\, E]$. 
}
\label{table:Litvin20collinear}

\end{table*}

\begin{table*}
\caption{
IRREPs of multipoles in the collinear SPG
$\mathcal{G}_{\mathrm{SO}}^\infty \times {}^{\bar{1}} 2 / {}^{\bar{1}} m 
= \mathrm{SO(2)}\times( {}^1 \bar{1} + [\theta \,\|\, C_{2y}] {}^1 \bar{1} + [\theta 2_{\perp\bm{n}} \,\|\, E]( {}^1 \bar{1} + [\theta \,\|\, C_{2y}] {}^1 \bar{1} ) )$.
All the symmetry operations of ${}^1 \bar{1} + [\theta \,\|\, C_{2y}] {}^1 \bar{1}$ are listed.
The other operations are omitted except for $[\theta 2_{\perp \bm{n}} \,\|\, E]$.
The absence [presence] of a bar over the IRREPs indicates parity +1 [$-1$] under the antiunitary operation $[\theta 2_{\perp\bm{n}} \,\|\, E]$. 
}
\label{table:Litvin23collinear}

\end{table*}

\begin{table*}
\caption{
IRREPs of multipoles in the collinear SPG
$\mathcal{G}_{\mathrm{SO}}^\infty \times {}^{\bar{1}} m {}^{\bar{1}} m {}^1 2 
= \mathrm{SO(2)}\times( {}^1 2 + [\theta \,\|\, IC_{2x}] {}^1 2 + [\theta 2_{\perp\bm{n}} \,\|\, E]( {}^1 2 + [\theta \,\|\, IC_{2x}] {}^1 2 ) )$.
All the symmetry operations of ${}^1 2 + [\theta \,\|\, IC_{2x}] {}^1 2$ are listed.
The other operations are omitted except for $[\theta 2_{\perp \bm{n}} \,\|\, E]$.
The absence [presence] of a bar over the IRREPs indicates parity +1 [$-1$] under the antiunitary operation $[\theta 2_{\perp\bm{n}} \,\|\, E]$. 
}
\label{table:Litvin37collinear}

\end{table*}

\begin{table*}
\caption{
IRREPs of multipoles in the collinear SPG
$\mathcal{G}_{\mathrm{SO}}^\infty \times {}^{\bar{1}} m {}^1 m {}^{\bar{1}} 2 
= \mathrm{SO(2)}\times( {}^1 m + [\theta \,\|\, C_{2z}] {}^1 m + [\theta 2_{\perp\bm{n}} \,\|\, E]( {}^1 m + [\theta \,\|\, C_{2z}] {}^1 m ) )$.
All the symmetry operations of ${}^1 m + [\theta \,\|\, C_{2z}] {}^1 m$ are listed.
The other operations are omitted except for $[\theta 2_{\perp \bm{n}} \,\|\, E]$.
The absence [presence] of a bar over the IRREPs indicates parity +1 [$-1$] under the antiunitary operation $[\theta 2_{\perp\bm{n}} \,\|\, E]$. 
}
\label{table:Litvin40collinear}

\end{table*}

\begin{table*}
\caption{
IRREPs of multipoles in the collinear SPG
$\mathcal{G}_{\mathrm{SO}}^\infty \times {}^{\bar{1}} 2 {}^{\bar{1}} 2 {}^1 2 
= \mathrm{SO(2)}\times( {}^1 2 + [\theta \,\|\, C_{2x}] {}^1 2 + [\theta 2_{\perp\bm{n}} \,\|\, E]( {}^1 2 + [\theta \,\|\, C_{2x}] {}^1 2 ) )$.
All the symmetry operations of ${}^1 2 + [\theta \,\|\, C_{2x}] {}^1 2$ are listed.
The other operations are omitted except for $[\theta 2_{\perp \bm{n}} \,\|\, E]$.
The absence [presence] of a bar over the IRREPs indicates parity +1 [$-1$] under the antiunitary operation $[\theta 2_{\perp\bm{n}} \,\|\, E]$. 
}
\label{table:Litvin50collinear}

\end{table*}

\begin{table*}
\caption{
IRREPs of multipoles in the collinear SPG
$\mathcal{G}_{\mathrm{SO}}^\infty \times {}^{\bar{1}} m {}^{\bar{1}} m {}^1 m 
= \mathrm{SO(2)}\times( {}^1 2 / {}^1 m + [\theta \,\|\, C_{2x}] {}^1 2 / {}^1 m + [\theta 2_{\perp\bm{n}} \,\|\, E]( {}^1 2 / {}^1 m + [\theta \,\|\, C_{2x}] {}^1 2 / {}^1 m ) )$.
All the symmetry operations of ${}^1 2 / {}^1 m + [\theta \,\|\, C_{2x}] {}^1 2 / {}^1 m$ are listed.
The other operations are omitted except for $[\theta 2_{\perp \bm{n}} \,\|\, E]$.
The absence [presence] of a bar over the IRREPs indicates parity +1 [$-1$] under the antiunitary operation $[\theta 2_{\perp\bm{n}} \,\|\, E]$. 
}
\label{table:Litvin57collinear}

\end{table*}

\begin{table*}
\caption{
IRREPs of multipoles in the collinear SPG
$\mathcal{G}_{\mathrm{SO}}^\infty \times {}^1 m {}^1 m {}^{\bar{1}} m 
= \mathrm{SO(2)}\times( {}^1 m {}^1 m {}^1 2 + [\theta \,\|\, I] {}^1 m {}^1 m {}^1 2 + [\theta 2_{\perp\bm{n}} \,\|\, E]( {}^1 m {}^1 m {}^1 2 + [\theta \,\|\, I] {}^1 m {}^1 m {}^1 2 ) )$.
All the symmetry operations of ${}^1 m {}^1 m {}^1 2 + [\theta \,\|\, I] {}^1 m {}^1 m {}^1 2$ are listed.
The other operations are omitted except for $[\theta 2_{\perp \bm{n}} \,\|\, E]$.
The absence [presence] of a bar over the IRREPs indicates parity +1 [$-1$] under the antiunitary operation $[\theta 2_{\perp\bm{n}} \,\|\, E]$. 
}
\label{table:Litvin60collinear}

\end{table*}

\begin{table*}
\caption{
IRREPs of multipoles in the collinear SPG
$\mathcal{G}_{\mathrm{SO}}^\infty \times {}^{\bar{1}} m {}^{\bar{1}} m {}^{\bar{1}} m 
= \mathrm{SO(2)}\times( {}^1 2 {}^1 2 {}^1 2 + [\theta \,\|\, I] {}^1 2 {}^1 2 {}^1 2 + [\theta 2_{\perp\bm{n}} \,\|\, E]( {}^1 2 {}^1 2 {}^1 2 + [\theta \,\|\, I] {}^1 2 {}^1 2 {}^1 2 ) )$.
All the symmetry operations of ${}^1 2 {}^1 2 {}^1 2 + [\theta \,\|\, I] {}^1 2 {}^1 2 {}^1 2$ are listed.
The other operations are omitted except for $[\theta 2_{\perp \bm{n}} \,\|\, E]$.
The absence [presence] of a bar over the IRREPs indicates parity +1 [$-1$] under the antiunitary operation $[\theta 2_{\perp\bm{n}} \,\|\, E]$. 
}
\label{table:Litvin63collinear}

\end{table*}

\clearpage 

\begin{table*}
\caption{
IRREPs of multipoles in the collinear SPG
$\mathcal{G}_{\mathrm{SO}}^\infty \times {}^{\bar{1}} 4 
= \mathrm{SO(2)}\times( {}^1 2 + [\theta \,\|\, C_{4z}] {}^1 2 + [\theta 2_{\perp\bm{n}} \,\|\, E]( {}^1 2 + [\theta \,\|\, C_{4z}] {}^1 2 ) )$.
All the symmetry operations of ${}^1 2 + [\theta \,\|\, C_{4z}] {}^1 2$ are listed.
The other operations are omitted except for $[\theta 2_{\perp \bm{n}} \,\|\, E]$.
The absence [presence] of a bar over the IRREPs indicates parity +1 [$-1$] under the antiunitary operation $[\theta 2_{\perp\bm{n}} \,\|\, E]$. 
}
\label{table:Litvin93collinear}

\end{table*}

\begin{table*}
\caption{
IRREPs of multipoles in the collinear SPG
$\mathcal{G}_{\mathrm{SO}}^\infty \times {}^{\bar{1}} \bar{4} 
= \mathrm{SO(2)}\times( {}^1 2 + [\theta \,\|\, IC_{4z}] {}^1 2 + [\theta 2_{\perp\bm{n}} \,\|\, E]( {}^1 2 + [\theta \,\|\, IC_{4z}] {}^1 2 ) )$.
All the symmetry operations of ${}^1 2 + [\theta \,\|\, IC_{4z}] {}^1 2$ are listed.
The other operations are omitted except for $[\theta 2_{\perp \bm{n}} \,\|\, E]$.
The absence [presence] of a bar over the IRREPs indicates parity +1 [$-1$] under the antiunitary operation $[\theta 2_{\perp\bm{n}} \,\|\, E]$. 
}
\label{table:Litvin99collinear}

\end{table*}

\begin{table*}
\caption{
IRREPs of multipoles in the collinear SPG
$\mathcal{G}_{\mathrm{SO}}^\infty \times {}^{\bar{1}} 4 / {}^1 m 
= \mathrm{SO(2)}\times( {}^1 2 / {}^1 m + [\theta \,\|\, C_{4z}] {}^1 2 / {}^1 m + [\theta 2_{\perp\bm{n}} \,\|\, E]( {}^1 2 / {}^1 m + [\theta \,\|\, C_{4z}] {}^1 2 / {}^1 m ) )$.
All the symmetry operations of ${}^1 2 / {}^1 m + [\theta \,\|\, C_{4z}] {}^1 2 / {}^1 m$ are listed.
The other operations are omitted except for $[\theta 2_{\perp \bm{n}} \,\|\, E]$.
The absence [presence] of a bar over the IRREPs indicates parity +1 [$-1$] under the antiunitary operation $[\theta 2_{\perp\bm{n}} \,\|\, E]$. 
}
\label{table:Litvin105collinear}

\end{table*}

\begin{table*}
\caption{
IRREPs of multipoles in the collinear SPG
$\mathcal{G}_{\mathrm{SO}}^\infty \times {}^{\bar{1}} 4 / {}^{\bar{1}} m 
= \mathrm{SO(2)}\times( {}^1 \bar{4} + [\theta \,\|\, I] {}^1 \bar{4} + [\theta 2_{\perp\bm{n}} \,\|\, E]( {}^1 \bar{4} + [\theta \,\|\, I] {}^1 \bar{4} ) )$.
All the symmetry operations of ${}^1 \bar{4} + [\theta \,\|\, I] {}^1 \bar{4}$ are listed.
The other operations are omitted except for $[\theta 2_{\perp \bm{n}} \,\|\, E]$.
The absence [presence] of a bar over the IRREPs indicates parity +1 [$-1$] under the antiunitary operation $[\theta 2_{\perp\bm{n}} \,\|\, E]$. 
}
\label{table:Litvin108collinear}

\end{table*}

\begin{table*}
\caption{
IRREPs of multipoles in the collinear SPG
$\mathcal{G}_{\mathrm{SO}}^\infty \times {}^1 4 / {}^{\bar{1}} m 
= \mathrm{SO(2)}\times( {}^1 4 + [\theta \,\|\, I] {}^1 4 + [\theta 2_{\perp\bm{n}} \,\|\, E]( {}^1 4 + [\theta \,\|\, I] {}^1 4 ) )$.
All the symmetry operations of ${}^1 4 + [\theta \,\|\, I] {}^1 4$ are listed.
The other operations are omitted except for $[\theta 2_{\perp \bm{n}} \,\|\, E]$.
The absence [presence] of a bar over the IRREPs indicates parity +1 [$-1$] under the antiunitary operation $[\theta 2_{\perp\bm{n}} \,\|\, E]$. 
}
\label{table:Litvin111collinear}

\end{table*}

\begin{table*}
\caption{
IRREPs of multipoles in the collinear SPG
$\mathcal{G}_{\mathrm{SO}}^\infty \times {}^1 4 {}^{\bar{1}} 2 {}^{\bar{1}} 2 
= \mathrm{SO(2)}\times( {}^1 4 + [\theta \,\|\, C_{2x}] {}^1 4 + [\theta 2_{\perp\bm{n}} \,\|\, E]( {}^1 4 + [\theta \,\|\, C_{2x}] {}^1 4 ) )$.
All the symmetry operations of ${}^1 4 + [\theta \,\|\, C_{2x}] {}^1 4$ are listed.
The other operations are omitted except for $[\theta 2_{\perp \bm{n}} \,\|\, E]$. 
Among the listed operations, $2C^{\prime}_{2x}$ denotes the conjugacy class consisting of two twofold rotations, with $C_{2x}$ as a representative element ($2C^{\prime}_{2x} \equiv C_{2x}, C_{2y}$).
The absence [presence] of a bar over the IRREPs indicates parity +1 [$-1$] under the antiunitary operation $[\theta 2_{\perp\bm{n}} \,\|\, E]$. 
}
\label{table:Litvin133collinear}

\end{table*}

\begin{table*}
\caption{
IRREPs of multipoles in the collinear SPG
$\mathcal{G}_{\mathrm{SO}}^\infty \times {}^{\bar{1}} 4 {}^1 2 {}^{\bar{1}} 2 
= \mathrm{SO(2)}\times( {}^1 2 {}^1 2 {}^1 2 + [\theta \,\|\, C_{2[110]}] {}^1 2 {}^1 2 {}^1 2 + [\theta 2_{\perp\bm{n}} \,\|\, E]( {}^1 2 {}^1 2 {}^1 2 + [\theta \,\|\, C_{2[110]}] {}^1 2 {}^1 2 {}^1 2 ) )$.
All the symmetry operations of ${}^1 2 {}^1 2 {}^1 2$ are listed.
The other operations are omitted except for $[\theta \,\|\, C_{2[110]}] $ and $[\theta 2_{\perp \bm{n}} \,\|\, E]$.
The absence [presence] of a bar over the IRREPs indicates parity +1 [$-1$] under the antiunitary operation $[\theta 2_{\perp\bm{n}} \,\|\, E]$. 
}
\label{table:Litvin136collinear}

\end{table*}

\begin{table*}
\caption{
IRREPs of multipoles in the collinear SPG
$\mathcal{G}_{\mathrm{SO}}^\infty \times {}^1 4 {}^{\bar{1}} m {}^{\bar{1}} m 
= \mathrm{SO(2)}\times( {}^1 4 + [\theta \,\|\, IC_{2x}] {}^1 4 + [\theta 2_{\perp\bm{n}} \,\|\, E]( {}^1 4 + [\theta \,\|\, IC_{2x}] {}^1 4 ) )$.
All the symmetry operations of ${}^1 4 + [\theta \,\|\, IC_{2x}] {}^1 4$ are listed.
The other operations are omitted except for $[\theta 2_{\perp \bm{n}} \,\|\, E]$. 
Among the listed operations, $2C^{\prime}_{2x}$ denotes the conjugacy class consisting of two twofold rotations, with $C_{2x}$ as a representative element ($2C^{\prime}_{2x} \equiv C_{2x}, C_{2y}$).
The absence [presence] of a bar over the IRREPs indicates parity +1 [$-1$] under the antiunitary operation $[\theta 2_{\perp\bm{n}} \,\|\, E]$. 
}
\label{table:Litvin149collinear}

\end{table*}

\begin{table*}
\caption{
IRREPs of multipoles in the collinear SPG
$\mathcal{G}_{\mathrm{SO}}^\infty \times {}^{\bar{1}} 4 {}^1 m {}^{\bar{1}} m 
= \mathrm{SO(2)}\times( {}^1 m {}^1 m {}^1 2 + [\theta \,\|\, IC_{2[110]}] {}^1 m {}^1 m {}^1 2 + [\theta 2_{\perp\bm{n}} \,\|\, E]( {}^1 m {}^1 m {}^1 2 + [\theta \,\|\, IC_{2[110]}] {}^1 m {}^1 m {}^1 2 ) )$.
All the symmetry operations of ${}^1 m {}^1 m {}^1 2$ are listed.
The other operations are omitted except for $[\theta \,\|\, IC_{2[110]}]$ and $[\theta 2_{\perp \bm{n}} \,\|\, E]$.
The absence [presence] of a bar over the IRREPs indicates parity +1 [$-1$] under the antiunitary operation $[\theta 2_{\perp\bm{n}} \,\|\, E]$. 
}
\label{table:Litvin152collinear}

\end{table*}

\begin{table*}
\caption{
IRREPs of multipoles in the collinear SPG
$\mathcal{G}_{\mathrm{SO}}^\infty \times {}^1 \bar{4} {}^{\bar{1}} 2 {}^{\bar{1}} m 
= \mathrm{SO(2)}\times( {}^1 \bar{4} + [\theta \,\|\, C_{2x}] {}^1 \bar{4} + [\theta 2_{\perp\bm{n}} \,\|\, E]( {}^1 \bar{4} + [\theta \,\|\, C_{2x}] {}^1 \bar{4} ) )$.
All the symmetry operations of ${}^1 \bar{4} + [\theta \,\|\, C_{2x}] {}^1 \bar{4}$ are listed.
The other operations are omitted except for $[\theta 2_{\perp \bm{n}} \,\|\, E]$. 
Among the listed operations, $2C^{\prime}_{2x}$ denotes the conjugacy class consisting of two twofold rotations, with $C_{2x}$ as a representative element ($2C^{\prime}_{2x} \equiv C_{2x}, C_{2y}$).
The absence [presence] of a bar over the IRREPs indicates parity +1 [$-1$] under the antiunitary operation $[\theta 2_{\perp\bm{n}} \,\|\, E]$. 
}
\label{table:Litvin165collinear}

\end{table*}

\begin{table*}
\caption{
IRREPs of multipoles in the collinear SPG
$\mathcal{G}_{\mathrm{SO}}^\infty \times {}^{\bar{1}} \bar{4} {}^1 m {}^{\bar{1}} 2 
= \mathrm{SO(2)}\times( {}^1 m {}^1 m {}^1 2 + [\theta \,\|\, C_{2[110]}] {}^1 m {}^1 m {}^1 2 + [\theta 2_{\perp\bm{n}} \,\|\, E]( {}^1 m {}^1 m {}^1 2 + [\theta \,\|\, C_{2[110]}] {}^1 m {}^1 m {}^1 2 ) )$.
All the symmetry operations of ${}^1 m {}^1 m {}^1 2$ are listed.
The other operations are omitted except for $[\theta \,\|\, C_{2[110]}]$ and $[\theta 2_{\perp \bm{n}} \,\|\, E]$.
The absence [presence] of a bar over the IRREPs indicates parity +1 [$-1$] under the antiunitary operation $[\theta 2_{\perp\bm{n}} \,\|\, E]$. 
}
\label{table:Litvin168collinear}

\end{table*}

\begin{table*}
\caption{
IRREPs of multipoles in the collinear SPG
$\mathcal{G}_{\mathrm{SO}}^\infty \times {}^{\bar{1}} \bar{4} {}^1 2 {}^{\bar{1}} m 
= \mathrm{SO(2)}\times( {}^1 2 {}^1 2 {}^1 2 + [\theta \,\|\, IC_{2[110]}] {}^1 2 {}^1 2 {}^1 2 + [\theta 2_{\perp\bm{n}} \,\|\, E]( {}^1 2 {}^1 2 {}^1 2 + [\theta \,\|\, IC_{2[110]}] {}^1 2 {}^1 2 {}^1 2 ) )$.
All the symmetry operations of ${}^1 2 {}^1 2 {}^1 2$ are listed.
The other operations are omitted except for $[\theta \,\|\, IC_{2[110]}]$ and $[\theta 2_{\perp \bm{n}} \,\|\, E]$.
The absence [presence] of a bar over the IRREPs indicates parity +1 [$-1$] under the antiunitary operation $[\theta 2_{\perp\bm{n}} \,\|\, E]$. 
}
\label{table:Litvin171collinear}

\end{table*}

\clearpage 

\begin{table*}
\caption{
IRREPs of multipoles in the collinear SPG
$\mathcal{G}_{\mathrm{SO}}^\infty \times {}^{\bar{1}} 4 / {}^{\bar{1}}  m {}^{\bar{1}} m {}^1 m 
= \mathrm{SO(2)}\times( {}^1 \bar{4} {}^1 2 {}^1 m + [\theta \,\|\, I] {}^1 \bar{4} {}^1 2 {}^1 m + [\theta 2_{\perp\bm{n}} \,\|\, E]( {}^1 \bar{4} {}^1 2 {}^1 m + [\theta \,\|\, I] {}^1 \bar{4} {}^1 2 {}^1 m ) )$.
All the symmetry operations of ${}^1 \bar{4} {}^1 2 {}^1 m + [\theta \,\|\, I] {}^1 \bar{4} {}^1 2 {}^1 m$ are listed.
The other operations are omitted except for $[\theta 2_{\perp \bm{n}} \,\|\, E]$. 
Among the listed operations, $2C^{\prime}_{2x}$ denotes the conjugacy class consisting of two twofold rotations, with $C_{2x}$ as a representative element ($2C^{\prime}_{2x} \equiv C_{2x}, C_{2y}$).
The absence [presence] of a bar over the IRREPs indicates parity +1 [$-1$] under the antiunitary operation $[\theta 2_{\perp\bm{n}} \,\|\, E]$. 
}
\label{table:Litvin189collinear}

\end{table*}

\begin{table*}
\caption{
IRREPs of multipoles in the collinear SPG
$\mathcal{G}_{\mathrm{SO}}^\infty \times {}^1 4 / {}^{\bar{1}} m {}^1 m {}^1 m 
= \mathrm{SO(2)}\times( {}^1 4 {}^1 m {}^1 m + [\theta \,\|\, I] {}^1 4 {}^1 m {}^1 m + [\theta 2_{\perp\bm{n}} \,\|\, E]( {}^1 4 {}^1 m {}^1 m + [\theta \,\|\, I] {}^1 4 {}^1 m {}^1 m ) )$.
All the symmetry operations of ${}^1 4 {}^1 m {}^1 m + [\theta \,\|\, I] {}^1 4 {}^1 m {}^1 m$ are listed.
The other operations are omitted except for $[\theta 2_{\perp \bm{n}} \,\|\, E]$. 
Among the listed operations, $2C^{\prime}_{2x}$ denotes the conjugacy class consisting of two twofold rotations, with $C_{2x}$ as a representative element ($2C^{\prime}_{2x} \equiv C_{2x}, C_{2y}$).
The absence [presence] of a bar over the IRREPs indicates parity +1 [$-1$] under the antiunitary operation $[\theta 2_{\perp\bm{n}} \,\|\, E]$. 
}
\label{table:Litvin192collinear}

\end{table*}

\begin{table*}
\caption{
IRREPs of multipoles in the collinear SPG
$\mathcal{G}_{\mathrm{SO}}^\infty \times {}^{\bar{1}} 4 / {}^1 m {}^1 m {}^{\bar{1}} m 
= \mathrm{SO(2)}\times( {}^1 m {}^1 m {}^1 m + [\theta \,\|\, C_{2[110]}] {}^1 m {}^1 m {}^1 m + [\theta 2_{\perp\bm{n}} \,\|\, E]( {}^1 m {}^1 m {}^1 m + [\theta \,\|\, C_{2[110]}] {}^1 m {}^1 m {}^1 m ) )$.
All the symmetry operations of ${}^1 m {}^1 m {}^1 m$ are listed.
The other operations are omitted except for $[\theta \,\|\, C_{2[110]}]$ and $[\theta 2_{\perp \bm{n}} \,\|\, E]$.
The absence [presence] of a bar over the IRREPs indicates parity +1 [$-1$] under the antiunitary operation $[\theta 2_{\perp\bm{n}} \,\|\, E]$. 
}
\label{table:Litvin195collinear}

\end{table*}

\begin{table*}
\caption{
IRREPs of multipoles in the collinear SPG
$\mathcal{G}_{\mathrm{SO}}^\infty \times {}^1 4 / {}^1 m {}^{\bar{1}} m {}^{\bar{1}} m 
= \mathrm{SO(2)}\times( {}^1 4 / {}^1 m + [\theta \,\|\, C_{2x}] {}^1 4 / {}^1 m + [\theta 2_{\perp\bm{n}} \,\|\, E]( {}^1 4 / {}^1 m + [\theta \,\|\, C_{2x}] {}^1 4 / {}^1 m ) )$.
All the symmetry operations of ${}^1 4 / {}^1 m + [\theta \,\|\, C_{2x}] {}^1 4 / {}^1 m$ are listed.
The other operations are omitted except for $[\theta 2_{\perp \bm{n}} \,\|\, E]$. 
Among the listed operations, $2C^{\prime}_{2x}$ denotes the conjugacy class consisting of two twofold rotations, with $C_{2x}$ as a representative element ($2C^{\prime}_{2x} \equiv C_{2x}, C_{2y}$).
The absence [presence] of a bar over the IRREPs indicates parity +1 [$-1$] under the antiunitary operation $[\theta 2_{\perp\bm{n}} \,\|\, E]$. 
}
\label{table:Litvin198collinear}

\end{table*}

\begin{table*}
\caption{
IRREPs of multipoles in the collinear SPG
$\mathcal{G}_{\mathrm{SO}}^\infty \times {}^1 4 / {}^{\bar{1}} m {}^{\bar{1}} m {}^{\bar{1}} m 
= \mathrm{SO(2)}\times( {}^1 4 {}^1 2 {}^1 2 + [\theta \,\|\, I] {}^1 4 {}^1 2 {}^1 2 + [\theta 2_{\perp\bm{n}} \,\|\, E]( {}^1 4 {}^1 2 {}^1 2 + [\theta \,\|\, I] {}^1 4 {}^1 2 {}^1 2 ) )$.
All the symmetry operations of ${}^1 4 {}^1 2 {}^1 2 + [\theta \,\|\, I] {}^1 4 {}^1 2 {}^1 2$ are listed.
The other operations are omitted except for $[\theta 2_{\perp \bm{n}} \,\|\, E]$. 
Among the listed operations, $2C^{\prime}_{2x}$ denotes the conjugacy class consisting of two twofold rotations, with $C_{2x}$ as a representative element ($2C^{\prime}_{2x} \equiv C_{2x}, C_{2y}$).
The absence [presence] of a bar over the IRREPs indicates parity +1 [$-1$] under the antiunitary operation $[\theta 2_{\perp\bm{n}} \,\|\, E]$. 
}
\label{table:Litvin201collinear}

\end{table*}

\clearpage 

\begin{table*}
\caption{
IRREPs of multipoles in the collinear SPG
$\mathcal{G}_{\mathrm{SO}}^\infty \times {}^1 2 / {}^{\bar{1}} m {}^{\bar{1}} \bar{3} 
= \mathrm{SO(2)}\times( {}^1 2 {}^1 3 + [\theta \,\|\, I] {}^1 2 {}^1 3 + [\theta 2_{\perp\bm{n}} \,\|\, E]( {}^1 2 {}^1 3 + [\theta \,\|\, I] {}^1 2 {}^1 3 ) )$.
All the symmetry operations of ${}^1 2 {}^1 3 + [\theta \,\|\, I] {}^1 2 {}^1 3$ are listed.
The other operations are omitted except for $[\theta 2_{\perp \bm{n}} \,\|\, E]$.
The absence [presence] of a bar over the IRREPs indicates parity +1 [$-1$] under the antiunitary operation $[\theta 2_{\perp\bm{n}} \,\|\, E]$. 
}
\label{table:Litvin544collinear}

\end{table*}

\begin{table*}
\caption{
IRREPs of multipoles in the collinear SPG
$\mathcal{G}_{\mathrm{SO}}^\infty \times {}^{\bar{1}} \bar{4} {}^1 3 {}^{\bar{1}} m 
= \mathrm{SO(2)}\times ( {}^1 2 {}^1 3 + [\theta \,\|\, IC_{2[110]}] {}^1 2 {}^1 3 + [\theta 2_{\perp\bm{n}} \,\|\, E]( {}^1 2 {}^1 3 + [\theta \,\|\, IC_{2[110]}] {}^1 2 {}^1 3 ) )$.
All the symmetry operations of ${}^1 2 {}^1 3 + [\theta \,\|\, IC_{2[110]}] {}^1 2 {}^1 3$ are listed.
The other operations are omitted except for $[\theta 2_{\perp \bm{n}} \,\|\, E]$.
The absence [presence] of a bar over the IRREPs indicates parity +1 [$-1$] under the antiunitary operation $[\theta 2_{\perp\bm{n}} \,\|\, E]$. 
}
\label{table:Litvin554collinear}

\end{table*}

\begin{table*}
\caption{
IRREPs of multipoles in the collinear SPG
$\mathcal{G}_{\mathrm{SO}}^\infty \times {}^{\bar{1}} 4 {}^1 3 {}^{\bar{1}} 2 
= \mathrm{SO(2)}\times ( {}^1 2 {}^1 3 + [\theta \,\|\, C_{2[110]}] {}^1 2 {}^1 3 + [\theta 2_{\perp\bm{n}} \,\|\, E]( {}^1 2 {}^1 3 + [\theta \,\|\, C_{2[110]}] {}^1 2 {}^1 3 ) )$.
All the symmetry operations of ${}^1 2 {}^1 3 + [\theta \,\|\, C_{2[110]}] {}^1 2 {}^1 3$ are listed.
The other operations are omitted except for $[\theta 2_{\perp \bm{n}} \,\|\, E]$.
The absence [presence] of a bar over the IRREPs indicates parity +1 [$-1$] under the antiunitary operation $[\theta 2_{\perp\bm{n}} \,\|\, E]$. 
}
\label{table:Litvin562collinear}

\end{table*}

\begin{table*}
\caption{
IRREPs of multipoles in the collinear SPG
$\mathcal{G}_{\mathrm{SO}}^\infty \times {}^{\bar{1}} 4 / {}^1 m {}^1 \bar{3} {}^{\bar{1}} 2 / {}^{\bar{1}} m  
= \mathrm{SO(2)}\times ({}^1 2 /  {}^1 m {}^1 \bar{3} + [\theta \,\|\, C_{2[110]}] {}^1 2 / {}^1 m {}^1 \bar{3} + [\theta 2_{\perp\bm{n}} \,\|\, E]( {}^1 2 / {}^1 m {}^1 \bar{3} + [\theta \,\|\, C_{2[110]}] {}^1 2 / {}^1 m {}^1 \bar{3} ) )$.
All the symmetry operations of ${}^1 2 / {}^1 m {}^1 \bar{3}$ are listed.
The other operations are omitted except for $[\theta \,\|\, C_{2[110]}]$ and $[\theta 2_{\perp \bm{n}} \,\|\, E]$.
The absence [presence] of a bar over the IRREPs indicates parity +1 [$-1$] under the antiunitary operation $[\theta 2_{\perp\bm{n}} \,\|\, E]$. 
}
\label{table:Litvin570collinear}

\end{table*}

\begin{table*}
\caption{
IRREPs of multipoles in the collinear SPG
$\mathcal{G}_{\mathrm{SO}}^\infty \times {}^{\bar{1}} 4 / {}^{\bar{1}} m {}^{\bar{1}} \bar{3} {}^{\bar{1}} 2 / {}^1 m  
= \mathrm{SO(2)}\times ( {}^1 \bar{4} {}^1 3 {}^1 m + [\theta \,\|\, I] {}^1 \bar{4} {}^1 3 {}^1 m + [\theta 2_{\perp\bm{n}} \,\|\, E]( {}^1 \bar{4} {}^1 3 {}^1 m + [\theta \,\|\, I] {}^1 \bar{4} {}^1 3 {}^1 m ) )$.
All the symmetry operations of ${}^1 \bar{4} {}^1 3 {}^1 m$ are listed.
The other operations are omitted except for $[\theta \,\|\, I]$ and $[\theta 2_{\perp \bm{n}} \,\|\, E]$.
The absence [presence] of a bar over the IRREPs indicates parity +1 [$-1$] under the antiunitary operation $[\theta 2_{\perp\bm{n}} \,\|\, E]$. 
}
\label{table:Litvin573collinear}

\end{table*}

\begin{table*}
\caption{
IRREPs of multipoles in the collinear SPG
$\mathcal{G}_{\mathrm{SO}}^\infty \times {}^1 4 / {}^{\bar{1}} m {}^{\bar{1}} \bar{3} {}^1 2 / {}^{\bar{1}} m 
= \mathrm{SO(2)}\times ( {}^1 4 {}^1 3 {}^1 2 + [\theta \,\|\, I] {}^1 4 {}^1 3 {}^1 2 + [\theta 2_{\perp\bm{n}} \,\|\, E]( {}^1 4 {}^1 3 {}^1 2 + [\theta \,\|\, I] {}^1 4 {}^1 3 {}^1 2 ) )$.
All the symmetry operations of ${}^1 4 {}^1 3 {}^1 2$ are listed.
The other operations are omitted except for $[\theta \,\|\, I] $ and $[\theta 2_{\perp \bm{n}} \,\|\, E]$.
The absence [presence] of a bar over the IRREPs indicates parity +1 [$-1$] under the antiunitary operation $[\theta 2_{\perp\bm{n}} \,\|\, E]$. 
}
\label{table:Litvin576collinear}

\end{table*}

\clearpage 

\begin{table*}
\caption{
IRREPs of multipoles in the collinear SPG
$\mathcal{G}_{\mathrm{SO}}^\infty \times {}^{\bar{1}} \bar{3} 
= \mathrm{SO(2)}\times( {}^1 3 + [\theta \,\|\, I] {}^1 3 + [\theta 2_{\perp\bm{n}} \,\|\, E]( {}^1 3 + [\theta \,\|\, I] {}^1 3 ) )$.
All the symmetry operations of ${}^1 3 + [\theta \,\|\, I] {}^1 3$ are listed.
The other operations are omitted except for $[\theta 2_{\perp \bm{n}} \,\|\, E]$.
The absence [presence] of a bar over the IRREPs indicates parity +1 [$-1$] under the antiunitary operation $[\theta 2_{\perp\bm{n}} \,\|\, E]$. 
}
\label{table:Litvin277collinear}

\end{table*}

\begin{table*}
\caption{
IRREPs of multipoles in the collinear SPG
$\mathcal{G}_{\mathrm{SO}}^\infty \times {}^1 3 {}^{\bar{1}} 2 
= \mathrm{SO(2)}\times ( {}^1 3 + [\theta \,\|\, C_{2y}] {}^1 3 + [\theta 2_{\perp\bm{n}} \,\|\, E]( {}^1 3 + [\theta \,\|\, C_{2y}] {}^1 3 ) )$.
All the symmetry operations of ${}^1 3 + [\theta \,\|\, C_{2y}] {}^1 3$ are listed.
The other operations are omitted except for $[\theta 2_{\perp \bm{n}} \,\|\, E]$.
Among the listed operations, $3C^{\prime}_{2y}$ denotes the conjugacy class consisting of three twofold rotations, with $C_{2y}$ as a representative element.
The absence [presence] of a bar over the IRREPs indicates parity +1 [$-1$] under the antiunitary operation $[\theta 2_{\perp\bm{n}} \,\|\, E]$. 
}
\label{table:Litvin285collinear}

\end{table*}

\begin{table*}
\caption{
IRREPs of multipoles in the collinear SPG
$\mathcal{G}_{\mathrm{SO}}^\infty \times {}^1 3 {}^{\bar{1}} m 
= \mathrm{SO(2)}\times ( {}^1 3 + [\theta \,\|\, IC_{2y}] {}^1 3 + [\theta 2_{\perp\bm{n}} \,\|\, E]( {}^1 3 + [\theta \,\|\, IC_{2y}] {}^1 3 ) )$.
All the symmetry operations of ${}^1 3 + [\theta \,\|\, IC_{2y}] {}^1 3$ are listed.
The other operations are omitted except for $[\theta 2_{\perp \bm{n}} \,\|\, E]$.
Among the listed operations, $3C^{\prime}_{2y}$ denotes the conjugacy class consisting of three twofold rotations, with $C_{2y}$ as a representative element.
The absence [presence] of a bar over the IRREPs indicates parity +1 [$-1$] under the antiunitary operation $[\theta 2_{\perp\bm{n}} \,\|\, E]$. 
}
\label{table:Litvin291collinear}

\end{table*}

\begin{table*}
\caption{
IRREPs of multipoles in the collinear SPG
$\mathcal{G}_{\mathrm{SO}}^\infty \times {}^1 \bar{3} {}^{\bar{1}} m 
= \mathrm{SO(2)}\times ( {}^1 \bar{3} + [\theta \,\|\, C_{2y}] {}^1 \bar{3} + [\theta 2_{\perp\bm{n}} \,\|\, E]( {}^1 \bar{3} + [\theta \,\|\, C_{2y}] {}^1 \bar{3} ) )$.
All the symmetry operations of ${}^1 \bar{3} + [\theta \,\|\, C_{2y}] {}^1 \bar{3}$ are listed.
The other operations are omitted except for $[\theta 2_{\perp \bm{n}} \,\|\, E]$.
Among the listed operations, $3C^{\prime}_{2y}$ denotes the conjugacy class consisting of three twofold rotations, with $C_{2y}$ as a representative element.
The absence [presence] of a bar over the IRREPs indicates parity +1 [$-1$] under the antiunitary operation $[\theta 2_{\perp\bm{n}} \,\|\, E]$. 
}
\label{table:Litvin297collinear}

\end{table*}

\begin{table*}
\caption{
IRREPs of multipoles in the collinear SPG
$\mathcal{G}_{\mathrm{SO}}^\infty \times {}^{\bar{1}} \bar{3} {}^1 m 
= \mathrm{SO(2)}\times ( {}^1 3 {}^1 m + [\theta \,\|\, I] {}^1 3 {}^1 m + [\theta 2_{\perp\bm{n}} \,\|\, E]( {}^1 3 {}^1 m + [\theta \,\|\, I] {}^1 3 {}^1 m ) )$.
All the symmetry operations of ${}^1 3 {}^1 m + [\theta \,\|\, I] {}^1 3 {}^1 m$ are listed.
The other operations are omitted except for $[\theta 2_{\perp \bm{n}} \,\|\, E]$.
Among the listed operations, $3C^{\prime}_{2y}$ denotes the conjugacy class consisting of three twofold rotations, with $C_{2y}$ as a representative element.
The absence [presence] of a bar over the IRREPs indicates parity +1 [$-1$] under the antiunitary operation $[\theta 2_{\perp\bm{n}} \,\|\, E]$. 
}
\label{table:Litvin300collinear}

\end{table*}

\begin{table*}
\caption{
IRREPs of multipoles in the collinear SPG
$\mathcal{G}_{\mathrm{SO}}^\infty \times {}^{\bar{1}} \bar{3} {}^{\bar{1}} m 
= \mathrm{SO(2)}\times ( {}^1 3 {}^1 2 + [\theta \,\|\, I] {}^1 3 {}^1 2 + [\theta 2_{\perp\bm{n}} \,\|\, E]( {}^1 3 {}^1 2 + [\theta \,\|\, I] {}^1 3 {}^1 2 ) )$.
All the symmetry operations of ${}^1 3 {}^1 2 + [\theta \,\|\, I] {}^1 3 {}^1 2$ are listed.
The other operations are omitted except for $[\theta 2_{\perp \bm{n}} \,\|\, E]$.
Among the listed operations, $3C^{\prime}_{2y}$ denotes the conjugacy class consisting of three twofold rotations, with $C_{2y}$ as a representative element.
The absence [presence] of a bar over the IRREPs indicates parity +1 [$-1$] under the antiunitary operation $[\theta 2_{\perp\bm{n}} \,\|\, E]$. 
}
\label{table:Litvin303collinear}

\end{table*}

\clearpage 

\begin{table*}
\caption{
IRREPs of multipoles in the collinear SPG
$\mathcal{G}_{\mathrm{SO}}^\infty \times {}^{\bar{1}} 6 
= \mathrm{SO(2)}\times ( {}^1 3 + [\theta \,\|\, C_{2z}] {}^1 3 + [\theta 2_{\perp\bm{n}} \,\|\, E]( {}^1 3 + [\theta \,\|\, C_{2z}] {}^1 3 ) )$.
All the symmetry operations of ${}^1 3 + [\theta \,\|\, C_{2z}] {}^1 3$ are listed.
The other operations are omitted except for $[\theta 2_{\perp \bm{n}} \,\|\, E]$.
The absence [presence] of a bar over the IRREPs indicates parity +1 [$-1$] under the antiunitary operation $[\theta 2_{\perp\bm{n}} \,\|\, E]$. 
}
\label{table:Litvin333collinear}

\end{table*}

\begin{table*}
\caption{
IRREPs of multipoles in the collinear SPG
$\mathcal{G}_{\mathrm{SO}}^\infty \times {}^{\bar{1}} \bar{6} 
= \mathrm{SO(2)}\times ( {}^1 3 + [\theta \,\|\, IC_{2z}] {}^1 3 + [\theta 2_{\perp\bm{n}} \,\|\, E]( {}^1 3 + [\theta \,\|\, IC_{2z}] {}^1 3 ) )$.
All the symmetry operations of ${}^1 3 + [\theta \,\|\, IC_{2z}] {}^1 3$ are listed.
The other operations are omitted except for $[\theta 2_{\perp \bm{n}} \,\|\, E]$.
The absence [presence] of a bar over the IRREPs indicates parity +1 [$-1$] under the antiunitary operation $[\theta 2_{\perp\bm{n}} \,\|\, E]$. 
}
\label{table:Litvin325collinear}

\end{table*}

\begin{table*}
\caption{
IRREPs of multipoles in the collinear SPG
$\mathcal{G}_{\mathrm{SO}}^\infty \times {}^{\bar{1}} 6 / {}^{\bar{1}} m 
= \mathrm{SO(2)}\times ( {}^1 \bar{3} + [\theta \,\|\, C_{2z}] {}^1 \bar{3} + [\theta 2_{\perp\bm{n}} \,\|\, E]( {}^1 \bar{3} + [\theta \,\|\, C_{2z}] {}^1 \bar{3} ) )$.
All the symmetry operations of ${}^1 \bar{3} + [\theta \,\|\, C_{2z}] {}^1 \bar{3}$ are listed.
The other operations are omitted except for $[\theta 2_{\perp \bm{n}} \,\|\, E]$.
The absence [presence] of a bar over the IRREPs indicates parity +1 [$-1$] under the antiunitary operation $[\theta 2_{\perp\bm{n}} \,\|\, E]$. 
}
\label{table:Litvin360collinear}

\end{table*}

\begin{table*}
\caption{
IRREPs of multipoles in the collinear SPG
$\mathcal{G}_{\mathrm{SO}}^\infty \times {}^{\bar{1}} 6 / {}^1 m 
= \mathrm{SO(2)}\times ( {}^1 \bar{6} + [\theta \,\|\, I] {}^1 \bar{6} + [\theta 2_{\perp\bm{n}} \,\|\, E]( {}^1 \bar{6} + [\theta \,\|\, I] {}^1 \bar{6} ) )$.
All the symmetry operations of ${}^1 \bar{6} + [\theta \,\|\, I] {}^1 \bar{6}$ are listed.
The other operations are omitted except for $[\theta 2_{\perp \bm{n}} \,\|\, E]$.
The absence [presence] of a bar over the IRREPs indicates parity +1 [$-1$] under the antiunitary operation $[\theta 2_{\perp\bm{n}} \,\|\, E]$. 
}
\label{table:Litvin363collinear}

\end{table*}

\begin{table*}
\caption{
IRREPs of multipoles in the collinear SPG
$\mathcal{G}_{\mathrm{SO}}^\infty \times {}^1 6 / {}^{\bar{1}} m 
= \mathrm{SO(2)}\times ( {}^1 6 + [\theta \,\|\, I] {}^1 6 + [\theta 2_{\perp\bm{n}} \,\|\, E]( {}^1 6 + [\theta \,\|\, I] {}^1 6 ) )$.
All the symmetry operations of ${}^1 6 + [\theta \,\|\, I] {}^1 6$ are listed.
The other operations are omitted except for $[\theta 2_{\perp \bm{n}} \,\|\, E]$.
The absence [presence] of a bar over the IRREPs indicates parity +1 [$-1$] under the antiunitary operation $[\theta 2_{\perp\bm{n}} \,\|\, E]$. 
}
\label{table:Litvin366collinear}

\end{table*}

\begin{table*}
\caption{
IRREPs of multipoles in the collinear SPG
$\mathcal{G}_{\mathrm{SO}}^\infty \times {}^1 6 {}^{\bar{1}} 2 {}^{\bar{1}} 2 
= \mathrm{SO(2)}\times ( {}^1 6 + [\theta \,\|\, C_{2x}] {}^1 6 + [\theta 2_{\perp\bm{n}} \,\|\, E]( {}^1 6 + [\theta \,\|\, C_{2x}] {}^1 6 ) )$.
All the symmetry operations of ${}^1 6 + [\theta \,\|\, C_{2x}] {}^1 6$ are listed.
The other operations are omitted except for $[\theta 2_{\perp \bm{n}} \,\|\, E]$.
Among the listed operations, $3C^{\prime}_{2x} (3C^{\prime\prime}_{2y})$ denotes the conjugacy class consisting of three twofold rotations, with $C_{2x} (C_{2y})$ as a representative element.
The absence [presence] of a bar over the IRREPs indicates parity +1 [$-1$] under the antiunitary operation $[\theta 2_{\perp\bm{n}} \,\|\, E]$. 
}
\label{table:Litvin341collinear}

\end{table*}

\begin{table*}
\caption{
IRREPs of multipoles in the collinear SPG
$\mathcal{G}_{\mathrm{SO}}^\infty \times {}^{\bar{1}} 6 {}^1 2 {}^{\bar{1}} 2 
= \mathrm{SO(2)}\times ( {}^1 3 {}^1 2 + [\theta \,\|\, C_{2z}] {}^1 3 {}^1 2 + [\theta 2_{\perp\bm{n}} \,\|\, E]( {}^1 3 {}^1 2 + [\theta \,\|\, C_{2z}] {}^1 3 {}^1 2 ) )$.
All the symmetry operations of ${}^1 3 {}^1 2 + [\theta \,\|\, C_{2z}] {}^1 3 {}^1 2$ are listed.
The other operations are omitted except for $[\theta 2_{\perp \bm{n}} \,\|\, E]$.
Among the listed operations, $3C^{\prime}_{2x} (3C^{\prime\prime}_{2y})$ denotes the conjugacy class consisting of three twofold rotations, with $C_{2x} (C_{2y})$ as a representative element.
The absence [presence] of a bar over the IRREPs indicates parity +1 [$-1$] under the antiunitary operation $[\theta 2_{\perp\bm{n}} \,\|\, E]$. 
}
\label{table:Litvin344collinear}

\end{table*}

\begin{table*}
\caption{
IRREPs of multipoles in the collinear SPG
$\mathcal{G}_{\mathrm{SO}}^\infty \times {}^1 6 {}^{\bar{1}} m {}^{\bar{1}} m 
= \mathrm{SO(2)}\times ( {}^1 6 + [\theta \,\|\, IC_{2x}] {}^1 6 + [\theta 2_{\perp\bm{n}} \,\|\, E]( {}^1 6 + [\theta \,\|\, IC_{2x}] {}^1 6 ) )$.
All the symmetry operations of ${}^1 6 + [\theta \,\|\, IC_{2x}] {}^1 6$ are listed.
The other operations are omitted except for $[\theta 2_{\perp \bm{n}} \,\|\, E]$.
Among the listed operations, $3C^{\prime}_{2x} (3C^{\prime\prime}_{2y})$ denotes the conjugacy class consisting of three twofold rotations, with $C_{2x} (C_{2y})$ as a representative element.
The absence [presence] of a bar over the IRREPs indicates parity +1 [$-1$] under the antiunitary operation $[\theta 2_{\perp\bm{n}} \,\|\, E]$. 
}
\label{table:Litvin396collinear}

\end{table*}

\begin{table*}
\caption{
IRREPs of multipoles in the collinear SPG
$\mathcal{G}_{\mathrm{SO}}^\infty \times {}^{\bar{1}} 6 {}^1 m {}^{\bar{1}} m 
= \mathrm{SO(2)}\times ( {}^1 3 {}^1 m + [\theta \,\|\, C_{2z}] {}^1 3 {}^1 m + [\theta 2_{\perp\bm{n}} \,\|\, E]( {}^1 3 {}^1 m + [\theta \,\|\, C_{2z}] {}^1 3 {}^1 m ) )$.
All the symmetry operations of ${}^1 3 {}^1 m + [\theta \,\|\, C_{2z}] {}^1 3 {}^1 m$ are listed.
The other operations are omitted except for $[\theta 2_{\perp \bm{n}} \,\|\, E]$.
Among the listed operations, $3C^{\prime}_{2x} (3C^{\prime\prime}_{2y})$ denotes the conjugacy class consisting of three twofold rotations, with $C_{2x} (C_{2y})$ as a representative element.
The absence [presence] of a bar over the IRREPs indicates parity +1 [$-1$] under the antiunitary operation $[\theta 2_{\perp\bm{n}} \,\|\, E]$. 
}
\label{table:Litvin399collinear}

\end{table*}

\begin{table*}
\caption{
IRREPs of multipoles in the collinear SPG
$\mathcal{G}_{\mathrm{SO}}^\infty \times {}^1 \bar{6} {}^{\bar{1}} m {}^{\bar{1}} 2 
= \mathrm{SO(2)}\times ( {}^1 \bar{6} + [\theta \,\|\, C_{2y}] {}^1 \bar{6} + [\theta 2_{\perp\bm{n}} \,\|\, E]( {}^1 \bar{6} + [\theta \,\|\, C_{2y}] {}^1 \bar{6} ) )$.
All the symmetry operations of ${}^1 \bar{6} + [\theta \,\|\, C_{2y}] {}^1 \bar{6}$ are listed.
The other operations are omitted except for $[\theta 2_{\perp \bm{n}} \,\|\, E]$.
Among the listed operations, $3C^{\prime}_{2x} (3C^{\prime\prime}_{2y})$ denotes the conjugacy class consisting of three twofold rotations, with $C_{2x} (C_{2y})$ as a representative element.
The absence [presence] of a bar over the IRREPs indicates parity +1 [$-1$] under the antiunitary operation $[\theta 2_{\perp\bm{n}} \,\|\, E]$. 
}
\label{table:Litvin415collinear}

\end{table*}

\begin{table*}
\caption{
IRREPs of multipoles in the collinear SPG
$\mathcal{G}_{\mathrm{SO}}^\infty \times {}^{\bar{1}} \bar{6} {}^1 m {}^{\bar{1}} 2 
= \mathrm{SO(2)}\times ( {}^1 3 {}^1 m + [\theta \,\|\, IC_{2z}] {}^1 3 {}^1 m + [\theta 2_{\perp\bm{n}} \,\|\, E]( {}^1 3 {}^1 m + [\theta \,\|\, IC_{2z}] {}^1 3 {}^1 m ) )$.
All the symmetry operations of ${}^1 3 {}^1 m + [\theta \,\|\, IC_{2z}] {}^1 3 {}^1 m$ are listed.
The other operations are omitted except for $[\theta 2_{\perp \bm{n}} \,\|\, E]$.
Among the listed operations, $3C^{\prime}_{2x} (3C^{\prime\prime}_{2y})$ denotes the conjugacy class consisting of three twofold rotations, with $C_{2x} (C_{2y})$ as a representative element.
The absence [presence] of a bar over the IRREPs indicates parity +1 [$-1$] under the antiunitary operation $[\theta 2_{\perp\bm{n}} \,\|\, E]$. 
}
\label{table:Litvin418collinear}

\end{table*}

\begin{table*}
\caption{
IRREPs of multipoles in the collinear SPG
$\mathcal{G}_{\mathrm{SO}}^\infty \times {}^{\bar{1}} \bar{6} {}^{\bar{1}} m {}^1 2 
= \mathrm{SO(2)}\times ( {}^1 3 {}^1 2 + [\theta \,\|\, IC_{2z}] {}^1 3 {}^1 2 + [\theta 2_{\perp\bm{n}} \,\|\, E]( {}^1 3 {}^1 2 + [\theta \,\|\, IC_{2z}] {}^1 3 {}^1 2 ) )$.
All the symmetry operations of ${}^1 3 {}^1 2 + [\theta \,\|\, IC_{2z}] {}^1 3 {}^1 2$ are listed.
The other operations are omitted except for $[\theta 2_{\perp \bm{n}} \,\|\, E]$.
Among the listed operations, $3C^{\prime}_{2x} (3C^{\prime\prime}_{2y})$ denotes the conjugacy class consisting of three twofold rotations, with $C_{2x} (C_{2y})$ as a representative element.
The absence [presence] of a bar over the IRREPs indicates parity +1 [$-1$] under the antiunitary operation $[\theta 2_{\perp\bm{n}} \,\|\, E]$. 
}
\label{table:Litvin421collinear}

\end{table*}

\begin{table*}
\caption{
IRREPs of multipoles in the collinear SPG
$\mathcal{G}_{\mathrm{SO}}^\infty \times {}^{\bar{1}} 6 / {}^{\bar{1}} m {}^1 m {}^{\bar{1}} m 
= \mathrm{SO(2)}\times ( {}^1 \bar{3} {}^1 m + [\theta \,\|\, C_{2z}] {}^1 \bar{3} {}^1 m + [\theta 2_{\perp\bm{n}} \,\|\, E]( {}^1 \bar{3} {}^1 m + [\theta \,\|\, C_{2z}] {}^1 \bar{3} {}^1 m ) )$.
All the symmetry operations of ${}^1 \bar{3} {}^1 m$ are listed.
The other operations are omitted except for $[\theta \,\|\, C_{2z}]$ and $[\theta 2_{\perp \bm{n}} \,\|\, E]$.
Among the listed operations, $3C^{\prime}_{2x}$ denotes the conjugacy class consisting of three twofold rotations, with $C_{2x}$ as a representative element.
The absence [presence] of a bar over the IRREPs indicates parity +1 [$-1$] under the antiunitary operation $[\theta 2_{\perp\bm{n}} \,\|\, E]$. 
}
\label{table:Litvin443collinear}

\end{table*}

\begin{table*}
\caption{
IRREPs of multipoles in the collinear SPG
$\mathcal{G}_{\mathrm{SO}}^\infty \times {}^{\bar{1}} 6 / {}^1 m {}^1 m {}^{\bar{1}} m 
= \mathrm{SO(2)}\times ( {}^1 \bar{6} {}^1 m {}^1 2 + [\theta \,\|\, I] {}^1 \bar{6} {}^1 m {}^1 2 + [\theta 2_{\perp\bm{n}} \,\|\, E]( {}^1 \bar{6} {}^1 m {}^1 2 + [\theta \,\|\, I] {}^1 \bar{6} {}^1 m {}^1 2 ) )$.
All the symmetry operations of ${}^1 \bar{6} {}^1 m {}^1 2$ are listed.
The other operations are omitted except for $[\theta \,\|\, I]$ and $[\theta 2_{\perp \bm{n}} \,\|\, E]$.
Among the listed operations, $3C^{\prime}_{2x} (3C^{\prime\prime}_{2y})$ denotes the conjugacy class consisting of three twofold rotations, with $C_{2x} (C_{2y})$ as a representative element.
The absence [presence] of a bar over the IRREPs indicates parity +1 [$-1$] under the antiunitary operation $[\theta 2_{\perp\bm{n}} \,\|\, E]$. 
}
\label{table:Litvin446collinear}

\end{table*}

\begin{table*}
\caption{
IRREPs of multipoles in the collinear SPG
$\mathcal{G}_{\mathrm{SO}}^\infty \times {}^1 6 / {}^1 m {}^{\bar{1}} m {}^{\bar{1}} m 
= \mathrm{SO(2)}\times ( {}^1 6 / {}^1 m + [\theta \,\|\, C_{2x}] {}^1 6 / {}^1 m + [\theta 2_{\perp\bm{n}} \,\|\, E]( {}^1 6 / {}^1 m + [\theta \,\|\, C_{2x}] {}^1 6 / {}^1 m ) )$.
All the symmetry operations of ${}^1 6 / {}^1 m$ are listed.
The other operations are omitted except for $[\theta \,\|\, C_{2x}]$ and $[\theta 2_{\perp \bm{n}} \,\|\, E]$.
The absence [presence] of a bar over the IRREPs indicates parity +1 [$-1$] under the antiunitary operation $[\theta 2_{\perp\bm{n}} \,\|\, E]$. 
}
\label{table:Litvin449collinear}

\end{table*}

\begin{table*}
\caption{
IRREPs of multipoles in the collinear SPG
$\mathcal{G}_{\mathrm{SO}}^\infty \times {}^1 6 / {}^{\bar{1}} m {}^1 m {}^1 m 
= \mathrm{SO(2)}\times ( {}^1 6 {}^1 m {}^1 m + [\theta \,\|\, I] {}^1 6 {}^1 m {}^1 m + [\theta 2_{\perp\bm{n}} \,\|\, E]( {}^1 6 {}^1 m {}^1 m + [\theta \,\|\, I] {}^1 6 {}^1 m {}^1 m ) )$.
All the symmetry operations of ${}^1 6 {}^1 m {}^1 m$ are listed.
The other operations are omitted except for $[\theta \,\|\, I]$ and $[\theta 2_{\perp \bm{n}} \,\|\, E]$.
Among the listed operations, $3C^{\prime}_{2x} (3C^{\prime\prime}_{2y})$ denotes the conjugacy class consisting of three twofold rotations, with $C_{2x} (C_{2y})$ as a representative element.
The absence [presence] of a bar over the IRREPs indicates parity +1 [$-1$] under the antiunitary operation $[\theta 2_{\perp\bm{n}} \,\|\, E]$. 
}
\label{table:Litvin452collinear}

\end{table*}

\begin{table*}
\caption{
IRREPs of multipoles in the collinear SPG
$\mathcal{G}_{\mathrm{SO}}^\infty \times {}^1 6 / {}^{\bar{1}} m {}^{\bar{1}}  m {}^{\bar{1}}  m 
= \mathrm{SO(2)}\times ( {}^1 6 {}^1 2 {}^1 2 + [\theta \,\|\, I] {}^1 6 {}^1 2 {}^1 2 + [\theta 2_{\perp\bm{n}} \,\|\, E]( {}^1 6 {}^1 2 {}^1 2 + [\theta \,\|\, I] {}^1 6 {}^1 2 {}^1 2 ) )$.
All the symmetry operations of ${}^1 6 {}^1 2 {}^1 2$ are listed.
The other operations are omitted except for $[\theta \,\|\, I]$ and $[\theta 2_{\perp \bm{n}} \,\|\, E]$.
Among the listed operations, $3C^{\prime}_{2x} (3C^{\prime\prime}_{2y})$ denotes the conjugacy class consisting of three twofold rotations, with $C_{2x} (C_{2y})$ as a representative element.
The absence [presence] of a bar over the IRREPs indicates parity +1 [$-1$] under the antiunitary operation $[\theta 2_{\perp\bm{n}} \,\|\, E]$. 
}
\label{table:Litvin455collinear}

\end{table*}

\input{appendix_tables_Tsymmetric_collinearSPG}

\clearpage 

\subsection{Nonmagnetic spin point groups \label{sec:ap_classification_nonmagSPG}}

The complete classification of orbital and spin multipoles for the 32 nonmagnetic SPGs is summarized in Tables~\ref{table:Litvin2nonmag} -- ~\ref{table:Litvin440nonmag}.

\renewcommand{\arraystretch}{1.8}
\begin{table*}[t]
\begin{center}
\caption{IRREPs of multipoles in the nonmagnetic SPG 
$\mathcal{G}_{\mathrm{SO}}^{\rm NM} \times {}^1 \bar{1} = {\rm SO(3)} \times ({}^1 \bar{1} + [\theta \,\|\, E] {}^1 \bar{1} )$.
All the symmetry operations of ${}^1 \bar{1} + [\theta \,\|\, E] {}^1 \bar{1}$ are listed.
The parity $\pm 1$ under the antiunitary operation $[\theta \,\|\, E]$ is attached on the IRREPs.
}
\label{table:Litvin2nonmag}

\end{center}
\end{table*}

\begin{table*}
\caption{
IRREPs of multipoles in the nonmagnetic SPG
$\mathcal{G}_{\mathrm{SO}}^{\rm NM} \times {}^1 2 
= {\rm SO(3)}\times ( {}^1 2 + [\theta \,\|\, E] {}^1 2 )$.
All the symmetry operations of ${}^1 2 + [\theta \,\|\, E] {}^1 2 $ are listed.
The parity $\pm 1$ under the antiunitary operation $[\theta \,\|\, E]$ is attached on the IRREPs.
}
\label{table:Litvin6nonmag}
%
\end{table*}

\begin{table*}
\caption{
IRREPs of multipoles in the nonmagnetic SPG
$\mathcal{G}_{\mathrm{SO}}^{\rm NM} \times {}^1 m 
= {\rm SO(3)}\times ( {}^1 m + [\theta \,\|\, E] {}^1 m )$.
All the symmetry operations of ${}^1 m + [\theta \,\|\, E] {}^1 m $ are listed.
The parity $\pm 1$ under the antiunitary operation $[\theta \,\|\, E]$ is attached on the IRREPs.
}
\label{table:Litvin10nonmag}
%
\end{table*}

\begin{table*}
\caption{
IRREPs of multipoles in the nonmagnetic SPG
$\mathcal{G}_{\mathrm{SO}}^{\rm NM} \times {}^1 2 / {}^1 m 
= {\rm SO(3)}\times ( {}^1 2 / {}^1 m + [\theta \,\|\, E] {}^1 2 / {}^1 m )$.
All the symmetry operations of ${}^1 2 / {}^1 m + [\theta \,\|\, E] {}^1 2 / {}^1 m $ are listed.
The parity $\pm 1$ under the antiunitary operation $[\theta \,\|\, E]$ is attached on the IRREPs.
}
\label{table:Litvin14nonmag}
%
\end{table*}

\begin{table*}
\caption{
IRREPs of multipoles in the nonmagnetic SPG
$\mathcal{G}_{\mathrm{SO}}^{\rm NM} \times {}^1 m {}^1 m {}^1 2 
= {\rm SO(3)}\times ( {}^1 m {}^1 m {}^1 2 + [\theta \,\|\, E] {}^1 m {}^1 m {}^1 2 )$.
All the symmetry operations of ${}^1 m {}^1 m {}^1 2 + [\theta \,\|\, E] {}^1 m {}^1 m {}^1 2 $ are listed.
The parity $\pm 1$ under the antiunitary operation $[\theta \,\|\, E]$ is attached on the IRREPs.
}
\label{table:Litvin34nonmag}
%
\end{table*}

\begin{table*}
\caption{
IRREPs of multipoles in the nonmagnetic SPG
$\mathcal{G}_{\mathrm{SO}}^{\rm NM} \times {}^1 2 {}^1 2 {}^1 2 
= {\rm SO(3)}\times ( {}^1 2 {}^1 2 {}^1 2 + [\theta \,\|\, E] {}^1 2 {}^1 2 {}^1 2 )$.
All the symmetry operations of ${}^1 2 {}^1 2 {}^1 2 + [\theta \,\|\, E] {}^1 2 {}^1 2 {}^1 2 $ are listed.
The parity $\pm 1$ under the antiunitary operation $[\theta \,\|\, E]$ is attached on the IRREPs.
}
\label{table:Litvin47nonmag}
%
\end{table*}

\begin{table*}
\caption{
IRREPs of multipoles in the nonmagnetic SPG
$\mathcal{G}_{\mathrm{SO}}^{\rm NM} \times {}^1 m {}^1 m {}^1 m 
= {\rm SO(3)}\times ( {}^1 m {}^1 m {}^1 m + [\theta \,\|\, E] {}^1 m {}^1 m {}^1 m )$.
All the symmetry operations of ${}^1 m {}^1 m {}^1 m$ are listed.
The other operations are omitted except for $[\theta \,\|\, E]$.
The parity $\pm 1$ under the antiunitary operation $[\theta \,\|\, E]$ is attached on the IRREPs.
}
\label{table:Litvin54nonmag}
%
\end{table*}

\clearpage 

\begin{table*}
\caption{
IRREPs of multipoles in the nonmagnetic SPG
$\mathcal{G}_{\mathrm{SO}}^{\rm NM} \times {}^1 4 
= {\rm SO(3)}\times ( {}^1 4 + [\theta \,\|\, E] {}^1 4 )$.
All the symmetry operations of ${}^1 4 + [\theta \,\|\, E] {}^1 4$ are listed.
The parity $\pm 1$ under the antiunitary operation $[\theta \,\|\, E]$ is attached on the IRREPs.
}
\label{table:Litvin90nonmag}
%
\end{table*}

\begin{table*}
\caption{
IRREPs of multipoles in the nonmagnetic SPG
$\mathcal{G}_{\mathrm{SO}}^{\rm NM} \times {}^1 \bar{4} 
= {\rm SO(3)}\times ( {}^1 \bar{4} + [\theta \,\|\, E] {}^1 \bar{4} )$.
All the symmetry operations of ${}^1 \bar{4} + [\theta \,\|\, E] {}^1 \bar{4}$ are listed.
The parity $\pm 1$ under the antiunitary operation $[\theta \,\|\, E]$ is attached on the IRREPs.
}
\label{table:Litvin96nonmag}
%
\end{table*}

\begin{table*}
\caption{
IRREPs of multipoles in the nonmagnetic SPG
$\mathcal{G}_{\mathrm{SO}}^{\rm NM} \times {}^1 4 / {}^1 m 
= {\rm SO(3)}\times ( {}^1 4 / {}^1 m + [\theta \,\|\, E] {}^1 4 / {}^1 m )$.
All the symmetry operations of ${}^1 4 / {}^1 m + [\theta \,\|\, E] {}^1 4 / {}^1 m$ are listed.
The parity $\pm 1$ under the antiunitary operation $[\theta \,\|\, E]$ is attached on the IRREPs.
}
\label{table:Litvin102nonmag}
%
\end{table*}

\begin{table*}
\caption{
IRREPs of multipoles in the nonmagnetic SPG
$\mathcal{G}_{\mathrm{SO}}^{\rm NM} \times {}^1 4 {}^1 2 {}^1 2 
= {\rm SO(3)}\times ( {}^1 4 {}^1 2 {}^1 2 + [\theta \,\|\, E] {}^1 4 {}^1 2 {}^1 2 )$.
All the symmetry operations of ${}^1 4 {}^1 2 {}^1 2 + [\theta \,\|\, E] {}^1 4 {}^1 2 {}^1 2$ are listed.
Among the listed operations, $2C^{\prime}_{2x}$ denotes the conjugacy class consisting of two twofold rotations, with $C_{2x}$ as a representative element ($2C^{\prime}_{2x} \equiv C_{2x}, C_{2y}$).
The parity $\pm 1$ under the antiunitary operation $[\theta \,\|\, E]$ is attached on the IRREPs.
}
\label{table:Litvin130nonmag}

\end{table*}

\begin{table*}
\caption{
IRREPs of multipoles in the nonmagnetic SPG
$\mathcal{G}_{\mathrm{SO}}^{\rm NM} \times {}^1 4 {}^1 m {}^1 m 
= {\rm SO(3)}\times ( {}^1 4 {}^1 m {}^1 m + [\theta \,\|\, E] {}^1 4 {}^1 m {}^1 m )$.
All the symmetry operations of ${}^1 4 {}^1 m {}^1 m + [\theta \,\|\, E] {}^1 4 {}^1 m {}^1 m$ are listed. 
Among the listed operations, $2C^{\prime}_{2x}$ denotes the conjugacy class consisting of two twofold rotations, with $C_{2x}$ as a representative element ($2C^{\prime}_{2x} \equiv C_{2x}, C_{2y}$).
The parity $\pm 1$ under the antiunitary operation $[\theta \,\|\, E]$ is attached on the IRREPs.
}
\label{table:Litvin146nonmag}

\end{table*}

\begin{table*}
\caption{
IRREPs of multipoles in the nonmagnetic SPG
$\mathcal{G}_{\mathrm{SO}}^{\rm NM} \times {}^1 \bar{4} {}^1 2 {}^1 m 
= {\rm SO(3)}\times ( {}^1 \bar{4} {}^1 2 {}^1 m + [\theta \,\|\, E] {}^1 \bar{4} {}^1 2 {}^1 m )$.
All the symmetry operations of ${}^1 \bar{4} {}^1 2 {}^1 m + [\theta \,\|\, E] {}^1 \bar{4} {}^1 2 {}^1 m$ are listed.
Among the listed operations, $2C^{\prime}_{2x}$ denotes the conjugacy class consisting of two twofold rotations, with $C_{2x}$ as a representative element ($2C^{\prime}_{2x} \equiv C_{2x}, C_{2y}$).
The parity $\pm 1$ under the antiunitary operation $[\theta \,\|\, E]$ is attached on the IRREPs.
}
\label{table:Litvin162nonmag}

\end{table*}

\begin{table*}
\caption{
IRREPs of multipoles in the nonmagnetic SPG
$\mathcal{G}_{\mathrm{SO}}^{\rm NM} \times {}^1 4 / {}^1 m {}^1 m {}^1 m 
= {\rm SO(3)}\times ( {}^1 4 / {}^1 m {}^1 m {}^1 m + [\theta \,\|\, E] {}^1 4 / {}^1 m {}^1 m {}^1 m )$.
All the symmetry operations of ${}^1 4 / {}^1 m {}^1 m {}^1 m$ are listed.
The other operations are omitted except for $[\theta \,\|\, E]$.
Among the listed operations, $2C^{\prime}_{2x}$ denotes the conjugacy class consisting of two twofold rotations, with $C_{2x}$ as a representative element ($2C^{\prime}_{2x} \equiv C_{2x}, C_{2y}$).
The parity $\pm 1$ under the antiunitary operation $[\theta \,\|\, E]$ is attached on the IRREPs.
}
\label{table:Litvin186nonmag}

\end{table*}

\clearpage 

\begin{table*}
\caption{
IRREPs of multipoles in the nonmagnetic SPG
$\mathcal{G}_{\mathrm{SO}}^{\rm NM} \times {}^1 2 {}^1 3 
= {\rm SO(3)}\times ( {}^1 2 {}^1 3 + [\theta \,\|\, E]  {}^1 2 {}^1 3 )$.
All the symmetry operations of ${}^1 2 {}^1 3 + [\theta \,\|\, E]  {}^1 2 {}^1 3 $ are listed.
The parity $\pm 1$ under the antiunitary operation $[\theta \,\|\, E]$ is attached on the IRREPs.
}
\label{table:Litvin538nonmag}
%
\end{table*}

\begin{table*}
\caption{
IRREPs of multipoles in the nonmagnetic SPG
$\mathcal{G}_{\mathrm{SO}}^{\rm NM} \times {}^1 2 / {}^1 m {}^1 \bar{3} 
= {\rm SO(3)}\times ( {}^1 m {}^1 \bar{3} + [\theta \,\|\, E]  {}^1 m {}^1 \bar{3} )$.
All the symmetry operations of ${}^1 2 / {}^1 m {}^1 \bar{3} + [\theta \,\|\, E]  {}^1 2 / {}^1 m {}^1 \bar{3}$ are listed.
The parity $\pm 1$ under the antiunitary operation $[\theta \,\|\, E]$ is attached on the IRREPs.
}
\label{table:Litvin541nonmag}
%
\end{table*}

\begin{table*}
\caption{
IRREPs of multipoles in the nonmagnetic SPG
$\mathcal{G}_{\mathrm{SO}}^{\rm NM} \times {}^1 \bar{4} {}^1 3 {}^1 m 
= {\rm SO(3)}\times ( {}^1 \bar{4} {}^1 3 {}^1 m + [\theta \,\|\, E] {}^1 \bar{4} {}^1 3 {}^1 m )$.
All the symmetry operations of ${}^1 \bar{4} {}^1 3 {}^1 m + [\theta \,\|\, E] {}^1 \bar{4} {}^1 3 {}^1 m$ are listed.
The parity $\pm 1$ under the antiunitary operation $[\theta \,\|\, E]$ is attached on the IRREPs.
}
\label{table:Litvin551nonmag}
%
\end{table*}

\begin{table*}
\caption{
IRREPs of multipoles in the nonmagnetic SPG
$\mathcal{G}_{\mathrm{SO}}^{\rm NM} \times {}^1 4 {}^1 3 {}^1 2 
= {\rm SO(3)}\times ( {}^1 4 {}^1 3 {}^1 2 + [\theta \,\|\, E] {}^1 4 {}^1 3 {}^1 2 )$.
All the symmetry operations of ${}^1 4 {}^1 3 {}^1 2 + [\theta \,\|\, E] {}^1 4 {}^1 3 {}^1 2$ are listed.
The parity $\pm 1$ under the antiunitary operation $[\theta \,\|\, E]$ is attached on the IRREPs.
}
\label{table:Litvin559nonmag}
%
\end{table*}

\begin{table*}
\caption{
IRREPs of multipoles in the nonmagnetic SPG
$\mathcal{G}_{\mathrm{SO}}^{\rm NM} \times {}^1 4 / {}^1 m {}^1 \bar{3} {}^1 2 / {}^{1} m
= {\rm SO(3)}\times ( {}^1 4 / {}^1 m {}^1 \bar{3} {}^1 2 / {}^1 m + [\theta \,\|\, E] {}^1 4 / {}^1 m {}^1 \bar{3} {}^1 2 / {}^1 m )$.
All the symmetry operations of ${}^1 4 / {}^1 m {}^1 \bar{3} {}^1 2 / {}^1 m$ are listed.
The other operations are omitted except for $[\theta \,\|\, E]$.
The parity $\pm 1$ under the antiunitary operation $[\theta \,\|\, E]$ is attached on the IRREPs.
}
\label{table:Litvin567nonmag}
%
\end{table*}

\clearpage 

\begin{table*}
\caption{
IRREPs of multipoles in the nonmagnetic SPG
$\mathcal{G}_{\mathrm{SO}}^{\rm NM} \times {}^1 3 
= {\rm SO(3)}\times ( {}^1 3 + [\theta \,\|\, E] {}^1 3 )$.
All the symmetry operations of ${}^1 3 + [\theta \,\|\, E] {}^1 3$ are listed.
The parity $\pm 1$ under the antiunitary operation $[\theta \,\|\, E]$ is attached on the IRREPs.
}
\label{table:Litvin272nonmag}
%
\end{table*}

\begin{table*}
\caption{
IRREPs of multipoles in the nonmagnetic SPG
$\mathcal{G}_{\mathrm{SO}}^{\rm NM} \times {}^1 \bar{3} 
= {\rm SO(3)}\times ( {}^1 \bar{3} + [\theta \,\|\, E] {}^1 \bar{3} )$.
All the symmetry operations of ${}^1 \bar{3} + [\theta \,\|\, E] {}^1 \bar{3}$ are listed.
The parity $\pm 1$ under the antiunitary operation $[\theta \,\|\, E]$ is attached on the IRREPs.
}
\label{table:Litvin274nonmag}
%
\end{table*}

\begin{table*}
\caption{
IRREPs of multipoles in the nonmagnetic SPG
$\mathcal{G}_{\mathrm{SO}}^{\rm NM} \times {}^1 3 {}^1 2 
= {\rm SO(3)}\times ( {}^1 3 {}^1 2 + [\theta \,\|\, E] {}^1 3 {}^1 2 )$.
All the symmetry operations of ${}^1 3 {}^1 2 + [\theta \,\|\, E] {}^1 3 {}^1 2$ are listed.
Among the listed operations, $3C^{\prime}_{2y}$ denotes the conjugacy class consisting of three twofold rotations, with $C_{2y}$ as a representative element.
The parity $\pm 1$ under the antiunitary operation $[\theta \,\|\, E]$ is attached on the IRREPs.
}
\label{table:Litvin282nonmag}

\end{table*}

\begin{table*}
\caption{
IRREPs of multipoles in the nonmagnetic SPG
$\mathcal{G}_{\mathrm{SO}}^{\rm NM} \times {}^1 3 {}^1 m 
= {\rm SO(3)}\times ( {}^1 3 {}^1 m + [\theta \,\|\, E] {}^1 3 {}^1 m )$.
All the symmetry operations of ${}^1 3 {}^1 m + [\theta \,\|\, E] {}^1 3 {}^1 m$ are listed.
Among the listed operations, $3C^{\prime}_{2y}$ denotes the conjugacy class consisting of three twofold rotations, with $C_{2y}$ as a representative element.
The parity $\pm 1$ under the antiunitary operation $[\theta \,\|\, E]$ is attached on the IRREPs.
}
\label{table:Litvin288nonmag}

\end{table*}

\begin{table*}
\caption{
IRREPs of multipoles in the nonmagnetic SPG
$\mathcal{G}_{\mathrm{SO}}^{\rm NM} \times {}^1 \bar{3} {}^1 m 
= {\rm SO(3)}\times ( {}^1 \bar{3} {}^1 m + [\theta \,\|\, E] {}^1 \bar{3} {}^1 m )$.
All the symmetry operations of ${}^1 \bar{3} {}^1 m + [\theta \,\|\, E] {}^1 \bar{3} {}^1 m$ are listed.
Among the listed operations, $3C^{\prime}_{2y}$ denotes the conjugacy class consisting of three twofold rotations, with $C_{2y}$ as a representative element.
The parity $\pm 1$ under the antiunitary operation $[\theta \,\|\, E]$ is attached on the IRREPs.
}
\label{table:Litvin294nonmag}

\end{table*}

\clearpage 

\begin{table*}
\caption{
IRREPs of multipoles in the nonmagnetic SPG
$\mathcal{G}_{\mathrm{SO}}^{\rm NM} \times {}^1 6 
= {\rm SO(3)}\times ( {}^1 6 + [\theta \,\|\, E] {}^1 6 )$.
All the symmetry operations of ${}^1 6 + [\theta \,\|\, E] {}^1 6$ are listed.
The parity $\pm 1$ under the antiunitary operation $[\theta \,\|\, E]$ is attached on the IRREPs.
}
\label{table:Litvin330nonmag}
%
\end{table*}

\begin{table*}
\caption{
IRREPs of multipoles in the nonmagnetic SPG
$\mathcal{G}_{\mathrm{SO}}^{\rm NM} \times {}^1 \bar{6} 
= {\rm SO(3)}\times ( {}^1 \bar{6} + [\theta \,\|\, E] {}^1 \bar{6} )$.
All the symmetry operations of ${}^1 \bar{6} + [\theta \,\|\, E] {}^1 \bar{6}$ are listed.
The parity $\pm 1$ under the antiunitary operation $[\theta \,\|\, E]$ is attached on the IRREPs.
}
\label{table:Litvin322nonmag}
%
\end{table*}

\begin{table*}
\caption{
IRREPs of multipoles in the nonmagnetic SPG
$\mathcal{G}_{\mathrm{SO}}^{\rm NM} \times {}^1 6 / {}^1 m 
= {\rm SO(3)}\times ( {}^1 6 / {}^1 m + [\theta \,\|\, E] {}^1 6 / {}^1 m )$.
All the symmetry operations of ${}^1 6 / {}^1 m + [\theta \,\|\, E] {}^1 6 / {}^1 m$ are listed.
The parity $\pm 1$ under the antiunitary operation $[\theta \,\|\, E]$ is attached on the IRREPs.
}
\label{table:Litvin357nonmag}
%
\end{table*}

\begin{table*}
\caption{
IRREPs of multipoles in the nonmagnetic SPG
$\mathcal{G}_{\mathrm{SO}}^{\rm NM} \times {}^1 6 {}^1 2 {}^1 2 
= {\rm SO(3)}\times ( {}^1 6 {}^1 2 {}^1 2 + [\theta \,\|\, E] {}^1 6 {}^1 2 {}^1 2 )$.
All the symmetry operations of ${}^1 6 {}^1 2 {}^1 2 + [\theta \,\|\, E] {}^1 6 {}^1 2 {}^1 2$ are listed.
Among the listed operations, $3C^{\prime}_{2x} (3C^{\prime\prime}_{2y})$ denotes the conjugacy class consisting of three twofold rotations, with $C_{2x} (C_{2y})$ as a representative element.
The parity $\pm 1$ under the antiunitary operation $[\theta \,\|\, E]$ is attached on the IRREPs.
}
\label{table:Litvin338nonmag}

\end{table*}

\begin{table*}
\caption{
IRREPs of multipoles in the nonmagnetic SPG
$\mathcal{G}_{\mathrm{SO}}^{\rm NM} \times {}^1 6 {}^1 m {}^1 m 
= {\rm SO(3)}\times ( {}^1 6 {}^1 m {}^1 m + [\theta \,\|\, E] {}^1 6 {}^1 m {}^1 m )$.
All the symmetry operations of ${}^1 6 {}^1 m {}^1 m + [\theta \,\|\, E] {}^1 6 {}^1 m {}^1 m$ are listed.
Among the listed operations, $3C^{\prime}_{2x} (3C^{\prime\prime}_{2y})$ denotes the conjugacy class consisting of three twofold rotations, with $C_{2x} (C_{2y})$ as a representative element.
The parity $\pm 1$ under the antiunitary operation $[\theta \,\|\, E]$ is attached on the IRREPs.
}
\label{table:Litvin393nonmag}

\end{table*}

\begin{table*}
\caption{
IRREPs of multipoles in the nonmagnetic SPG
$\mathcal{G}_{\mathrm{SO}}^{\rm NM} \times {}^1 \bar{6} {}^1 m {}^1 2 
= {\rm SO(3)}\times ( {}^1 \bar{6} {}^1 m {}^1 2 + [\theta \,\|\, E] {}^1 \bar{6} {}^1 m {}^1 2 )$.
All the symmetry operations of ${}^1 \bar{6} {}^1 m {}^1 2 + [\theta \,\|\, E] {}^1 \bar{6} {}^1 m {}^1 2$ are listed.
Among the listed operations, $3C^{\prime}_{2x} (3C^{\prime\prime}_{2y})$ denotes the conjugacy class consisting of three twofold rotations, with $C_{2x} (C_{2y})$ as a representative element.
The parity $\pm 1$ under the antiunitary operation $[\theta \,\|\, E]$ is attached on the IRREPs.
}
\label{table:Litvin412nonmag}

\end{table*}

\begin{table*}
\caption{
IRREPs of multipoles in the nonmagnetic SPG
$\mathcal{G}_{\mathrm{SO}}^{\rm NM} \times {}^1 6 / {}^1 m {}^1 m {}^1 m 
= {\rm SO(3)}\times ( {}^1 6 / {}^1 m {}^1 m {}^1 m + [\theta \,\|\, E] {}^1 6 / {}^1 m {}^1 m {}^1 m )$.
All the symmetry operations of ${}^1 6 / {}^1 m {}^1 m {}^1 m$ are listed.
The other operations are omitted except for $[\theta \,\|\, E]$.
Among the listed operations, $3C^{\prime}_{2x} (3C^{\prime\prime}_{2y})$ denotes the conjugacy class consisting of three twofold rotations, with $C_{2x} (C_{2y})$ as a representative element.
The parity $\pm 1$ under the antiunitary operation $[\theta \,\|\, E]$ is attached on the IRREPs.
}
\label{table:Litvin440nonmag}

\end{table*}

\clearpage


\section{Collinear spin Laue group \label{sec:ap_Laue}}

The correspondence between collinear spin Laue groups and collinear spin point groups is listed in Tables~\ref{table:SLG_withPT} and \ref{SLG_withoutPT}.
Table~\ref{table:SLG_withPT} presents the spin Laue group with the $\mathcal{T}$ and/or $\mathcal{PT}$ symmetry, whereas Table~\ref{SLG_withoutPT} presents that without the $\mathcal{PT}$ symmetry.

\renewcommand{\arraystretch}{1.5}
\begin{table*}[h!]
\centering
\caption{
Collinear spin Laue group (SLG) for the $\mathcal{T}$-symmetric collinear spin point group and the $\mathcal{PT}$-symmetric collinear spin point group. 
 \label{table:SLG_withPT}}
\vspace{2mm}

\end{table*}

\begin{table*}[h!]
\centering
\caption{
Collinear spin Laue group (SLG) for the $\mathcal{PT}$-breaking collinear spin point group.
 \label{SLG_withoutPT}}
\vspace{2mm}
%
\end{table*}

\clearpage

\bibliography{ref}
\end{document}